\documentclass[aps, pra, twocolumn, superscriptaddress, amsmath,  tightenlines, longbibliography]{revtex4-1}

\usepackage{booktabs}
\usepackage{mathtools}
\usepackage{dcolumn}
\usepackage{graphicx}
\usepackage{mathrsfs}
\usepackage{subfigure}
\usepackage{booktabs}
\usepackage{amsmath}
\usepackage{physics}
\usepackage{dsfont}
\usepackage{amstext}
\usepackage{amssymb}
\usepackage{amsbsy}
\usepackage{bbm}
\usepackage{amsthm}
\usepackage{graphicx}
\usepackage{color}
\usepackage[colorlinks,citecolor=blue]{hyperref}

\usepackage{url}
\usepackage[colorlinks]{hyperref}
\hypersetup{%
	plainpages=true,
	breaklinks=true,       
	hypertexnames=false,  
	pageanchor=true,
	colorlinks=true,
	linkcolor={blue},
	citecolor={red},
	urlcolor={blue},
	anchorcolor={black}
}

 \makeatletter

\newcommand{\Rmnum}[1]{\expandafter\@slowromancap\romannumeral #1@}
\makeatother

\begin{document}

\title{Quantum-Squeezing-Induced Algebraic Non-Hermitian Skin  \\ Effects and Ultra Spectral  Sensitivity}
\author{Zhao-Fan Cai}
\affiliation{School of Physics and Optoelectronics, South China University of Technology,  Guangzhou 510640, China}
\author{Tao Liu}
\email[E-mail: ]{liutao0716@scut.edu.cn}
\affiliation{School of Physics and Optoelectronics, South China University of Technology,  Guangzhou 510640, China}

\date{{\small \today}}


\begin{abstract}
The well-established non-Bloch band theory predicts exponential localization of skin-mode eigenstates in one-dimensional (1D) non-Hermitian systems. Recent studies, however, have uncovered anomalous algebraic localization in higher dimensions. Here, we extend these ideas to Hermitian bosonic quadratic Hamiltonians incorporating quantum squeezing, offering a genuine quantum framework to explore non-Hermitian phenomena without external reservoirs. We construct a two-dimensional (2D) bosonic lattice model with two-mode squeezing and study the spectral properties of its bosonic excitation within the Bogoliubov–de Gennes (BdG) formalism. We demonstrate an algebraic non-Hermitian skin effect (NHSE), characterized by quasi-long-range power-law localization of complex eigenstates. The system shows ultra spectral sensitivity to double infinitesimal on-site and long-range hopping impurities, while remaining insensitive to single impurity. Analytical treatment via the Green’s function reveals that this sensitivity originates from the divergence of the nonlocal Green’s function associated with the formation of nonlocal bound states. Our study establishes a framework for realizing novel higher-dimensional non-Hermitian physics in Hermitian bosonic platforms such as superconducting circuits, photonic lattices, and optomechanical arrays, with the demonstrated ultra spectral sensitivity enabling quantum sensing and amplification via bosonic squeezing.
\end{abstract}

\maketitle

\section{Introduction}

In recent years, non-Hermitian Hamiltonians have attracted considerable interest for their ability to host exotic physical phenomena absent in their Hermitian counterparts \cite{Ashida2020, PhysRevA.101.062112,RevModPhys.93.015005, PhysRevResearch.5.043105,Leefmans2022, PhysRevA.109.023317,PhysRevB.110.024205,Reisenbauer2024, Ochkan2024, PhysRevLett.132.050402,PhysRevLett.134.196302, PhysRevB.109.165127,z9m1-3mwb,PhysRevResearch.7.L022037,PhysRevLett.133.136503,PhysRevLett.133.076502, Yang2025,  PhysRevResearch.5.033058,PhysRevLett.131.076401,PhysRevLett.131.116601,PhysRevLett.129.223601,Cai2025,PhysRevB.111.235412,PhysRevLett.134.156601, q6wr-2rt9,Hashemi2025,arxiv.2507.09447,q4nh-m1jh,arxiv.2504.18926,Wang2025,dl59-vl7v,vs7x-clqd}. These effects have been extensively investigated across diverse classical and quantum platforms, including ultracold atoms \cite{Ren2022}, superconducting qubits \cite{Naghiloo2019}, quantum \cite{Shen2025} and electrical \cite{Imhof2018} circuits, optical waveguides \cite{ElGanainy2018}, and acoustic metamaterials \cite{PhysRevLett.134.176601}. Among the defining features of non-Hermitian systems is the non-Hermitian skin effect (NHSE), characterized by the accumulation of a macroscopic number of eigenstates at the boundaries \cite{ShunyuYao2018,PhysRevLett.123.066404,PhysRevLett.125.126402,arxiv.2407.01296, PhysRevLett.122.076801, PhysRevLett.121.026808,PhysRevB.111.205418,arXiv:2408.12451,arXiv:1802.07964,PhysRevLett.125.126402,PhysRevLett.123.066404, YaoarXiv:1804.04672,PhysRevLett.121.026808,PhysRevLett.122.076801,PhysRevLett.123.170401, PhysRevLett.123.206404,PhysRevLett.123.066405,PhysRevLett.123.206404, ZhangJ2018,  PhysRevB.100.054105, PhysRevA.100.062131,Zhao2019, PhysRevX.9.041015,PhysRevLett.124.056802, PhysRevB.97.121401,   PhysRevA.102.033715,  PhysRevLett.124.086801, PhysRevLett.127.196801, PhysRevLett.128.223903,   PhysRevLett.129.093001,arXiv:2505.05058,PhysRevAL061701, PhysRevLett.131.036402,haowang2025, arXiv:2403.07459,arXiv:2311.03777,  Parto2023, PhysRevX.14.021011,  Guo2024,1bvp-p2cz}, giving rise to an anomalous spectral response that is extraordinarily sensitive to boundary conditions. The NHSE has been shown to give rise to a variety of critical phenomena \cite{Li2020,PhysRevB.108.L161409,PhysRevA.109.063329,PhysRevX.13.021007} and anomalous dynamical behaviors \cite{ PhysRevLett.128.120401,arxiv.2503.13671,PhysRevLett.133.136602,PhysRevB.107.L140302,PhysRevLett.133.070801,arxiv.2503.11505} in one-dimensional (1D) systems, lying beyond the scope of the conventional Bloch band description. These effects are now well understood within the generalized Bloch framework, formulated in terms of the generalized Brillouin zone (GBZ) \cite{ShunyuYao2018,PhysRevLett.123.066404,PhysRevLett.125.126402,arxiv.2407.01296}.

The NHSE has been thoroughly explored in 1D systems. When generalized to higher dimensions, however, recent studies have revealed an even richer set of unconventional phenomena. Notable examples include the geometry-dependent skin effect \cite{Zhang2022,PhysRevLett.131.207201,Zhou2023,Wan2023}, where boundary accumulation is dictated by the system’s geometry, and the algebraic NHSE \cite{arxiv.2407.01296,cwwd-bclc, arXiv:2501.13440}, in which skin modes exhibit quasi-long-range behavior with power-law decaying spatial profiles, in sharp contrast to the exponential localization characteristic of 1D systems. Uniquely emerging in higher-dimensional non-Hermitian systems, the algebraic NHSE leads to striking consequences, such as ultra spectral sensitivity to impurities, wherein the eigenenergy spectrum is dramatically altered by two spatially separated  impurities with infinitesimally weak onsite potentials \cite{arXiv:2409.13623}. This sensitive effect arises from the formation of non-local bound states between the impurities and is closely associated with a divergence in the nonlocal back-and-forth Green’s function \cite{arXiv:2409.13623}. In marked contrast to the perturbative robustness inherent to Hermitian systems and 1D nonreciprocal non-Hermitian counterparts, this phenomenon reveals the genuinely new physics that arises only in higher-dimensional non-Hermitian Hamiltonians. 

Non-Hermitian Hamiltonians have been widely investigated in both classical and quantum settings, typically through engineered couplings to external dissipative baths \cite{Ashida2020}, which poses significant experimental challenges, particularly in quantum systems. As a complementary approach, bosonic quadratic Hamiltonians with quantum squeezing remain fully Hermitian yet can emulate effective non-Hermitian dynamics, thereby offering a genuine quantum framework to explore non-Hermitian phenomena without relying on external reservoirs \cite{PhysRevA.99.063834,PhysRevX.8.041031,Flynn2020, PhysRevB.106.024301,PhysRevB.105.224301,PhysRevLett.128.173602,PhysRevD.110.084039, PhysRevLett.127.245701,PhysRevLett.130.203605, PhysRevB.103.165123, Wanjura2020, PhysRevB.98.115135,arxiv.2012.03333,PhysRevResearch.7.013309,arxiv.2508.14560, PhysRevB.103.165123,arxiv.2505.02776,Slim2024,PhysRevLett.130.123602}. Thus far, studies of the NHSE in Hermitian bosonic quadratic systems have been largely restricted to 1D settings. A particularly intriguing open question is whether two-dimensional (2D) Hermitian bosonic quadratic Hamiltonians with quantum squeezing can host the algebraic NHSE and exhibit ultra spectral sensitivity to arbitrarily weak external perturbations.

In this study, we investigate Hermitian quadratic many-body bosonic Hamiltonians that do not conserve particle number. The bosonic quadratic Hamiltonian is implemented through quantum squeezing in a 2D lattice model. By applying the Bogoliubov–de Gennes (BdG) transformation, the many-body problem is mapped onto an effective single-particle Hamiltonian, whose quasiparticle excitations exhibit a complex eigenvalue spectrum—a hallmark of non-Hermitian physics. Unlike their one-dimensional counterparts, the 2D bosonic quadratic Hamiltonian on a square lattice hosts the algebraic NHSE, characterized by quasi-long-range, power-law localization of excitations away from the boundary. Remarkably, while the complex eigenspectrum remains robust against a single weak on-site impurity, it displays extreme spectral sensitivity in the presence of two infinitesimal on-site impurities or a long-range hopping impurity. In Sec.~\Rmnum{2}, we construct the 2D Hermitian bosonic quadratic Hamiltonian, incorporating both on-site and off-site quantum squeezing. In Sec.~\Rmnum{3}, we demonstrate the emergence of quasi-long-range localization in this 2D Hermitian bosonic quadratic model. Finally, in Sec.~\Rmnum{4}, we illustrate the ultra spectral sensitivity to two infinitesimal impurities and elucidate its origin using the Green’s function formalism.

\section{Model}

\begin{figure}[!tb]
	\centering
	\includegraphics[width=7.0cm]{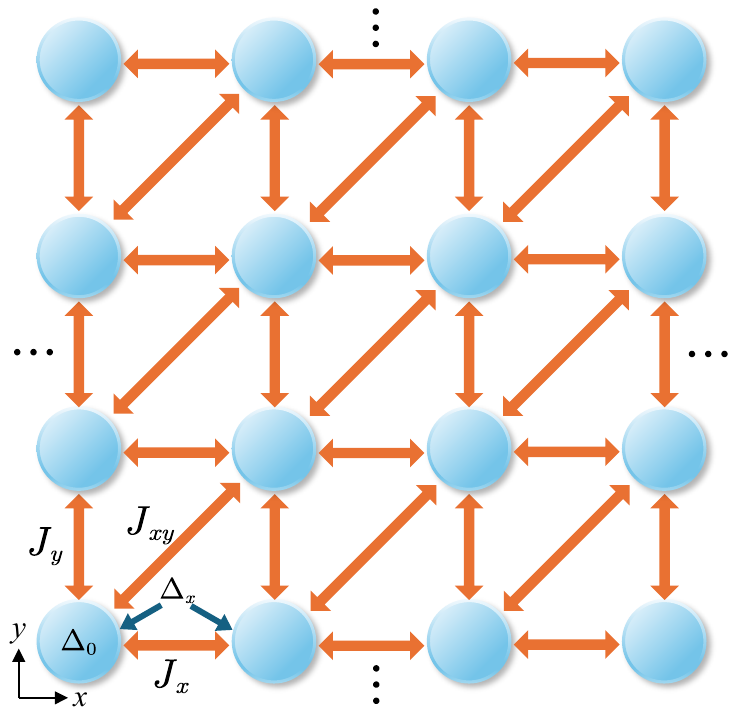}
	\caption{Schematic of a two-dimensional quadratic bosonic Hermitian lattice system.  $J_x$, $J_y$, and $J_{xy}$ represent single-particle hopping amplitudes along the $x$ direction, $y$ direction, and anti-diagonal directions, respectively. $\Delta_0$ characterizes on-site quantum squeezing, and $\Delta_x$ accounts for off-site quantum squeezing between adjacent sites along the $x$ direction. }\label{Fig1}
\end{figure}

We consider a  quadratic Hamiltonian describing bosons on a 2D lattice. As schematically illustrated in Fig.~\ref{Fig1}, the lattice is subjected to on-site and off-site parametric driving. The system Hamiltonian in real space is written as
\begin{align}\label{Hamil}
	\hat{H}_\text{R} = & \sum_{x,y} \omega_0 \hat{a}^\dagger_{x,y} \hat{a}_{x,y} + \sum_{x,y} \left(J_{xy} \hat{a}^\dagger_{x+1,y+1} \hat{a}_{x,y} +\text{H.c.}\right) \notag \\ 
	&  + \sum_{x,y} \left(J_{x} \hat{a}^\dagger_{x+1,y} \hat{a}_{x,y} + J_{y} \hat{a}^\dagger_{x,y+1} \hat{a}_{x,y}  + \text{H.c.}\right)\notag \\
	&  + \sum_{x,y} \left(\Delta_0 \hat{a}^\dagger_{x,y} \hat{a}^\dagger_{x,y} + \Delta_x \hat{a}^\dagger_{x+1,y} \hat{a}^\dagger_{x,y} +\text{H.c.}\right),
\end{align}
where $\hat{a}^\dagger_{x,y}$ ($\hat{a}_{x,y}$) denotes the bosonic creation (annihilation) operator at the lattice site $(x,y)$, $J_x$ and $J_y$ represent the nearest-neighbor hopping amplitudes along the $x$  and $y$ directions, respectively, and $J_{xy}$ corresponds to next-nearest-neighbor hopping along the anti-diagonal direction. The parameter  $\Delta_0$ characterizes on-site quantum squeezing, while $\Delta_x$ accounts for off-site quantum squeezing between adjacent sites along the $x$ direction. These terms break global $U(1)$ symmetry, violating particle number conservation. In experimental platforms, such pairing terms can be realized in systems such as quantum superconducting circuits \cite{PhysRevB.87.014508, Busnaina2024,  PRXQuantum.5.020306}, optomechanical setups \cite{SafaviNaeini2013,Slim2024, Marti2024} and nanophotonic platforms \cite{Vaidya2020,Nehra2022}.   The Hamiltonian $\hat{H}_\text{R}$ is Hermitian, and its parameters can, in general, take real or complex values. Unless otherwise specified, we set  $\omega_0 = 0$ throughout,  which corresponds to working in the rotating frame. 

The eigenvalue problem associated with the quadratic bosonic Hamiltonian $\hat{H}_\text{R}$ in Eq.~(\ref{Hamil}) can be reformulated in terms of a  BdG  formalism \cite{PhysRevB.87.174427, PhysRevB.98.115135, PhysRevB.103.165123}, as
\begin{align}\label{HamilBdG}
	\hat{H}_\text{R} = \frac{1}{2} \hat{\Psi}^\dagger \mathcal{H}_\text{BdG} \hat{\Psi},
\end{align}
where $\hat{\Psi} = (\hat{\mathbf{a}},~\hat{\mathbf{a}}^\dagger)^T$, and $\mathbf{\hat{a}}$ is a column vector of bosonic annihilation operator, defined as 
\begin{align}\label{a}
	\mathbf{\hat{a}} = \left(\hat{a}_{1,1},\hat{a}_{2,1}\dots,\hat{a}_{L_x,1},\hat{a}_{1,2}\dots,\hat{a}_{x,y},\dots,\hat{a}_{L_x,L_y}\right).
\end{align}
Here, $L_x$ and $L_y$ denote the number of lattice sites along $x$  and $y$ directions (see Fig.~\ref{Fig1}), respectively. The BdG Hamiltonian $\mathcal{H}_\text{BdG}$ takes the following matrix form with
\begin{align}\label{HamilBdG2}
	\mathcal{H}_\text{BdG} = \begin{pmatrix}
		\mathbf{h} & \Delta \\
		\Delta^\dagger & \mathbf{h}^\text{T}
	\end{pmatrix},
\end{align}
where $\mathbf{h} = \mathbf{h}^\dagger$ is a Hermitian matrix describing the single-particle hopping and on-site energy terms, and $\Delta=\Delta^\text{T}$ is a symmetric matrix that encodes the  pairing interactions. This BdG Hamiltonian exhibits   the bosonic analogy of particle-hole symmetry \cite{PhysRevB.105.224301} with
\begin{align}\label{PHS}
	\tau_x \mathcal{H}_\text{BdG}^\ast \tau_x = \mathcal{H}_\text{BdG},
\end{align}
where $\tau_\mu$ ($\mu=x,y,z$) are generalized Pauli matrices  defined in Nambu space with 
\begin{align}\label{tau}
	\tau_x = \begin{pmatrix}
		\mathcal{O} & \mathbf{I} \\
		\mathbf{I} & \mathcal{O}
	\end{pmatrix},
	\tau_y = i\begin{pmatrix}
		\mathcal{O} & -\mathbf{I} \\
		\mathbf{I} & \mathcal{O}
	\end{pmatrix},
	\tau_z = \begin{pmatrix}
		\mathbf{I} & \mathcal{O} \\
		\mathcal{O}  & -\mathbf{I}
	\end{pmatrix}.
\end{align}	%
Here, $\mathcal{O}$ denotes the zero matrix and $\mathbf{I}$ is the identity matrix.

When periodic boundary conditions (PBCs) are imposed, the system can be transformed into momentum space and described by the Bloch Hamiltonian
$\hat{H}_\text{B} = \frac{1}{2} \sum{\mathbf{k}} \hat{\Psi}^\dagger_\mathbf{k} \mathcal{H}_\text{BdG}(\mathbf{k}) \hat{\Psi}_\mathbf{k} + C$,
where $\hat{\Psi}_\mathbf{k} = (\hat{a}_\mathbf{k},~\hat{a}^\dagger_{-\mathbf{k}})^T$, and $C = -\text{Tr}[\mathcal{H}_0(\mathbf{k})]/2$ is a constant energy offset. The momentum-space BdG Hamiltonian $\mathcal{H}_\text{BdG}(\mathbf{k})$ is given by
\begin{align}\label{HamilB}
	\mathcal{H}_\text{BdG}(\mathbf{k}) = \begin{pmatrix}
		\mathcal{H}_0(\mathbf{k}) & \Delta(\mathbf{k}) \\
		\Delta^\ast(\mathbf{-k}) & \mathcal{H}^\ast_0(\mathbf{-k})
	\end{pmatrix}.
\end{align}
Here, $\mathcal{H}_0(\mathbf{k})$ is single-particle hopping term, with
\begin{align}\label{Hamil0k}
	\mathcal{H}_0(\mathbf{k}) = J_x e^{-ik_x} + J_y e^{-ik_y} + J_{xy} e^{-i(k_x+k_y)} + \textrm{H.c.},
\end{align}
and $\Delta(\mathbf{k}) = \Delta_0+\Delta_x e^{-ik_x} + \text{H.c.}$ denotes the pairing term. 

Within this formalism, although $\mathcal{H}_\text{BdG}(\mathbf{k})$ is Hermitian, the system’s dynamics exhibit a subtle non-Hermitian character \cite{PhysRevB.98.115135, PhysRevB.103.165123}. This emergent non-Hermiticity becomes evident when analyzing the time evolution of the Nambu spinor in the Heisenberg picture, which is governed by
\begin{align}\label{Dynamics}
	i \frac{\partial}{\partial t} \hat{\Psi}_{\mathbf{k}}(t) = [\hat{\Psi}_{\mathbf{k}}(t), \mathcal{H}_\text{BdG}(\mathbf{k})] = \mathcal{M}_\text{B}(\mathbf{k}) \hat{\Psi}_{\mathbf{k}}(t),
\end{align}
where the dynamical matrix, in the momentum space, is $\mathcal{M}_\text{B}(\mathbf{k}) = \sigma_z \mathcal{H}_\text{BdG}(\mathbf{k})$, with $\sigma_\mu$ denoting Pauli matrices acting on the particle-hole degrees of freedom. 

\section{Quantum Squeezing-induced Algebraic Non-Hermitian Skin Effect}

\begin{figure*}[!tb]
	\centering
	\includegraphics[width=18cm]{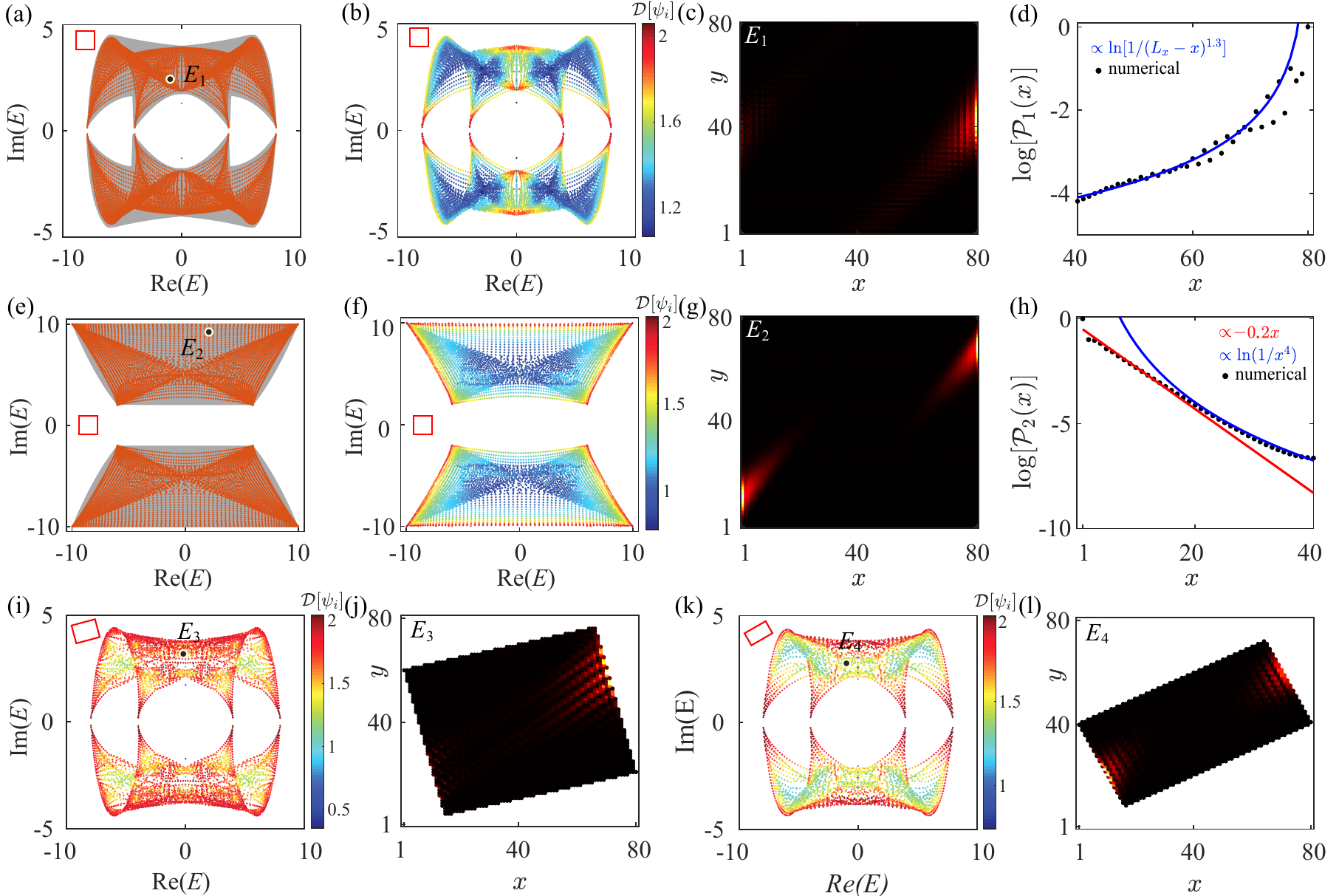}
	\caption{(a,e) Complex eigenenergies (orange dots) of quasiparticle excitations  computed from $\mathcal{M}_\text{B}$ under  the square geometry with OBCs for (a) $(J_x,J_y,J_{xy},\Delta_0,\Delta_x)=(1i,1,3i,-1,2i)$, and (e) $(J_x,J_y,J_{xy},\Delta_0,\Delta_x)=(0,1i,4i,3,2)$. Gray regions mark the PBC spectra. The corresponding eigenenergy-resolved fractal dimensions $\mathcal{D}[\psi_i]$ are shown in (b,f). (c,g) Probability densities $\abs{\Psi_i(\mathbf{r})}^2 = \abs{\psi_{p,i}(\mathbf{r})}^2 + \abs{\psi_{h,i}(\mathbf{r})}^2$ for the eigenstates with $E_1=-0.97+2.43i$ from (a) and $E_2=2.09+9.23i$ from (e). 	(d,h) Layer-resolved densities $\mathcal{P}_1(x)$ and $\mathcal{P}_2(x)$ for $E_1$ and $E_2$, shown on a logarithmic scale (black dots). Red and blue curves represent exponential and power-law fits, respectively.	(i,k) $\mathcal{D}[\psi_i]$ under oblique-square geometries with varying oblique angles for $(J_x,J_y,J_{xy},\Delta_0,\Delta_x)=(1i,1,3i,-1,2i)$.	(j,l) Corresponding probability densities $\abs{\Psi_i(\mathbf{r})}^2$ for eigenstates with $E_3=-0.18+3.18i$ from (i), and $E_4=-0.97+2.77i$ from (k). }\label{Fig2}
\end{figure*}

The non-Hermitian skin effect in 1D systems is well understood within the framework of non-Bloch band theory \cite{ShunyuYao2018,PhysRevLett.123.066404}. In this framework, a complex deformation of the Bloch wavevector, $k \rightarrow \tilde{k} = k + i \mu$ ($ \mu \in \mathbb{R}$), gives rise to the generalized Brillouin zone (GBZ), where $\beta = e^{i\tilde{k}}$ encodes the non-Bloch nature of the eigenstates. Under open boundary conditions (OBCs), the eigenstates of 1D non-Hermitian systems take thus the form of modified Bloch waves dressed with an exponential factor $e^ \mu$. This results in the characteristic exponential accumulation of eigenstates toward the system boundaries.

A straightforward extension of the 1D concept to higher-dimensional nonreciprocal systems leads to exponentially localized eigenstates along multiple spatial directions, characterized by a single vectorial attenuation factor, e.g., $\boldsymbol{ \mu} = (\mu_x,  \mu_y)$ in 2D system \cite{PhysRevB.107.195112,PhysRevX.14.021011}. However, recent studies have revealed an unusual form of algebraic NHSE in reciprocal non-Hermitian systems \cite{cwwd-bclc,arXiv:2501.13440}, where eigenstates exhibit power-law rather than exponential localization. In such systems, the anomalous localization behavior renders the conventional  GBZ  description, based on a single attenuation vector $\boldsymbol{ \mu}$, inadequate. Instead, a distribution or set of attenuation factors may be needed to fully characterize the system \cite{cwwd-bclc,arXiv:2501.13440}. The algebraic localization further suggests the presence of long-range correlations, offering new opportunities for manipulating quantum states. Despite these recent developments, the manifestation of such algebraic NHSE in Hermitian systems remains largely unexplored. In the following, we investigate the emergence of algebraic NHSE in bosonic quadratic Hermitian systems.

To demonstrate the emergence of the algebraic NHSE in Hermitian systems with the bosonic quadratic Hamiltonian, we analyze the eigenspectrum of its elementary excitation (or quasiparticle), governed by the dynamical matrix  $\mathcal{M}_\text{B} = \tau_z \mathcal{H}_\text{BdG}$ in real space. Figures \ref{Fig2}(a) and (e) show the complex eigenenergies (orange dots) of $\mathcal{M}_\textrm{B}$ under OBCs for the square    geometry with different parameters, while the gray regions indicate the spectra under PBCs. The localization behavior of the full eigenspectrum can be quantitatively characterized by evaluating the fractal dimension   of the   eigenstate  $\psi_i(\mathbf{r})$ under OBCs \cite{PhysRevA.109.023317,PhysRevLett.123.180601}, defined as
\begin{align}\label{FD}
	\mathcal{D}[\psi_i] = -\frac{\ln \left[\sum_{\mathbf{r}}\left(\abs{\psi_{p,i}(\mathbf{r})}^4 + \abs{\psi_{h,i}(\mathbf{r})}^4  \right)\right]}{\ln \sqrt{L_x L_y}},
\end{align}
where $\mathbf{r} = (x, y)$, $\psi_{p,i}(\mathbf{r})$ and $\psi_{h,i}(\mathbf{r})$, in the Nambu space, represent the particle and hole components of the normalized wavefunction $\psi_i(\mathbf{r})$ of $\mathcal{M}_\text{B}$, and $i$ labels specific eigenstate with eigenvalue  $E_i$. Here, $\mathcal{D}[\psi_i] = 2$ signifies a fully extended state, $\mathcal{D}[\psi_i] = 1$ indicates strong localization along the boundary, and $\mathcal{D}[\psi_i] = 0$ denotes a state localized well at a corner.

Figures \ref{Fig2}(b) and (f) present the eigenenergy-resolved fractal dimension $\mathcal{D}[\psi_i]$ corresponding to the parameters used in panels (a) and (e), respectively. In both cases, a small fraction of eigenstates correspond to well-localized 1D edge modes with $\mathcal{D}[\psi_i] \simeq 1$, while others form 2D extended states with $\mathcal{D}[\psi_i] \simeq 2$. Strikingly, the majority of eigenstates exhibit intermediate FD values between 1 and 2, indicating a hybrid localization character associated with quasi-long-range localization. The probability density distributions of eigenstates, given by $\abs{\Psi_i(\mathbf{r})}^2 = \abs{\psi_{p,i}(\mathbf{r})}^2 + \abs{\psi_{h,i}(\mathbf{r})}^2$ with $\mathbf{r} = (x, y)$, are shown in Figs.~\ref{Fig2}(c) and (g). The chosen eigenstates have  fractal dimension with values  $\mathcal{D}[\psi_i] \in [1.4,1.6]$, with their spatial profiles exhibiting long tails along the $x$ direction. This quasi-long-range localization is further confirmed by examining the layer density of eigenstates along the $x$ direction \cite{Zhang2022}, defined as
\begin{align}\label{layery}
	\mathcal{P}_i(x) = \sum_{y}\left(\abs{\psi_{p,i}(x,y)}^2 + \abs{\psi_{h,i}(x,y)}^2  \right).
\end{align}

Figures \ref{Fig2}(d) and (h) display the layer-resolved densities $\mathcal{P}_1(x)$ and $\mathcal{P}_2(x)$ on a logarithmic scale (black dots) for the eigenstate with energy $E_1$ from Fig.~\ref{Fig2}(a) and $E_2$ from Fig.~\ref{Fig2}(e). For the eigenstate with $E_1$, a power-law decay fit to $\mathcal{P}_1(x)$ reveals a quasi-long-range localization of algebraic form rather than exponential. For the eigenstate with $E_2$, a comparison between exponential (red line) and power-law (blue line) fits reveals a crossover behavior: near the boundary, the decay follows an exponential form, whereas farther away, the profile develops a power-law tail. This power-law tail signifies quasi-long-range localization.

To further examine the quasi-long-range localization in different lattice geometry, we calculate the fractal dimension $\mathcal{D}[\psi_i]$ and probability density $\abs{\Psi(\mathbf{r})}^2$ for typical eigenstates in two oblique-square lattices with different tilt angles, shown in Fig.~\ref{Fig2}(i,j) and (k,l). The intermediate fractal dimension values $\mathcal{D}[\psi_i] \in [1.4,~1.6]$ indicate that strong quasi-long-range localization persists for different lattice geometries. These results demonstrate the emergence of the algebraic NHSE independent of lattice geometry in bosonic quadratic Hermitian systems.

\section{Ultra spectral Sensitivity to  Impurities}

\subsection{Numerical results}

\begin{figure*}[!tb]
	\centering
	\includegraphics[width=17.8cm]{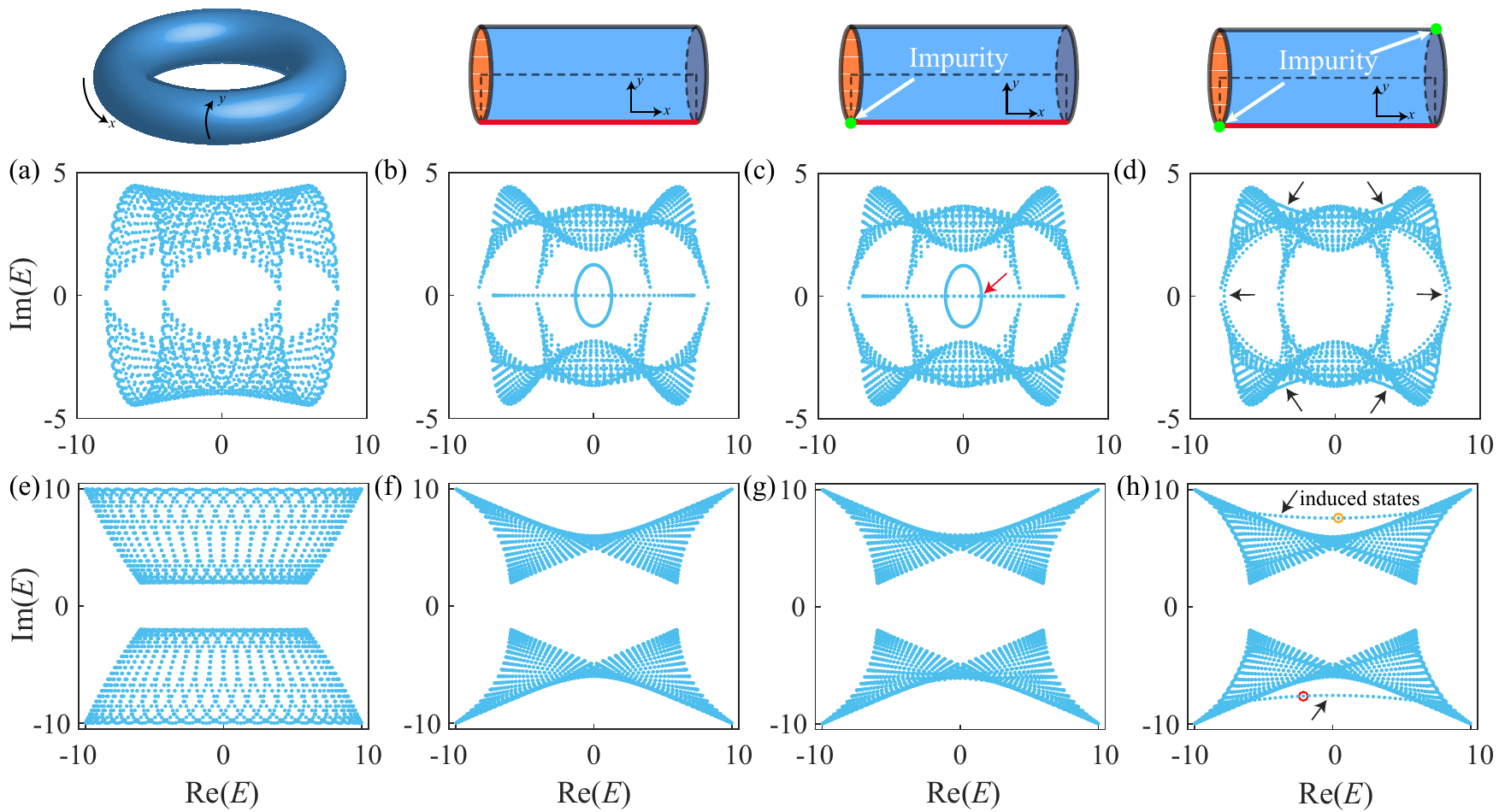}
	\caption{ Complex eigenenergies of quasiparticle excitations  (a,e) under PBCs along both the $x$ and $y$ directions, (b–d) and (f–h) under OBC along the $x$ direction and PBC along the $y$ direction. Panels (c,g) correspond to a single  impurity with sufficiently weak onsite potential $V_1=0.01$ ($V_2=0$) coupled to the left edge, while panels (d,h) correspond to two distant  impurities with sufficiently weak onsite potentials $V_1=0.01$ and $V_2=0.01$, each coupled to the left and right edges, respectively, as indicated by green dots in the top schematic plot. In (d), the red arrow marks states that disappear when two impurities are present, whereas in (d,h), the black arrows denote states induced solely by the two impurities. The parameters are $(J_x,J_y,J_{xy},\Delta_0,\Delta_x)=(1i,1,3i,-1,2i)$ for (a–d) and $(J_x,J_y,J_{xy},\Delta_0,\Delta_x)=(0,1i,4i,3,2)$ for (e–h). The lattice size is $L_x\times L_y=50\times 50$, with impurities positioned at $\mathbf{r}_1=(1,1)$ and $\mathbf{r}_2=(L_x,L_y/2)$.}\label{Fig3}
\end{figure*}

We now demonstrate the ultra spectral sensitivity to impurities in bosonic quadratic Hermitian systems, a phenomenon that was previously identified only in higher-dimensional non-Hermitian systems \cite{arXiv:2409.13623}.

The quasiparticle excitation spectrum is strongly influenced by boundary conditions, as seen by the contrast between Fig.~\ref{Fig3}(a,e) under PBCs along both $x$ and $y$ and (b,f) under OBC in $x$ and PBC in $y$. Even more strikingly, the eigenspectrum of the bosonic quadratic Hermitian system shows remarkable sensitivity to impurities with sufficiently weak strength. To illustrate this, we introduce on-site impurities at positions $\mathbf{r}_1$ and $\mathbf{r}_2$, where the Hamiltonian takes the form $\hat{H} = \hat{H}_\text{R} + \hat{V}$, with
\begin{align}\label{onsiteimpurities2} 
	\hat{V} = V_1 \hat{a}^\dagger_{\mathbf{r}_1} \hat{a}_{\mathbf{r}_1} + V_2 \hat{a}^\dagger_{\mathbf{r}_2} \hat{a}_{\mathbf{r}_2},
\end{align}
where $V_1$ and $V_2$ denote the on-site potentials. The full Hamiltonian of the impurity lattice, expressed in the Nambu spinor basis, is written as  
\begin{align}
	\hat{H} = \hat{H}_\text{R} + \hat{V} = \frac{1}{2} \hat{\Psi}^\dagger \mathcal{H} \hat{\Psi}, 
\end{align} 
where  $\mathcal{H} = \mathcal{H}_\text{BdG} + \mathcal{V}$, with $\mathcal{V}$  being the  BdG representation of the impurity potential.  Unless otherwise specified, here, we consider OBC along the $x$ direction and PBC along the $y$ direction.

We plot the quasiparticle excitation spectra for two sets of parameters in Fig.~\ref{Fig3}(c,d) and  (g,h), corresponding to a single  impurity with sufficiently weak strength [$V_1=0.01$, $V_2=0$, see Fig.~\ref{Fig3}(c,g)] and two distant  impurities with sufficiently weak strength [$V_1=V_2=0.01$, see Fig.~\ref{Fig3}(d,h)]. Whereas a single  impurity with sufficiently weak on-site potential leaves the eigenspectrum almost unchanged [see Fig.~\ref{Fig3}(c,g)], introducing two   impurities with sufficiently weak strength produces a marked alteration of the quasiparticle excitations [see Fig.~\ref{Fig3}(d,h)],  as evident from the comparison with the single-impurity case. Specifically, the two infinitesimal impurities not only create additional eigenstates [black arrows in Fig.~\ref{Fig3}(d,h)], but also eliminate existing ones, such as the state indicated by the red arrow in Fig.~\ref{Fig3}(d). This behavior demonstrates that the bosonic quadratic Hermitian system, studied here, exhibits an ultra spectral sensitivity to infinitesimal impurities, a feature that is generally absent in particle-conserving Hermitian systems.

\begin{figure}[!tb]
	\centering
	\includegraphics[width=8.7cm]{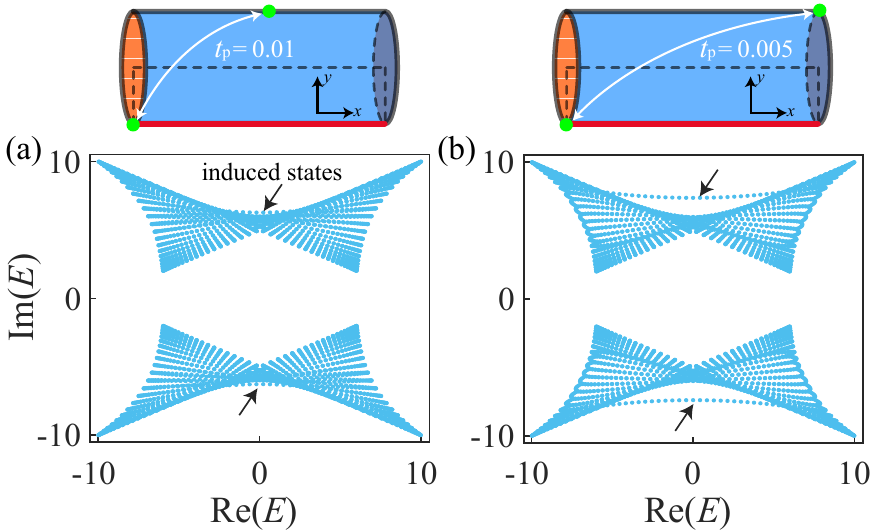}
	\caption{ Complex quasiparticle eigenenergies in the presence of a long-range hopping impurity under OBC along the $x$ direction and PBC along the $y$ direction.  Panels (a) and (b) correspond to different hopping ranges: (a) $(x_1,y_1)=(1,1)$, $(x_2,y_2)=(3L_x/5, L_y/2)$ with $t_\textrm{p}=0.01$, and (b) $(x_1,y_1)=(1,1)$, $(x_2,y_2)=(L_x, L_y/2)$ with $t_\textrm{p}=0.005$. The arrows indicate the new states arising from the long-range hopping impurity. The other parameters are  $(J_x,J_y,J_{xy},\Delta_0,\Delta_x) = (0,1i,4i,3,2)$ with  $L_x \times   L_y = 50 \time 50$. }\label{Fig4}
\end{figure}

In addition to the on-site impurity, the bosonic quadratic Hermitian system can also exhibit ultra spectral sensitivity to a long-range hopping impurity. To illustrate this, we consider the following form of long-range hopping impurity
\begin{align}\label{onsiteimpurities3} 
	\hat{V} = t_\textrm{p}( \hat{a}^\dagger_{x_1,y_1} \hat{a}_{x_2,y_2} + \text{H.c.}).
\end{align}

With OBC imposed along the $x$ direction and PBC along the $y$ direction, we compute the complex quasiparticle eigenenergies by fixing the impurity position at $(x_1,y_1)=(1,1)$ and varying the second site $(x_2,y_2)$ to explore different hopping ranges. The results are shown in Fig.~\ref{Fig4}(a) for $(x_2,y_2)=(3L_x/5, L_y/2)$ and in Fig.~\ref{Fig4}(b) for $(x_2, y_2)=(L_x, L_y/2)$, both obtained under a sufficiently weak hopping strength $t_\text{p}$. Despite the weakness of the hopping strength, the eigenspectrum exhibits impurity-induced states, indicated by the black arrows in Fig.~\ref{Fig4}(a,b), once the hopping range exceeds a critical length [see Fig.~\ref{Fig4}(a)].

\subsection{Physical mechanism}

To gain deeper insight into the ultra spectral sensitivity of bosonic quadratic Hermitian systems to impurities, we employ the Green’s function formalism, which offers a powerful framework for analyzing their response to perturbations \cite{arXiv:2409.13623}. 

The eigenvalue equation of the full system in the presence of impurities is written as 
\begin{align}\label{HEM32}
	(E\tau_0-\mathcal{M})\Psi =0,
\end{align}
where $\tau_0$ is identity matrix, and $\mathcal{M} = \tau_z \mathcal{H}$. The Green's function in a bosonic BdG system can be defined as the inverse of this linear operator $E\tau_0-\mathcal{M}$. Explicitly, the full Green's function in real space takes the form   as
\begin{align}\label{GM2}
	\mathcal{G} = (E\tau_0-\mathcal{M})^{-1} = (\tau_0-\mathcal{G}_0 \mathcal{V}_z)^{-1} \mathcal{G}_0,
\end{align}
where $\mathcal{V}_z = \tau_z \mathcal{V}$, and $\mathcal{G}_0 = (E\tau_0 - \mathcal{M}_\text{BdG})^{-1}$   denotes the Green’s function of the pristine lattice in the absence of impurities. In this work, we restrict our analysis to energies $E$ outside the spectrum of $\mathcal{M}_\text{BdG}$, for which $\mathcal{G}_0$ remains generally non-singular.

The effect of sufficiently weak impurities is captured by $(\tau_0 - \mathcal{G}_0 \mathcal{V}_z)^{-1}$ in Eq.~(\ref{GM2}), which renormalizes the unperturbed propagator $\mathcal{G}_0$. When all eigenvalues of the response matrix $\mathcal{G}_0 \mathcal{V}_z$ remain much smaller than unity, $\mathcal{G}$ stays close to $\mathcal{G}_0$ and the perturbed spectrum of $\mathcal{M}$ is essentially unchanged. In contrast, once even a single eigenvalue approaches or exceeds unity, the system enters a non-perturbative regime in which both the Green’s function $\mathcal{G}$ and the associated spectrum of $\mathcal{M}$ exhibit strong departures from their unperturbed counterparts $\mathcal{G}_0$ and $\mathcal{M}_\text{BdG}$. 

By defining the spectral radius $\rho$ of $\mathcal{G}_0 \mathcal{V}_z$ \cite{arXiv:2409.13623}, a quantitative criterion for the spectral stability against infinitesimal impurities can be formulated as
\begin{align}\label{SpectralStability}
	\rho (\mathcal{G}_0 \mathcal{V}_z) = \max \abs{\sigma(\mathcal{G}_0 \mathcal{V}_z)} \ll 1,
\end{align}
where $\sigma(\cdot)$ denotes the set of eigenvalues. Violation of this criterion can lead to non-perturbative changes, even for weak impurities.

To obtain an analytical result for the Green’s function $\mathcal{G}$ in Eq.~(\ref{GM2}) and gain an intuitive understanding of the ultra spectral sensitivity, we set $J_x=0$, $J_y=it_y$, and $J_{xy}=it_{xy}$ with real pairing amplitudes $\Delta_0$ and $\Delta_x$, where $t_y$ and $t_{xy}$ are real. Under these conditions, the dynamical matrix $\mathcal{M}_\text{BdG}$ can be unitarily transformed into a block-diagonal form (see details in Appendix.~\ref{AppendixA1}) with
\begin{align}\label{UnitaryRM2}
	U \mathcal{M}_\text{BdG} U^\dagger = \begin{pmatrix}
		\mathcal{M}_p & \mathcal{O} \\
		\mathcal{O} & \mathcal{M}_m
	\end{pmatrix},
\end{align}
where $U = (\tau_0-i\tau_x)/\sqrt{2}$ with $U^\dagger = U^{-1}$, $\mathcal{M}_p$ and $\mathcal{M}_m$ represent the  decoupled matrices in real space.

We now apply the same unitary transformation to the dynamical matrix $\mathcal{M}$ in the presence of impurities, yielding $\bar{\mathcal{M}} = U \mathcal{M} U^\dagger$. Under this transformation, the Green’s function $\mathcal{G}$ in Eq.~(\ref{GM2}) becomes
\begin{align}\label{GbarM2}
	\bar{\mathcal{G}} = (E \tau_0-\bar{\mathcal{M}})^{-1} = (\tau_0-\bar{\mathcal{G}}_0 \bar{\mathcal{V}}_z)^{-1} \bar{\mathcal{G}}_0,
\end{align}
where $\bar{\mathcal{G}}_0 = (E\tau_0 - U\mathcal{M}_\text{BdG}U^\dagger)^{-1}$, and $\bar{\mathcal{V}}_z = U \mathcal{V}_z U^\dagger$.

The unitary transformation of the Green’s function in Eq.~(\ref{GbarM2}) gives rise to a  modified spectral stability criterion for infinitesimal impurities in Eq.~(\ref{SpectralStability}), expressed via the reduced spectral radius   with 
\begin{align}\label{SpectralStability2}
	\rho (\bar{\mathcal{G}}_0 \bar{\mathcal{V}}_z)\ll 1.
\end{align}

As shown in the Appendix.~\ref{AppendixA2}, for a single on-site impurity with $\hat{V} = V_1 \hat{a}^\dagger_{\mathbf{r}_1} \hat{a}_{\mathbf{r}_1}$,  the spectral radius is obtained as
\begin{align}\label{SRSMain}
	\rho(\bar{\mathcal{G}}_0 \bar{\mathcal{V}}_z) =\abs{\sqrt{\xi}} = \abs{V_1 \sqrt{\bar{\mathcal{G}}_0^p (\mathbf{r}_1,\mathbf{r}_1;E) \bar{\mathcal{G}}_0^m (\mathbf{r}_1,\mathbf{r}_1;E)}},
\end{align}
where   the unperturbed Green's function $\bar{\mathcal{G}}_0^{p/m}(\mathbf{r}_i,\mathbf{r}_j;E)$ is given by
\begin{align}\label{W}
	\bar{\mathcal{G}}_0^{p/m}(\mathbf{r}_i,\mathbf{r}_j;E) = \bra{\mathbf{r}_i}(E-\mathcal{M}_{p/m})^{-1}\ket{\mathbf{r}_j}.
\end{align}

We explicitly evaluate the Green's function $\bar{\mathcal{G}}_0^{p/m}(\mathbf{r}_i,\mathbf{r}_j;E)$ under OBC along the $x$ direction and PBC along the $y$ direction for the cylindrical geometry. To capture the NHSE and perform the momentum-space integration in GBZ, we introduce the non-Bloch wavevector $\beta_x = e^{ik_x+ \mu_x}$  along the $x$ direction, while keeping the Bloch wavevector $k_y$ along the $y$ direction \cite{arXiv:2409.13623}. The Green's function then reads (see Appendix \ref{AppendixB})

\begin{align}\label{Gbar0phr1r25}
	&  \bar{\mathcal{G}}_0^{p/m}(\mathbf{r}_i,\mathbf{r}_j;E) \nonumber \\ 
	\addlinespace[0.8em]
	& = \int_{0}^{2\pi} dk_y \oint_{\beta_x \in \text{GBZ}} \frac{\text{d} \beta_x}{\beta_x} \frac{e^{ik_y(y_i-y_j)} \beta_x^{x_i-x_j}}{E-E_\pm(\beta_x,k_y)}, 
\end{align}
where $E_\pm(\beta_x,k_y)$ reads
\begin{align}\label{Epm200}
	E_\pm(\beta_x, k_y) = & \pm i\Delta_x\left(\beta_x+\beta_x^{-1}\right) + it_y \left(e^{ik_y}-e^{-ik_y}\right)  \nonumber \\
	& \pm 2i\Delta_0 + it_{xy} \left(\beta_x e^{ik_y}-\beta_x^{-1}e^{-ik_y}\right) \nonumber \\
	&  + it_y \left(e^{ik_y}-e^{-ik_y}\right).
\end{align}

According to Eq.~(\ref{Gbar0phr1r25}), the local Green’s functions $\bar{\mathcal{G}}_0^{p/m} (\mathbf{r}_1,\mathbf{r}_1;E)$ in Eq.~(\ref{SRSMain}) remain finite because the energy $E$ is chosen off-resonant from all eigenstates of the unperturbed dynamical matrix $\mathcal{M}_\text{BdG}$. Consequently, the spectral radius in Eq.~(\ref{SRSMain}) vanishes in the limit $V_1 \to 0$, corresponding to an infinitesimal impurity. This indicates that the system’s spectrum remains robust against a single sufficiently-weak impurity perturbation, with no emergence of additional eigenvalues, consistent with the numerical results in Fig.~\ref{Fig3}(c,g).

Furthermore, in the presence of double sufficiently-weak onsite impurities with    $\hat{V} = V_1 \hat{a}^\dagger_{\mathbf{r}_1} \hat{a}_{\mathbf{r}_1} + V_2 \hat{a}^\dagger_{\mathbf{r}_2} \hat{a}_{\mathbf{r}_2}$, the spectral radius (see Appendix~\ref{AppendixA2}) is written as
\begin{align}\label{SRS20}
	\rho(\bar{\mathcal{G}}_0 \bar{\mathcal{V}}_z) = \max (\abs{\sqrt{\xi_\pm}}),
\end{align}
where
\begin{align}\label{xiAppendixA001}
	\xi_\pm =& \frac{1}{2} \left(\mathcal{BC}_{11} + \mathcal{BC}_{22} \pm \sqrt{(\mathcal{BC}_{11}-\mathcal{BC}_{22})^2 + 4 \mathcal{BC}_{12} \mathcal{BC}_{21}}\right),
\end{align}
and $\mathcal{BC}$ is the $2\times2$ matrix, with its elements given by
\begin{align}\label{equationAppendixA01} 
	\mathcal{BC}_{11} = ~& V_1^2 \bar{\mathcal{G}}_0^p (\mathbf{r}_1,\mathbf{r}_1;E) \bar{\mathcal{G}}_0^m (\mathbf{r}_1,\mathbf{r}_1;E) \nonumber \\ & + V_1V_2 \bar{\mathcal{G}}_0^p (\mathbf{r}_1,\mathbf{r}_2;E) \bar{\mathcal{G}}_0^m (\mathbf{r}_2,\mathbf{r}_1;E),   
\end{align}
\begin{align}\label{equationAppendixA02}
	\mathcal{BC}_{12} = ~& V_1V_2 \bar{\mathcal{G}}_0^p (\mathbf{r}_1,\mathbf{r}_1;E) \bar{\mathcal{G}}_0^m (\mathbf{r}_1,\mathbf{r}_2;E) \nonumber \\& + V^2_2 \bar{\mathcal{G}}_0^p (\mathbf{r}_1,\mathbf{r}_2;E) \bar{\mathcal{G}}_0^m (\mathbf{r}_2,\mathbf{r}_2;E), 		
\end{align}
\begin{align}\label{equationAppendixA03}
	\mathcal{BC}_{21} = ~& V_1^2 \bar{\mathcal{G}}_0^p (\mathbf{r}_2,\mathbf{r}_1;E) \bar{\mathcal{G}}_0^m (\mathbf{r}_1,\mathbf{r}_1;E) \nonumber \\& + V_1V_2 \bar{\mathcal{G}}_0^p (\mathbf{r}_2,\mathbf{r}_2;E) \bar{\mathcal{G}}_0^m (\mathbf{r}_2,\mathbf{r}_1;E), 		
\end{align}
\begin{align}\label{equationAppendixA04}	
	\mathcal{BC}_{22} = ~& V_1V_2 \bar{\mathcal{G}}_0^p (\mathbf{r}_2,\mathbf{r}_1;E) \bar{\mathcal{G}}_0^m (\mathbf{r}_1,\mathbf{r}_2;E) \nonumber \\ &+ V^2_2 \bar{\mathcal{G}}_0^p (\mathbf{r}_2,\mathbf{r}_2;E) \bar{\mathcal{G}}_0^m (\mathbf{r}_2,\mathbf{r}_2;E).
\end{align}

As shown in Eqs.~(\ref{SRS20})-(\ref{equationAppendixA04}), unlike the single impurity case, the spectral radius for two impurities depend not only on the local Green's function $\bar{\mathcal{G}}_0^{p/m} (\mathbf{r}_i,\mathbf{r}_i;E)$ ($i=1,2$) , but also on the propagators $\bar{\mathcal{G}}_0^{p/m}(\mathbf{r}_i,\mathbf{r}_j;E)$ with $i\neq j$.

To carry out the contour integral for the propagator in Eq.~(\ref{Gbar0phr1r25}),  we solve  characteristic equations $E-E_\pm(\beta_x,k_y)=0$  in the subspace   defined in Eq.~(\ref{UnitaryRM2}), where $E$ is outside the eigenspectrum   of $\mathcal{M}_{p/m}$. The corresponding roots of the  characteristic equations, $\beta_{1,2}^\pm(k_y)$, are ordered such that $\abs{\beta_2^\pm(k_y)}>\abs{\beta_1^\pm(k_y)}$.  Then, using the residue theorem and the thermodynamic limit, the propagators $\bar{\mathcal{G}}_0^{p/m}(\mathbf{r}_i,\mathbf{r}_j;E)$ ($i\neq j$) are approximated  (see details in Appendix~\ref{AppendixB}) as
\begin{align}\label{Gbar0phr2r1E2New0}
	\bar{\mathcal{G}}_0^{p/m}(\mathbf{r}_2,\mathbf{r}_1;E) \simeq \int_{0}^{2\pi} dk_y [\beta^\pm_1(k_y)]^{L} e^{ik_y\delta_y},
\end{align}
\begin{align}\label{Gbar0phr1r2E2New0}
	\bar{\mathcal{G}}_0^{p/m}(\mathbf{r}_1,\mathbf{r}_2;E) \simeq \int_{0}^{2\pi} dk_y [\frac{1}{\beta^\pm_2(k_y)}]^{L} e^{-ik_y\delta_y},
\end{align}
where  $L=x_2-x_1$ and $\delta_y=y_2-y_1$.

\begin{figure*}[!ht]
	\centering
	\includegraphics[width=18cm]{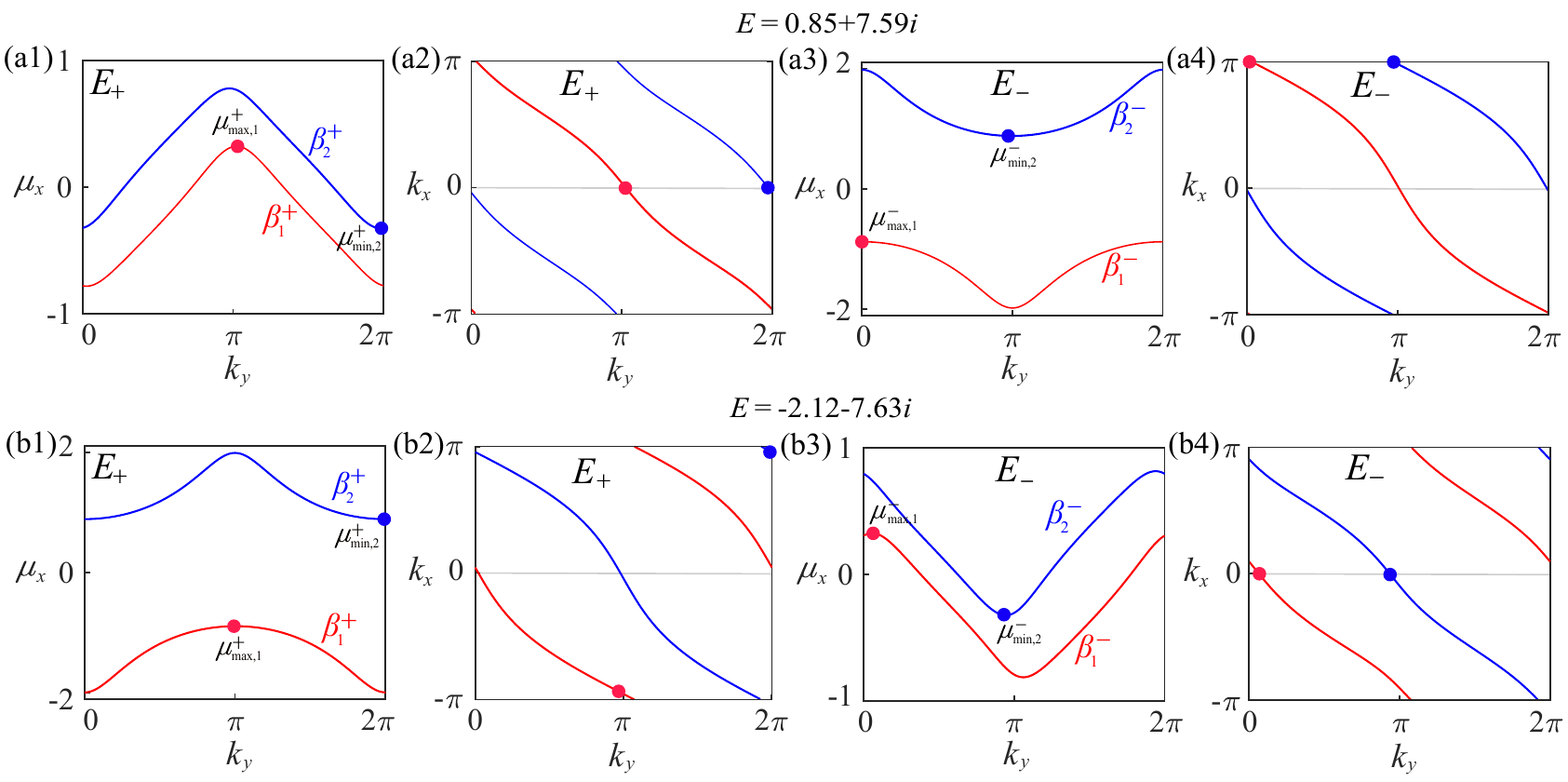}
	\caption{Two roots $\beta_{1,2}^+(k_y)$ and $\beta_{1,2}^-(k_y)$ of characteristic equations $E-E_+(\beta_x,k_y)=0$ and $E-E_-(\beta_x,k_y)=0$ with $\beta_x = e^{i k_x + \mu_x}$, where $E$ is outside the eigenspectrum   of $\mathcal{M}_{p/m}$ with (a1-a4) $E = 0.85+7.59i$ indicated by the orange circle in Fig.~\ref{Fig3}(h), and  (b1-b4) $E = -2.12-7.63i$ indicated by the red circle in Fig.~\ref{Fig3}(h). The red and blue dots represent the maximum values $\mu_\text{max,1}^\pm   = \max_{k_y\in[0,2\pi]} \ln [\beta_1^\pm(k_y)]$, and minimum values of $\mu_\text{min,2}^\pm   = \min_{k_y\in[0,2\pi]}\ln [\beta_2^\pm(k_y)]$.}\label{Fig5}
\end{figure*}

Our goal is to estimate the spectral radius $\rho(\bar{\mathcal{G}}_0 \bar{\mathcal{V}}_z)$ in Eq.~(\ref{SRS20}) for a large lattice size along the $x$ direction, and to demonstrate that this quantity deviates from the spectral stability criterion in Eq.~(\ref{SpectralStability2}) in the presence of two infinitesimal on-site impurities, thereby revealing the ultra spectral sensitivity of Hermitian bosonic quadratic systems. For large $L$, the integrals in Eqs.~(\ref{Gbar0phr2r1E2New0}) and (\ref{Gbar0phr1r2E2New0}) can be approximated by choosing values of $\delta_y$ that cancel the oscillatory phase in the integrands (see Appendix \ref{AppendixB}). These stationary-phase points provide the dominant contributions to the integrals.

For the new eigenvalue $E$ with positive imaginary part, for instance, the eigenvalue $E$ indicated by the orange circle in Fig.~\ref{Fig3}(h), the integrals in Eqs.~(\ref{Gbar0phr2r1E2New0}) and (\ref{Gbar0phr1r2E2New0}) can be estimated as
\begin{align}\label{PBintegralAppendix01}
	\begin{dcases}
		\bar{\mathcal{G}}_0^p(\mathbf{r}_2,\mathbf{r}_1;E) \sim e^{\mu_\text{max,1}^+ L}, \\
		\bar{\mathcal{G}}_0^m(\mathbf{r}_2,\mathbf{r}_1;E) \sim e^{\mu_\text{max,1}^- L} \sim 0, \\
		\bar{\mathcal{G}}_0^p(\mathbf{r}_1,\mathbf{r}_2;E) \sim e^{-\mu_\text{min,2}^+ L}, \\
		\bar{\mathcal{G}}_0^m(\mathbf{r}_1,\mathbf{r}_2;E) \sim e^{-\mu_\text{min,2}^- L} \sim 0,
	\end{dcases}
\end{align}
with 
\begin{align}\label{mu12101}
	\mu_\text{max,1}^\pm = \max_{k_y\in[0,2\pi]}\ln \abs{\beta_1^\pm(k_y)}, 
\end{align}
and
\begin{align}\label{mu12201}
	\mu_\text{min,2}^\pm = \min_{k_y\in[0,2\pi]}\ln \abs{\beta_2^\pm(k_y)}. 
\end{align}
Here, $\mu_{\text{max},1}^+ > 0$, $\mu_{\text{max},1}^- < 0$, $\mu_{\text{min},2}^+ < 0$, and $\mu_{\text{min},2}^- > 0$ are typically satisfied in the parameter regime of interest [see red and blue dots in Figs.~\ref{Fig5}(a1) and (a3)], thereby facilitating the demonstration of the ultra spectral sensitivity.

Therefore, for double onsite impurities, the spectral radius associated with the induced new eigenvalue $E$ with positive imaginary part can be approximated by
\begin{align}\label{SRS2New101}
	\rho(\bar{\mathcal{G}}_0 \bar{\mathcal{V}}_z) \sim V e^{\frac{\mu_\text{max,1}^+ L}{4}} e^{-\frac{\mu_\text{min,2}^+ L}{4}}.
\end{align}
Here, we consider the case where the two on-site impurities have identical on-site potentials with $V_1 = V_2 = V$.

For the new eigenvalue $E$ with negetive imaginary part, for instance, the eigenvalue $E$ marked by the red circle in Fig.~\ref{Fig3}(h), the integrals in Eqs.~(\ref{Gbar0phr2r1E2New0}), and (\ref{Gbar0phr1r2E2New0}) can be estimated as
\begin{align}\label{HBintegralAppendix01}
	\begin{dcases}
		\bar{\mathcal{G}}_0^p(\mathbf{r}_2,\mathbf{r}_1;E) \sim e^{\mu_\text{max,1}^+ L} \sim 0, \\
		\bar{\mathcal{G}}_0^m(\mathbf{r}_2,\mathbf{r}_1;E) \sim e^{\mu_\text{max,1}^- L},\\
		\bar{\mathcal{G}}_0^p(\mathbf{r}_1,\mathbf{r}_2;E) \sim e^{-\mu_\text{min,2}^+ L} \sim 0, \\
		\bar{\mathcal{G}}_0^m(\mathbf{r}_1,\mathbf{r}_2;E) \sim e^{-\mu_\text{min,2}^- L},
	\end{dcases}
\end{align}
where $\mu_{\text{max},1}^+ < 0$, $\mu_{\text{max},1}^- > 0$, $\mu_{\text{min},2}^+ > 0$, and $\mu_{\text{min},2}^- < 0$ are typically satisfied in the parameter regime of interest [see red and blue dots in Figs.~\ref{Fig5}(b1) and (b3)]. The corresponding the spectral radius associated with the induced new eigenvalue $E$ with negative imaginary part is
\begin{align}\label{SRS2New201}
	\rho(\bar{\mathcal{G}}_0 \bar{\mathcal{V}}_z) \sim V e^{\frac{\mu_\text{max,1}^- L}{4}} e^{-\frac{\mu_\text{min,2}^- L}{4}},	
\end{align}
where $V_1 = V_2 = V$ is assumed.

In contrast to the single-impurity case, the presence of two infinitesimal onsite impurities can cause the propagators to diverge for large lattice sizes [see Eqs.~(\ref{PBintegralAppendix01}) and (\ref{HBintegralAppendix01})]. Consequently, the spectral radius in Eqs.~(\ref{SRS2New101}) and (\ref{SRS2New201}) deviates from the spectral stability criterion in Eq.~(\ref{SpectralStability2}), thereby revealing the ultra spectral sensitivity of Hermitian bosonic quadratic systems. The underlying mechanism is directly analogous to that uncovered in the non-Hermitian case discussed in Ref.~\cite{arXiv:2409.13623}. Importantly, the higher experimental accessibility of Hermitian platforms paves the way for the direct observation of both the algebraic NHSE and the ultra spectral sensitivity to infinitesimal impurities, opening promising avenues for precision control and detection in bosonic systems.

\section{Conclusions}

In this work, we demonstrate the algebraic NHSE and non-perturbative spectral sensitivity to impurities in Hermitian bosonic quantum many-body systems. We construct a 2D Hermitian bosonic quadratic Hamiltonian on a square lattice, realized through both on-site and off-site quantum squeezing. By numerically investigating the bosonic excitations within the BdG framework, we reveal a quasi-long-range power-law localization of complex eigenstates, in stark contrast to the exponential localization observed in the 1D counterpart. Furthermore, the 2D Hermitian bosonic quadratic Hamiltonian is insensitive to infinitesimal single impurities, yet exhibits ultra spectral sensitivity to double infinitesimal on-site impurities and long-range hopping impurities. Specifically, for double infinitesimal on-site impurities, the eigenspectrum either develops new eigenstates or causes certain eigenstates to vanish. Using the Green’s function formalism, we analytically uncover the mechanism underlying this ultra spectral sensitivity, which arises from the formation of nonlocal bound states between impurities, leading to the divergence of the nonlocal Green’s function. Our findings establish a foundation for exploring exotic higher-dimensional non-Hermitian phenomena within Hermitian quantum systems, and they may inspire future realizations in bosonic platforms such as superconducting circuits, photonic lattices, and optomechanical arrays. Moreover, the demonstrated ultra spectral sensitivity could enable novel applications in quantum sensing and amplification, leveraging bosonic squeezing as an intrinsic physical resource.

\begin{acknowledgments}
	T.L. acknowledges the support from the National Natural
	Science Foundation of China (Grant No.~12274142),  the  Key Program of the National Natural Science Foundation of China (Grant No.~62434009), and Introduced Innovative Team Project of Guangdong Pearl River Talents Program (Grant No. 2021ZT09Z109).
\end{acknowledgments}

\appendix
\section{Details on Green's Function and Reduced Spectral Radius}\label{AppendixA}

\subsection{Green's function in bosonic BdG system}\label{AppendixA1}

We begin by writing the full Hamiltonian in the presence of impurities, expressed in the Nambu basis as
\begin{align}
	\label{HBdGM} \hat{H} = \hat{H}_\text{R} + \hat{V} = \frac{1}{2} \hat{\Psi}^\dagger \mathcal{H} \hat{\Psi}. 
\end{align} 
where  $\Psi = (\hat{\mathbf{a}},~\hat{\mathbf{a}}^\dagger)^T$.

According to Eq.~(\ref{Dynamics}), the eigenvalue equation of the full system  is written as 
\begin{align}\label{HEM3}
	(E\tau_0-\mathcal{M})\Psi =0.
\end{align}

The Green’s function of the bosonic BdG system is then defined as the inverse of the operator $E\tau_0 - \mathcal{M}$. Explicitly, the full Green’s function reads
\begin{align}\label{GM}
	\mathcal{G} = (E\tau_0-\mathcal{M})^{-1} = (\tau_0-\mathcal{G}_0 \mathcal{V}_z)^{-1} \mathcal{G}_0,
\end{align}
where $\mathcal{V}_z = \tau_z \mathcal{V}$, and $\mathcal{G}_0 = (E\tau_0 - \mathcal{M}_\text{BdG})^{-1}$ denotes the Green’s function of the unperturbed system.

To obtain an analytical expression for the Green’s function $\mathcal{G}$ and the spectral radius $\rho$, and thereby gain an intuitive understanding of the ultra 	spectral sensitivity, we set $J_x=0$, $J_y=it_y$, and $J_{xy}=it_{xy}$ with real pairing amplitudes $\Delta_0$ and $\Delta_x$, where $t_y$ and $t_{xy}$ are real. Under these conditions, the momentum-space BdG Hamiltonian is written as 
\begin{align}\label{HamilB2}
	\mathcal{H}_\text{BdG}(\mathbf{k}) = & [2t_y\sin k_y + 2t_{xy} \sin (k_x+k_y)]\sigma_z \notag \nonumber \\
	&+ (2\Delta_0 + 2\Delta_x\cos k_x )\sigma_x.
\end{align}
Consequently, the eigenvalues of the dynamical matrix $\mathcal{M}_\text{BdG}(\mathbf{k}) = \sigma_z \mathcal{H}_\text{BdG}(\mathbf{k})$, denoted $E_\pm(\mathbf{k})$, are readily obtained as
\begin{align}\label{Epm}
	E_\pm(\mathbf{k}) = & \pm 2i \Delta_0 \pm 2i \Delta_x \cos k_x + 2 t_y \sin k_y \notag \nonumber \\
	&  + 2 t_{xy} \sin (k_x+k_y).
\end{align}

Moreover, the dynamical matrix $\mathcal{M}_\text{BdG}(\mathbf{k})$ can be unitarily transformed into a block-diagonal form with
\begin{align}\label{UnitarykAppendix}
	U_\mathbf{k} \mathcal{M}_\text{BdG}(\mathbf{k}) U^\dagger_\mathbf{k} = \begin{pmatrix}
		E_+(\mathbf{k}) & 0 \\
		0 & E_-(\mathbf{k})
	\end{pmatrix},
\end{align}
where $U_\mathbf{k} = (\sigma_0-i\sigma_x)/\sqrt{2}$. 

In real space, the dynamical matrix $\mathcal{M}_\text{BdG}$ can likewise be expressed in a block-diagonal form 
\begin{align}\label{UnitaryRM}
	U \mathcal{M}_\text{BdG} U^\dagger = \begin{pmatrix}
		\mathcal{M}_p & \mathcal{O} \\
		\mathcal{O} & \mathcal{M}_m
	\end{pmatrix},
\end{align}
where $U = (\tau_0-i\tau_x)/\sqrt{2}$ with $U^\dagger = U^{-1}$, $\mathcal{M}_p$ and $\mathcal{M}_m$ denote the  decoupled matrices in real space.

We now apply the same unitary transformation to the dynamical matrix $\mathcal{M}$ in the presence of impurities, yielding $\bar{\mathcal{M}} = U \mathcal{M} U^\dagger$. Under this transformation, the Green’s function $\mathcal{G}$ in Eq.~(\ref{GM}) becomes
\begin{align}\label{GbarM}
	\bar{\mathcal{G}} = (E \tau_0-\bar{\mathcal{M}})^{-1} = (\tau_0-\bar{\mathcal{G}}_0 \bar{\mathcal{V}}_z)^{-1} \bar{\mathcal{G}}_0,
\end{align}
where $\bar{\mathcal{G}}_0 = (E\tau_0 - U\mathcal{M}_\text{BdG}U^\dagger)^{-1}$, and $\bar{\mathcal{V}}_z = U \mathcal{V}_z U^\dagger$.

\subsection{Spectral radius for on-site impurities}\label{AppendixA2}

Here, we focus on the spectral radius in a bosonic quadratic Hermitian system perturbed by on-site impurities $\hat{V} = V_1 \hat{a}^\dagger_{\mathbf{r}_1} \hat{a}_{\mathbf{r}_1} + V_2 \hat{a}^\dagger_{\mathbf{r}_2} \hat{a}_{\mathbf{r}_2}$ in Eq.~(\ref{onsiteimpurities2}). 

The unitary transformation of the Green’s function in Eq.~(\ref{GbarM}) gives rise to a  modified spectral stability criterion for infinitesimal impurities in Eq.~(\ref{SpectralStability}), expressed via the reduced spectral radius   with 
\begin{align}\label{SpectralStability22}
	\rho (\bar{\mathcal{G}}_0 \bar{\mathcal{V}}_z)\ll 1.
\end{align}

We now turn to a detailed discussion of the calculation of this spectral radius $\rho (\bar{\mathcal{G}}_0 \bar{\mathcal{V}}_z)$.

In position representation, the   response matrix $\bar{\mathcal{G}}_0 \bar{\mathcal{V}}_z$ can be written as   
\begin{widetext}
	\begin{align}\label{G0barVbar}
		\bar{\mathcal{G}}_0 \bar{\mathcal{V}}_z = &
		\begingroup
		\setlength{\tabcolsep}{4pt}             
		\renewcommand{\arraystretch}{2}        
		\begin{pmatrix}
			0 & 0 &  0 & \cdots  & 0 & iV_1 \bar{\mathcal{G}}^p_0(\mathbf{r}_1,\mathbf{r}_1;E) & iV_2 \bar{\mathcal{G}}^p_0(\mathbf{r}_1,\mathbf{r}_2;E) & 0 & \cdots & 0 \\	
			0 & 0 & 0 & \cdots  & 0 & iV_1 \bar{\mathcal{G}}^p_0(\mathbf{r}_2,\mathbf{r}_1;E) & iV_2 \bar{\mathcal{G}}^p_0(\mathbf{r}_2,\mathbf{r}_2;E) & 0 & \cdots & 0 \\
			\vdots & \vdots & \vdots & \ddots & \vdots & \vdots & \vdots & \vdots &  \ddots & \vdots \\	
			0 & 0 & 0 & \cdots & 0 & iV_1 \bar{\mathcal{G}}^p_0(\mathbf{r}_N,\mathbf{r}_1;E) & iV_2 \bar{\mathcal{G}}^p_0(\mathbf{r}_N,\mathbf{r}_2;E) & 0 & \cdots & 0 \\
			-iV_1 \bar{\mathcal{G}}^m_0(\mathbf{r}_1,\mathbf{r}_1;E) & -iV_2 \bar{\mathcal{G}}^m_0(\mathbf{r}_1,\mathbf{r}_2;E) &  0 &  \cdots & 0 &  0 & 0 & 0 & \cdots & 0 \\
			-iV_1 \bar{\mathcal{G}}^m_0(\mathbf{r}_2,\mathbf{r}_1;E) & -iV_2 \bar{\mathcal{G}}^m_0(\mathbf{r}_2,\mathbf{r}_2;E) &  0 & \cdots & 0 &  0 & 0 & 0 & \cdots & 0 \\
			\vdots & \vdots & \vdots & \ddots & \vdots & \vdots & \vdots & \vdots  & \ddots & \vdots \\	
			-iV_1 \bar{\mathcal{G}}^m_0(\mathbf{r}_N,\mathbf{r}_1;E) & -iV_2 \bar{\mathcal{G}}^m_0(\mathbf{r}_N,\mathbf{r}_2;E)  & 0 & \cdots & 0 & 0 & 0 & 0 & \cdots & 0 \\
		\end{pmatrix} \endgroup 
		 \nonumber \\  = &\begingroup 
		 	\setlength{\tabcolsep}{4pt}             
		 \renewcommand{\arraystretch}{2}        
		   \begin{pmatrix}
			\mathcal{O} & \mathcal{B} \\
			\mathcal{C} & \mathcal{O}
		\end{pmatrix},
		\endgroup
	\end{align}
\end{widetext}
where the element  of the unperturbed Green's function is given by
\begin{align}\label{Gbar0rirjM}
	\bar{\mathcal{G}}_0^{p/m}(\mathbf{r}_i,\mathbf{r}_j;E) = \bra{\mathbf{r}_i}(E-\mathcal{M}_{p/m})^{-1}\ket{\mathbf{r}_j}.
\end{align}

As discussed in detail in Appendix~\ref{AppendixA3}, the eigenvalues of the reduced response matrix $\bar{\mathcal{G}}_0 \bar{\mathcal{V}}_z$ in Eq.~(\ref{G0barVbar}) are determined by the matrix product $\mathcal{BC}$, where the response matrix possesses two (four) non-zero eigenvalues for single on-site impurity (double on-site impurities) [see Appendix~\ref{AppendixA3}].
 
When a single impurity is introduced ($V_1 \neq 0,, V_2 = 0$), the matrix $\mathcal{BC}$ has only one non-zero eigenvalue  
\begin{align}\label{BCSingle}
	\xi = V_1^2 \bar{\mathcal{G}}_0^p (\mathbf{r}_1,\mathbf{r}_1;E) \bar{\mathcal{G}}_0^m (\mathbf{r}_1,\mathbf{r}_1;E),
\end{align}
while the remaining $N-1$ eigenvalues are zero. Consequently, the response matrix $\bar{\mathcal{G}}_0 \bar{\mathcal{V}}_z$ has two non-zero eigenvalues, $\pm \sqrt{\xi}$, and $2N-2$ zeros. Then, the spectral radius is derived as
\begin{align}\label{SRS}
	\rho(\bar{\mathcal{G}}_0 \bar{\mathcal{V}}_z) = \abs{  V_1 \sqrt{\bar{\mathcal{G}}_0^p (\mathbf{r}_1,\mathbf{r}_1;E) \bar{\mathcal{G}}_0^m (\mathbf{r}_1,\mathbf{r}_1;E)}}.
\end{align}

Furthermore, when two impurities are introduced ($V_1, V_2 \neq 0$), the matrix $\mathcal{BC}$ has non-zero entries solely in the upper-left $2 \times 2$ block  with
\begin{align}\label{BCAppendixA}
	\mathcal{BC} = 		\begingroup
	\setlength{\tabcolsep}{5.8pt}             
	\renewcommand{\arraystretch}{1.5}        
	 \begin{pmatrix}
			\mathcal{BC}_{11} & \mathcal{BC}_{12} & 0 & \cdots & 0 \\
			\mathcal{BC}_{21} & \mathcal{BC}_{22} & 0 & \cdots & 0 \\
			0 & 0 & 0 &\cdots & 0 \\
			\vdots & \vdots & \vdots & \ddots & \vdots \\
			0 & 0 & 0 &\cdots & 0
	\end{pmatrix},
	\endgroup
\end{align}
where four non-zero elements of the matrix $\mathcal{BC}$ are written as
\begin{align}\label{equationAppendixA101} 
	\mathcal{BC}_{11} = ~& V_1^2 \bar{\mathcal{G}}_0^p (\mathbf{r}_1,\mathbf{r}_1;E) \bar{\mathcal{G}}_0^m (\mathbf{r}_1,\mathbf{r}_1;E) \nonumber \\ & + V_1V_2 \bar{\mathcal{G}}_0^p (\mathbf{r}_1,\mathbf{r}_2;E) \bar{\mathcal{G}}_0^m (\mathbf{r}_2,\mathbf{r}_1;E),   
\end{align}
\begin{align}\label{equationAppendixA102}
	\mathcal{BC}_{12} = ~& V_1V_2 \bar{\mathcal{G}}_0^p (\mathbf{r}_1,\mathbf{r}_1;E) \bar{\mathcal{G}}_0^m (\mathbf{r}_1,\mathbf{r}_2;E) \nonumber \\& + V^2_2 \bar{\mathcal{G}}_0^p (\mathbf{r}_1,\mathbf{r}_2;E) \bar{\mathcal{G}}_0^m (\mathbf{r}_2,\mathbf{r}_2;E), 		
\end{align}
\begin{align}\label{equationAppendixA103}
	\mathcal{BC}_{21} = ~& V_1^2 \bar{\mathcal{G}}_0^p (\mathbf{r}_2,\mathbf{r}_1;E) \bar{\mathcal{G}}_0^m (\mathbf{r}_1,\mathbf{r}_1;E) \nonumber \\& + V_1V_2 \bar{\mathcal{G}}_0^p (\mathbf{r}_2,\mathbf{r}_2;E) \bar{\mathcal{G}}_0^m (\mathbf{r}_2,\mathbf{r}_1;E), 		
\end{align}
\begin{align}\label{equationAppendixA104}	
	\mathcal{BC}_{22} = ~& V_1V_2 \bar{\mathcal{G}}_0^p (\mathbf{r}_2,\mathbf{r}_1;E) \bar{\mathcal{G}}_0^m (\mathbf{r}_1,\mathbf{r}_2;E) \nonumber \\ &+ V^2_2 \bar{\mathcal{G}}_0^p (\mathbf{r}_2,\mathbf{r}_2;E) \bar{\mathcal{G}}_0^m (\mathbf{r}_2,\mathbf{r}_2;E).
\end{align}

The matrix $\mathcal{BC}$ has two non-zero eigenvalues, which are  given by 
\begin{align}\label{xiAppendixA}
	\xi_\pm =& \frac{1}{2} \left(\mathcal{BC}_{11} + \mathcal{BC}_{22} \pm \sqrt{(\mathcal{BC}_{11}-\mathcal{BC}_{22})^2 + 4 \mathcal{BC}_{12} \mathcal{BC}_{21}}\right).
\end{align}
Therefore, the spectral radius of the response matrix for two onsite impurities is
\begin{align}\label{SRS2}
	\rho(\bar{\mathcal{G}}_0 \bar{\mathcal{V}}_z) = \max (\abs{\sqrt{\xi_\pm}}).
\end{align}

\subsection{Details on eigenvalues of  response matrix}\label{AppendixA3}

Here, we provide a detailed discussion of how the eigenvalues of the response matrix $\bar{\mathcal{G}}_0 \bar{\mathcal{V}}_z$ in Eq.~(\ref{G0barVbar}) are obtained, which is written in the off-diagonal block-matrix form as
\begin{align}\label{matrixBC}
	\bar{\mathcal{G}}_0 \bar{\mathcal{V}}_z = \begin{pmatrix}
		\mathcal{O} & \mathcal{B} \\
		\mathcal{C} & \mathcal{O}
	\end{pmatrix},
\end{align}
where $\mathcal{O}$ denotes the zero matrix, while $\mathcal{B}$ and $\mathcal{C}$ represent the upper-right and lower-left subblocks of the response matrix, respectively.

The   eigenvalue problem of the response matrix  is written as
\begin{align}\label{equationAppendixB2}
	\begin{pmatrix}
		\mathcal{O} & \mathcal{B} \\
		\mathcal{C} & \mathcal{O}
	\end{pmatrix} \begin{pmatrix}
	\mathbf{x} \\
	\mathbf{y}
	\end{pmatrix} = \lambda \begin{pmatrix}
	\mathbf{x} \\
	\mathbf{y}
	\end{pmatrix},
\end{align}
which yields the coupled equations
\begin{align}\label{equationAppendixB3}
	\begin{dcases}
		\mathcal{B} \mathbf{y} = \lambda \mathbf{x}, \\
		\mathcal{C} \mathbf{x} = \lambda \mathbf{y}.
	\end{dcases} 
\end{align}
By eliminating   either $\mathbf{x}$ or $\mathbf{y}$ from these equations, we obtain   reduced eigenvalue problems with  
\begin{align}\label{equation2AppendixB}
	\mathcal{BC}\mathbf{x}=\lambda^2\mathbf{x},~~~\mathcal{CB}\mathbf{y}=\lambda^2\mathbf{y}.
\end{align}

This relation shows that the nonzero eigenvalues $\lambda$ of the original block matrix must satisfy $\lambda^2 = \xi$, where $\xi$ is a nonzero eigenvalue of the matrix product $\mathcal{BC}$ (or $\mathcal{CB}$). As a result, for each nonzero $\xi$, the response operator has a pair of eigenvalues $\lambda = \pm \sqrt{\xi}$.

If either $\mathcal{BC}$ or $\mathcal{CB}$ possesses zero eigenvalues (i.e., $\xi = 0$), these translate directly into zero eigenvalues of the original off-diagonal block matrix in Eq.~(\ref{matrixBC}). The total number of such zero eigenvalues is determined by the nullity (dimension of the kernel) of the matrices $\mathcal{B}$ and $\mathcal{C}$. Specifically, $\dim (\ker (\mathcal{B})) + \dim (\ker (\mathcal{C}))$  gives the number of independent zero eigenvectors of the block matrix. For the matrix $\mathcal{B}$ and $\mathcal{C}$ defined in Eq.~(\ref{G0barVbar}), it follows that $\dim (\ker (\mathcal{B})) + \dim (\ker (\mathcal{C})) = 2N-2$ ($\dim (\ker (\mathcal{B})) + \dim (\ker (\mathcal{C})) = 2N-4$) for one impurity (two impurities). Therefore, the block matrix has exactly $2N - 2$ ($2N - 4$) zero eigenvalues for one impurity (two impurities).  These zero modes arise from the localized nature of the on-site impurities, which affect only a restricted subset of the response matrix. For spectral analysis, however, the focus is on the nonzero eigenvalues. The physically relevant part of the spectrum is governed by the matrix $\mathcal{BC}$, and all spectral radius calculations are therefore performed with respect to this matrix.

\section{Analytical Approximation of the Propagator and Spectral Radius}\label{AppendixB}

Here, we provide an analytical derivation of the propagator $\bar{\mathcal{G}}_0^{p/h}(\mathbf{r}_i,\mathbf{r}_j;E)$ for $i \neq j$ under suitable approximations, which enables an estimation of the spectral radius $\rho(\bar{\mathcal{G}}_0 \bar{\mathcal{V}}_z)$ and thereby offers physical insight into the system’s sensitivity to infinitesimal perturbations.

Using Eq.~(\ref{Gbar0rirjM}), we explicitly evaluate the propagator $\bar{\mathcal{G}}_0^{p/m}(\mathbf{r}_1,\mathbf{r}_2;E)$ under OBC along the $x$ direction and PBC along the $y$ direction for the cylindrical geometry. To capture the NHSE and perform the momentum-space integration in GBZ, we introduce the non-Bloch wavevector $\beta_x = e^{ik_x+ \mu_x}$  along the $x$ direction, while keeping the Bloch wavevector $k_y$ along the $y$ direction \cite{arXiv:2409.13623}. The propagator then reads
\begin{align}\label{Gbar0phr1r21}
	&  \bar{\mathcal{G}}_0^{p/m}(\mathbf{r}_1,\mathbf{r}_2;E) \nonumber \\ 
	\addlinespace[0.8em]
	& = \bra{\mathbf{r}_1}(E-\mathcal{M}_{p/m})^{-1}\ket{\mathbf{r}_2}    \nonumber  \\
	\addlinespace[0.8em]
	& = \int_{0}^{2\pi} dk_y \oint_{\beta_x \in \text{GBZ}} \frac{\text{d} \beta_x}{\beta_x} \frac{e^{ik_y(y_1-y_2)} \beta_x^{x_1-x_2}}{E-E_\pm(\beta_x,k_y)}. 
\end{align}
Here, the constant prefactor is omitted and  consistently neglected in the following derivations, and $E_\pm(\beta_x,k_y)$ reads
\begin{align}\label{Epm20}
	E_\pm(\beta_x, k_y) = & \pm i\Delta_x\left(\beta_x+\beta_x^{-1}\right) + it_y \left(e^{ik_y}-e^{-ik_y}\right)  \nonumber \\
	& \pm 2i\Delta_0 + it_{xy} \left(\beta_x e^{ik_y}-\beta_x^{-1}e^{-ik_y}\right) \nonumber \\
	&  + it_y \left(e^{ik_y}-e^{-ik_y}\right).
\end{align}

For the contour integral in Eq.~(\ref{Gbar0phr1r21}), we consider the new eigenvalue $E$ of the dynamical matrix $\mathcal{M}$, in the presence of onsite impurities, outside the eigenspectrum (indicated by $\sigma_\textrm{cyl}$) of $\mathcal{M}_{p/m}$  (i.e., $E \notin \sigma_\textrm{cyl}$) in the    cylindrical geometry. To carry out this  contour integral,  we solve  characteristic equations $E-E_\pm(\beta_x,k_y)=0$ for the propagators $\bar{\mathcal{G}}_0^{p/m}(\mathbf{r}_1,\mathbf{r}_2;E)$ in the subspace   defined in Eq.~(\ref{UnitaryRM}). The corresponding roots of the  characteristic equations, $\beta_{1,2}^\pm(k_y)$, are ordered such that $\abs{\beta_2^\pm(k_y)}>\abs{\beta_1^\pm(k_y)}$. Then, the equation (\ref{Epm20}) becomes
\begin{align}\label{Epm3}
	E - E_\pm(\beta_x,k_y) = \frac{\left(\beta_x - \beta^\pm_1(k_y)\right) \left(\beta_x - \beta^\pm_2(k_y)\right)}{\beta_x}.
\end{align}
Note that the GBZ is determined by standing-wave condition $\abs{\beta_1^\pm(k_y)}=\abs{\beta_2^\pm(k_y)}$.

Due to $\mathcal{M}_{p/m}^T \neq \mathcal{M}_{p/m}$, the propagator is non-reciprocal, $\bar{\mathcal{G}}_0^{p/m}(\mathbf{r}_1,\mathbf{r}_2;E) \neq \bar{\mathcal{G}}_0^{p/m}(\mathbf{r}_2,\mathbf{r}_1;E)$, and thus each expression must be evaluated separately.  Without loss of generality, we take $x_2 > x_1$, upon which Eq.~(\ref{Gbar0phr1r21}) can be further expanded as
\begin{align}\label{Gbar0phr1r22}
	\bar{\mathcal{G}}_0^{p/m}(&\mathbf{r}_2,\mathbf{r}_1;E)  = \int_{0}^{2\pi} dk_y e^{ik_y(y_2-y_1)} ~ \times   \nonumber \\
	\addlinespace[0.4em]
	&  \oint_{\beta_x \in \text{GBZ}} \text{d} \beta_x \frac{ \beta_x^{x_2-x_1}}{\left(\beta_x - \beta^\pm_1(k_y)\right) \left(\beta_x - \beta^\pm_2(k_y)\right)},    
\end{align}
and
\begin{align}\label{Gbar0phr1r23}
	  \bar{\mathcal{G}}_0^{p/m}(&\mathbf{r}_1,\mathbf{r}_2;E)  = \int_{0}^{2\pi} dk_y e^{ik_y(y_1-y_2)} ~ \times   \nonumber \\
	  \addlinespace[0.4em]
	&  \oint_{\beta_x \in \text{GBZ}} \text{d} \beta_x \frac{1}{\beta_x^{x_2-x_1} \left(\beta_x - \beta^\pm_1(k_y)\right) \left(\beta_x - \beta^\pm_2(k_y)\right)}.    
\end{align}

\begin{figure}[!tb]
	\centering
	\includegraphics[width=8cm]{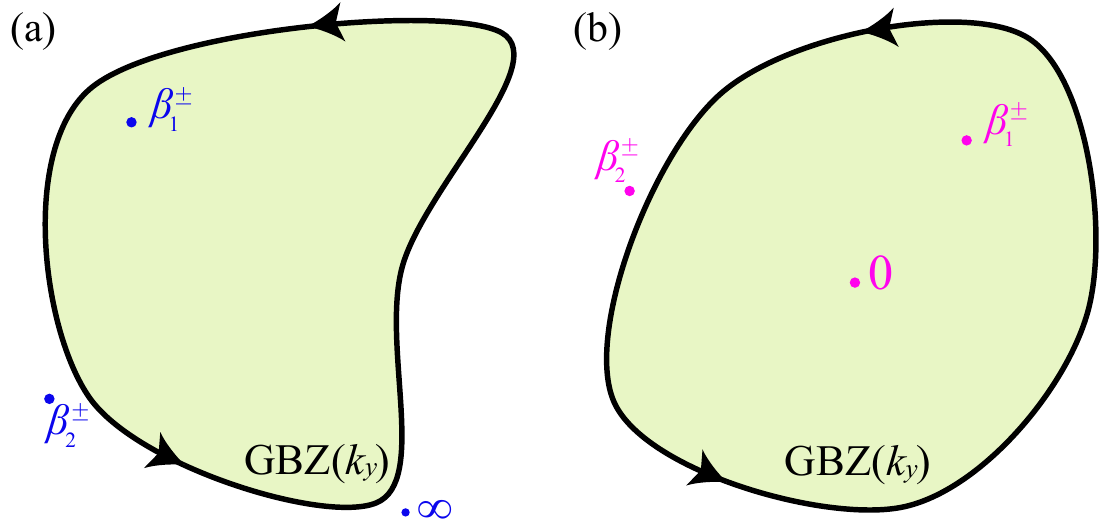}
	\caption{Schematic illustration of the isolated singularities associated with the contour integrals (a) for Eq.~(\ref{Gbar0phr1r22})  and (b) for Eq.~(\ref{Gbar0phr1r23}) . The GBZ trajectories (black curves) define the integration contours, while the polesof the integrand are indicated by dots. $\beta_1^\pm(k_y)$ lies inside the GBZ contour, whereas $\beta_2^\pm(k_y)$ lies outside it.}\label{FigS1}
\end{figure}

By applying Vieta’s formulas to the quadratic polynomial in Eq.~(\ref{Epm3}), we can verify   that $\beta_1^\pm(k_y)$ lies inside the GBZ contour, whereas $\beta_2^\pm(k_y)$ lies outside it, as shown in Fig.~\ref{FigS1}. When $x_2 > x_1$, the integrand of $\bar{\mathcal{G}}_0^{p/m}(\mathbf{r}_2,\mathbf{r}_1;E)$ therefore possesses  one pole $\beta_1^\pm$  inside the contour, and $\bar{\mathcal{G}}_0^{p/m}(\mathbf{r}_1,\mathbf{r}_2;E)$ has two poles $\beta_1^\pm$ [see Fig.~\ref{FigS1}(a)],   and $0$   inside the contour [see Fig.~\ref{FigS1}(b)]. Applying the residue theorem along the GBZ contour, we obtain
\begin{align}\label{Gbar0phr2r1E2Appendix}
	\bar{\mathcal{G}}_0^{p/m}(\mathbf{r}_2,\mathbf{r}_1;E) = \int_{0}^{2\pi} dk_y \frac{[\beta^\pm_1(k_y)]^{x_2-x_1} e^{ik_y(y_2-y_1)}}{\beta^\pm_1(k_y) - \beta^\pm_2(k_y)},    
\end{align}
and
\begin{align}\label{Gbar0phr1r2E2Appendix}
	\bar{\mathcal{G}}_0^{p/m}(\mathbf{r}_1,\mathbf{r}_2;E) = \int_{0}^{2\pi} dk_y \frac{[\beta^\pm_2(k_y)]^{x_1-x_2} e^{ik_y(y_1-y_2)}}{\beta^\pm_1(k_y) - \beta^\pm_2(k_y)}.    
\end{align}

In this section, our aim is to estimate the spectral radius $\rho(\bar{\mathcal{G}}_0 \bar{\mathcal{V}}_z)$ in Eq.~(\ref{SRS2}) for a large lattice size along   the $x$ direction. In the thermodynamic limit, the power-law scaling behavior in Eqs.~(\ref{Gbar0phr2r1E2Appendix}) and (\ref{Gbar0phr1r2E2Appendix}) enables us to approximate the propagators as
\begin{align}\label{Gbar0phr2r1E2New}
	\bar{\mathcal{G}}_0^{p/m}(\mathbf{r}_2,\mathbf{r}_1;E) \simeq \int_{0}^{2\pi} dk_y [\beta^\pm_1(k_y)]^{L} e^{ik_y\delta_y},
\end{align}
\begin{align}\label{Gbar0phr1r2E2New}
	\bar{\mathcal{G}}_0^{p/m}(\mathbf{r}_1,\mathbf{r}_2;E) \simeq \int_{0}^{2\pi} dk_y [\frac{1}{\beta^\pm_2(k_y)}]^{L} e^{-ik_y\delta_y},
\end{align}
where  $L=x_2-x_1$ and $\delta_y=y_2-y_1$.

For large $L$, the dominant contributions to Eqs.~(\ref{Gbar0phr2r1E2New}) and (\ref{Gbar0phr1r2E2New}) arise from values of $k_y$ where phase fluctuations in the integrands are effectively suppressed. This occurs when $\delta_y$ is tuned such that $[\beta^\pm_1(k_y)]^L e^{i k_y \delta_y}$ and $[\beta^\pm_2(k_y)]^{-L} e^{-i k_y \delta_y}$ vary slowly with $k_y$ \cite{arXiv:2409.13623}, thereby maximizing phase cancellation. Following this reasoning, we estimate the integrals in Eqs.~(\ref{Gbar0phr2r1E2New}) and (\ref{Gbar0phr1r2E2New}) by choosing values of $\delta_y$ for which the phase in the integrand is canceled \cite{arXiv:2409.13623}.

For the new eigenvalue $E$ with positive imaginary part, for instance, the eigenvalue $E$ indicated by the orange circle in Fig.~\ref{Fig3}(h), the integrals in Eqs.~(\ref{Gbar0phr2r1E2New}) and (\ref{Gbar0phr1r2E2New}) can be estimated as
\begin{align}\label{PBintegralAppendix}
	\begin{dcases}
		\bar{\mathcal{G}}_0^p(\mathbf{r}_2,\mathbf{r}_1;E) \sim e^{\mu_\text{max,1}^+ L}, \\
		\bar{\mathcal{G}}_0^m(\mathbf{r}_2,\mathbf{r}_1;E) \sim e^{\mu_\text{max,1}^- L} \sim 0, \\
		\bar{\mathcal{G}}_0^p(\mathbf{r}_1,\mathbf{r}_2;E) \sim e^{-\mu_\text{min,2}^+ L}, \\
		\bar{\mathcal{G}}_0^m(\mathbf{r}_1,\mathbf{r}_2;E) \sim e^{-\mu_\text{min,2}^- L} \sim 0,
	\end{dcases}
\end{align}
with 
\begin{align}\label{mu121}
\mu_\text{max,1}^\pm = \max_{k_y\in[0,2\pi]}\ln \abs{\beta_1^\pm(k_y)}, 
\end{align}
and
\begin{align}\label{mu122}
 \mu_\text{min,2}^\pm = \min_{k_y\in[0,2\pi]}\ln \abs{\beta_2^\pm(k_y)}. 
\end{align}
Here, $\mu_{\text{max},1}^+ > 0$, $\mu_{\text{max},1}^- < 0$, $\mu_{\text{min},2}^+ < 0$, and $\mu_{\text{min},2}^- > 0$ are typically satisfied in the parameter regime of interest [see red and blue dots in Figs.~\ref{Fig5}(a1) and (a3)], thereby facilitating the demonstration of the ultra spectral sensitivity.

Therefore, for double onsite impurities, the spectral radius associated with the induced new eigenvalue $E$ with positive imaginary part can be approximated by
\begin{align}\label{SRS2New1}
	\rho(\bar{\mathcal{G}}_0 \bar{\mathcal{V}}_z) \sim V e^{\frac{\mu_\text{max,1}^+ L}{4}} e^{-\frac{\mu_\text{min,2}^+ L}{4}}.
\end{align}
Here, we consider the case where the two onsite impurities have identical onsite potentials with $V_1 = V_2 = V$.

For the new eigenvalue $E$ with negetive imaginary part, for instance, the eigenvalue $E$ marked by the red circle in Fig.~\ref{Fig3}(h), the integrals in Eqs.~(\ref{Gbar0phr2r1E2New}), and (\ref{Gbar0phr1r2E2New}) can be estimated as
\begin{align}\label{HBintegralAppendix}
	\begin{dcases}
		\bar{\mathcal{G}}_0^p(\mathbf{r}_2,\mathbf{r}_1;E) \sim e^{\mu_\text{max,1}^+ L} \sim 0, \\
		\bar{\mathcal{G}}_0^m(\mathbf{r}_2,\mathbf{r}_1;E) \sim e^{\mu_\text{max,1}^- L},\\
		\bar{\mathcal{G}}_0^p(\mathbf{r}_1,\mathbf{r}_2;E) \sim e^{-\mu_\text{min,2}^+ L} \sim 0, \\
		\bar{\mathcal{G}}_0^m(\mathbf{r}_1,\mathbf{r}_2;E) \sim e^{-\mu_\text{min,2}^- L},
	\end{dcases}
\end{align}
where $\mu_{\text{max},1}^+ < 0$, $\mu_{\text{max},1}^- > 0$, $\mu_{\text{min},2}^+ > 0$, and $\mu_{\text{min},2}^- < 0$ are typically satisfied in the parameter regime of interest [see red and blue dots in Figs.~\ref{Fig5}(b1) and (b3)]. The corresponding the spectral radius associated with the induced new eigenvalue $E$ with negative imaginary part is
\begin{align}\label{SRS2New2}
	\rho(\bar{\mathcal{G}}_0 \bar{\mathcal{V}}_z) \sim V e^{\frac{\mu_\text{max,1}^- L}{4}} e^{-\frac{\mu_\text{min,2}^- L}{4}},	
\end{align}
where $V_1 = V_2 = V$ is assumed.


\begin{thebibliography}{118}%
	\makeatletter
	\providecommand \@ifxundefined [1]{%
		\@ifx{#1\undefined}
	}%
	\providecommand \@ifnum [1]{%
		\ifnum #1\expandafter \@firstoftwo
		\else \expandafter \@secondoftwo
		\fi
	}%
	\providecommand \@ifx [1]{%
		\ifx #1\expandafter \@firstoftwo
		\else \expandafter \@secondoftwo
		\fi
	}%
	\providecommand \natexlab [1]{#1}%
	\providecommand \enquote  [1]{``#1''}%
	\providecommand \bibnamefont  [1]{#1}%
	\providecommand \bibfnamefont [1]{#1}%
	\providecommand \citenamefont [1]{#1}%
	\providecommand \href@noop [0]{\@secondoftwo}%
	\providecommand \href [0]{\begingroup \@sanitize@url \@href}%
	\providecommand \@href[1]{\@@startlink{#1}\@@href}%
	\providecommand \@@href[1]{\endgroup#1\@@endlink}%
	\providecommand \@sanitize@url [0]{\catcode `\\12\catcode `\$12\catcode
		`\&12\catcode `\#12\catcode `\^12\catcode `\_12\catcode `\%12\relax}%
	\providecommand \@@startlink[1]{}%
	\providecommand \@@endlink[0]{}%
	\providecommand \url  [0]{\begingroup\@sanitize@url \@url }%
	\providecommand \@url [1]{\endgroup\@href {#1}{\urlprefix }}%
	\providecommand \urlprefix  [0]{URL }%
	\providecommand \Eprint [0]{\href }%
	\providecommand \doibase [0]{http://dx.doi.org/}%
	\providecommand \selectlanguage [0]{\@gobble}%
	\providecommand \bibinfo  [0]{\@secondoftwo}%
	\providecommand \bibfield  [0]{\@secondoftwo}%
	\providecommand \translation [1]{[#1]}%
	\providecommand \BibitemOpen [0]{}%
	\providecommand \bibitemStop [0]{}%
	\providecommand \bibitemNoStop [0]{.\EOS\space}%
	\providecommand \EOS [0]{\spacefactor3000\relax}%
	\providecommand \BibitemShut  [1]{\csname bibitem#1\endcsname}%
	\let\auto@bib@innerbib\@empty
	\bibitem [{\citenamefont {Ashida}\ \emph {et~al.}(2020)\citenamefont {Ashida},
		\citenamefont {Gong},\ and\ \citenamefont {Ueda}}]{Ashida2020}%
	\BibitemOpen
	\bibfield  {author} {\bibinfo {author} {\bibfnamefont {Y.}~\bibnamefont
			{Ashida}}, \bibinfo {author} {\bibfnamefont {Z.}~\bibnamefont {Gong}}, \ and\
		\bibinfo {author} {\bibfnamefont {M.}~\bibnamefont {Ueda}},\ }\bibfield
	{title} {\enquote {\bibinfo {title} {Non-{H}ermitian physics},}\ }\href
	{\doibase 10.1080/00018732.2021.1876991} {\bibfield  {journal} {\bibinfo
			{journal} {Adv. Phys.}\ }\textbf {\bibinfo {volume} {69}},\ \bibinfo {pages}
		{249} (\bibinfo {year} {2020})}\BibitemShut {NoStop}%
	\bibitem [{\citenamefont {Minganti}\ \emph {et~al.}(2020)\citenamefont
		{Minganti}, \citenamefont {Miranowicz}, \citenamefont {Chhajlany},
		\citenamefont {Arkhipov},\ and\ \citenamefont {Nori}}]{PhysRevA.101.062112}%
	\BibitemOpen
	\bibfield  {author} {\bibinfo {author} {\bibfnamefont {F.}~\bibnamefont
			{Minganti}}, \bibinfo {author} {\bibfnamefont {A.}~\bibnamefont
			{Miranowicz}}, \bibinfo {author} {\bibfnamefont {R.~W.}\ \bibnamefont
			{Chhajlany}}, \bibinfo {author} {\bibfnamefont {I.~I.}\ \bibnamefont
			{Arkhipov}}, \ and\ \bibinfo {author} {\bibfnamefont {F.}~\bibnamefont
			{Nori}},\ }\bibfield  {title} {\enquote {\bibinfo {title}
			{Hybrid-{L}iouvillian formalism connecting exceptional points of
				non-{H}ermitian {H}amiltonians and {L}iouvillians via postselection of
				quantum trajectories},}\ }\href {\doibase 10.1103/PhysRevA.101.062112}
	{\bibfield  {journal} {\bibinfo  {journal} {Phys. Rev. A}\ }\textbf {\bibinfo
			{volume} {101}},\ \bibinfo {pages} {062112} (\bibinfo {year}
		{2020})}\BibitemShut {NoStop}%
	\bibitem [{\citenamefont {Bergholtz}\ \emph {et~al.}(2021)\citenamefont
		{Bergholtz}, \citenamefont {Budich},\ and\ \citenamefont
		{Kunst}}]{RevModPhys.93.015005}%
	\BibitemOpen
	\bibfield  {author} {\bibinfo {author} {\bibfnamefont {E.~J.}\ \bibnamefont
			{Bergholtz}}, \bibinfo {author} {\bibfnamefont {J.~C.}\ \bibnamefont
			{Budich}}, \ and\ \bibinfo {author} {\bibfnamefont {F.~K.}\ \bibnamefont
			{Kunst}},\ }\bibfield  {title} {\enquote {\bibinfo {title} {Exceptional
				topology of non-{H}ermitian systems},}\ }\href {\doibase
		10.1103/RevModPhys.93.015005} {\bibfield  {journal} {\bibinfo  {journal}
			{Rev. Mod. Phys.}\ }\textbf {\bibinfo {volume} {93}},\ \bibinfo {pages}
		{015005} (\bibinfo {year} {2021})}\BibitemShut {NoStop}%
	\bibitem [{\citenamefont {Micallo}\ \emph {et~al.}(2023)\citenamefont
		{Micallo}, \citenamefont {Lehmann},\ and\ \citenamefont
		{Budich}}]{PhysRevResearch.5.043105}%
	\BibitemOpen
	\bibfield  {author} {\bibinfo {author} {\bibfnamefont {T.}~\bibnamefont
			{Micallo}}, \bibinfo {author} {\bibfnamefont {C.}~\bibnamefont {Lehmann}}, \
		and\ \bibinfo {author} {\bibfnamefont {J.~C.}\ \bibnamefont {Budich}},\
	}\bibfield  {title} {\enquote {\bibinfo {title} {Correlation-induced
				sensitivity and non-{H}ermitian skin effect of quasiparticles},}\ }\href
	{\doibase 10.1103/PhysRevResearch.5.043105} {\bibfield  {journal} {\bibinfo
			{journal} {Phys. Rev. Res.}\ }\textbf {\bibinfo {volume} {5}},\ \bibinfo
		{pages} {043105} (\bibinfo {year} {2023})}\BibitemShut {NoStop}%
	\bibitem [{\citenamefont {Leefmans}\ \emph {et~al.}(2022)\citenamefont
		{Leefmans}, \citenamefont {Dutt}, \citenamefont {Williams}, \citenamefont
		{Yuan}, \citenamefont {Parto}, \citenamefont {Nori}, \citenamefont {Fan},\
		and\ \citenamefont {Marandi}}]{Leefmans2022}%
	\BibitemOpen
	\bibfield  {author} {\bibinfo {author} {\bibfnamefont {C.}~\bibnamefont
			{Leefmans}}, \bibinfo {author} {\bibfnamefont {A.}~\bibnamefont {Dutt}},
		\bibinfo {author} {\bibfnamefont {J.}~\bibnamefont {Williams}}, \bibinfo
		{author} {\bibfnamefont {L.}~\bibnamefont {Yuan}}, \bibinfo {author}
		{\bibfnamefont {M.}~\bibnamefont {Parto}}, \bibinfo {author} {\bibfnamefont
			{F.}~\bibnamefont {Nori}}, \bibinfo {author} {\bibfnamefont {S.}~\bibnamefont
			{Fan}}, \ and\ \bibinfo {author} {\bibfnamefont {A.}~\bibnamefont
			{Marandi}},\ }\bibfield  {title} {\enquote {\bibinfo {title} {Topological
				dissipation in a time-multiplexed photonic resonator network},}\ }\href
	{\doibase 10.1038/s41567-021-01492-w} {\bibfield  {journal} {\bibinfo
			{journal} {Nat. Phys.}\ }\textbf {\bibinfo {volume} {18}},\ \bibinfo {pages}
		{442} (\bibinfo {year} {2022})}\BibitemShut {NoStop}%
	\bibitem [{\citenamefont {Qin}\ \emph {et~al.}(2024)\citenamefont {Qin},
		\citenamefont {Zhang},\ and\ \citenamefont {Li}}]{PhysRevA.109.023317}%
	\BibitemOpen
	\bibfield  {author} {\bibinfo {author} {\bibfnamefont {Y.}~\bibnamefont
			{Qin}}, \bibinfo {author} {\bibfnamefont {K.}~\bibnamefont {Zhang}}, \ and\
		\bibinfo {author} {\bibfnamefont {L.}~\bibnamefont {Li}},\ }\bibfield
	{title} {\enquote {\bibinfo {title} {Geometry-dependent skin effect and
				anisotropic {B}loch oscillations in a non-{H}ermitian optical lattice},}\
	}\href {\doibase 10.1103/PhysRevA.109.023317} {\bibfield  {journal} {\bibinfo
			{journal} {Phys. Rev. A}\ }\textbf {\bibinfo {volume} {109}},\ \bibinfo
		{pages} {023317} (\bibinfo {year} {2024})}\BibitemShut {NoStop}%
	\bibitem [{\citenamefont {Xiao}\ and\ \citenamefont
		{Zeng}(2024)}]{PhysRevB.110.024205}%
	\BibitemOpen
	\bibfield  {author} {\bibinfo {author} {\bibfnamefont {H.}~\bibnamefont
			{Xiao}}\ and\ \bibinfo {author} {\bibfnamefont {Q.-B.}\ \bibnamefont
			{Zeng}},\ }\bibfield  {title} {\enquote {\bibinfo {title} {Coexistence of
				non-{H}ermitian skin effect and extended states in one-dimensional
				nonreciprocal lattices},}\ }\href {\doibase 10.1103/PhysRevB.110.024205}
	{\bibfield  {journal} {\bibinfo  {journal} {Phys. Rev. B}\ }\textbf {\bibinfo
			{volume} {110}},\ \bibinfo {pages} {024205} (\bibinfo {year}
		{2024})}\BibitemShut {NoStop}%
	\bibitem [{\citenamefont {Reisenbauer}\ \emph {et~al.}(2024)\citenamefont
		{Reisenbauer}, \citenamefont {Rudolph}, \citenamefont {Egyed}, \citenamefont
		{Hornberger}, \citenamefont {Zasedatelev}, \citenamefont {Abuzarli},
		\citenamefont {Stickler},\ and\ \citenamefont {Delić}}]{Reisenbauer2024}%
	\BibitemOpen
	\bibfield  {author} {\bibinfo {author} {\bibfnamefont {M.}~\bibnamefont
			{Reisenbauer}}, \bibinfo {author} {\bibfnamefont {H.}~\bibnamefont
			{Rudolph}}, \bibinfo {author} {\bibfnamefont {L.}~\bibnamefont {Egyed}},
		\bibinfo {author} {\bibfnamefont {K.}~\bibnamefont {Hornberger}}, \bibinfo
		{author} {\bibfnamefont {A.~V.}\ \bibnamefont {Zasedatelev}}, \bibinfo
		{author} {\bibfnamefont {M.}~\bibnamefont {Abuzarli}}, \bibinfo {author}
		{\bibfnamefont {B.~A.}\ \bibnamefont {Stickler}}, \ and\ \bibinfo {author}
		{\bibfnamefont {U.}~\bibnamefont {Delić}},\ }\bibfield  {title} {\enquote
		{\bibinfo {title} {Non-{H}ermitian dynamics and non-reciprocity of optically
				coupled nanoparticles},}\ }\href {\doibase 10.1038/s41567-024-02589-8}
	{\bibfield  {journal} {\bibinfo  {journal} {Nat. Phys.}\ }\textbf {\bibinfo
			{volume} {20}},\ \bibinfo {pages} {1629} (\bibinfo {year}
		{2024})}\BibitemShut {NoStop}%
	\bibitem [{\citenamefont {Ochkan}\ \emph {et~al.}(2024)\citenamefont {Ochkan},
		\citenamefont {Chaturvedi}, \citenamefont {K\"{o}nye}, \citenamefont
		{Veyrat}, \citenamefont {Giraud}, \citenamefont {Mailly}, \citenamefont
		{Cavanna}, \citenamefont {Gennser}, \citenamefont {Hankiewicz}, \citenamefont
		{B\"{u}chner}, \citenamefont {van~den Brink}, \citenamefont {Dufouleur},\
		and\ \citenamefont {Fulga}}]{Ochkan2024}%
	\BibitemOpen
	\bibfield  {author} {\bibinfo {author} {\bibfnamefont {K.}~\bibnamefont
			{Ochkan}}, \bibinfo {author} {\bibfnamefont {R.}~\bibnamefont {Chaturvedi}},
		\bibinfo {author} {\bibfnamefont {V.}~\bibnamefont {K\"{o}nye}}, \bibinfo
		{author} {\bibfnamefont {L.}~\bibnamefont {Veyrat}}, \bibinfo {author}
		{\bibfnamefont {R.}~\bibnamefont {Giraud}}, \bibinfo {author} {\bibfnamefont
			{D.}~\bibnamefont {Mailly}}, \bibinfo {author} {\bibfnamefont
			{A.}~\bibnamefont {Cavanna}}, \bibinfo {author} {\bibfnamefont
			{U.}~\bibnamefont {Gennser}}, \bibinfo {author} {\bibfnamefont {E.~M.}\
			\bibnamefont {Hankiewicz}}, \bibinfo {author} {\bibfnamefont
			{B.}~\bibnamefont {B\"{u}chner}}, \bibinfo {author} {\bibfnamefont
			{J.}~\bibnamefont {van~den Brink}}, \bibinfo {author} {\bibfnamefont
			{J.}~\bibnamefont {Dufouleur}}, \ and\ \bibinfo {author} {\bibfnamefont
			{I.~C.}\ \bibnamefont {Fulga}},\ }\bibfield  {title} {\enquote {\bibinfo
			{title} {Non-{H}ermitian topology in a multi-terminal quantum hall device},}\
	}\href {\doibase 10.1038/s41567-023-02337-4} {\bibfield  {journal} {\bibinfo
			{journal} {Nat. Phys.}\ }\textbf {\bibinfo {volume} {20}},\ \bibinfo {pages}
		{395} (\bibinfo {year} {2024})}\BibitemShut {NoStop}%
	\bibitem [{\citenamefont {Hu}\ \emph {et~al.}(2024)\citenamefont {Hu},
		\citenamefont {Wang}, \citenamefont {Wang},\ and\ \citenamefont
		{Song}}]{PhysRevLett.132.050402}%
	\BibitemOpen
	\bibfield  {author} {\bibinfo {author} {\bibfnamefont {Y.-M.}\ \bibnamefont
			{Hu}}, \bibinfo {author} {\bibfnamefont {H.-Y.}\ \bibnamefont {Wang}},
		\bibinfo {author} {\bibfnamefont {Z.}~\bibnamefont {Wang}}, \ and\ \bibinfo
		{author} {\bibfnamefont {F.}~\bibnamefont {Song}},\ }\bibfield  {title}
	{\enquote {\bibinfo {title} {Geometric origin of non-{B}loch
				$\mathcal{P}\mathcal{T}$ symmetry breaking},}\ }\href {\doibase
		10.1103/PhysRevLett.132.050402} {\bibfield  {journal} {\bibinfo  {journal}
			{Phys. Rev. Lett.}\ }\textbf {\bibinfo {volume} {132}},\ \bibinfo {pages}
		{050402} (\bibinfo {year} {2024})}\BibitemShut {NoStop}%
	\bibitem [{\citenamefont {Longhi}(2025)}]{PhysRevLett.134.196302}%
	\BibitemOpen
	\bibfield  {author} {\bibinfo {author} {\bibfnamefont {S.}~\bibnamefont
			{Longhi}},\ }\bibfield  {title} {\enquote {\bibinfo {title} {Erratic
				non-{H}ermitian skin localization},}\ }\href {\doibase
		10.1103/PhysRevLett.134.196302} {\bibfield  {journal} {\bibinfo  {journal}
			{Phys. Rev. Lett.}\ }\textbf {\bibinfo {volume} {134}},\ \bibinfo {pages}
		{196302} (\bibinfo {year} {2025})}\BibitemShut {NoStop}%
	\bibitem [{\citenamefont {Zhang}\ \emph {et~al.}(2024)\citenamefont {Zhang},
		\citenamefont {Yang},\ and\ \citenamefont {Sun}}]{PhysRevB.109.165127}%
	\BibitemOpen
	\bibfield  {author} {\bibinfo {author} {\bibfnamefont {K.}~\bibnamefont
			{Zhang}}, \bibinfo {author} {\bibfnamefont {Z.}~\bibnamefont {Yang}}, \ and\
		\bibinfo {author} {\bibfnamefont {K.}~\bibnamefont {Sun}},\ }\bibfield
	{title} {\enquote {\bibinfo {title} {Edge theory of non-{H}ermitian skin
				modes in higher dimensions},}\ }\href {\doibase 10.1103/PhysRevB.109.165127}
	{\bibfield  {journal} {\bibinfo  {journal} {Phys. Rev. B}\ }\textbf {\bibinfo
			{volume} {109}},\ \bibinfo {pages} {165127} (\bibinfo {year}
		{2024})}\BibitemShut {NoStop}%
	\bibitem [{\citenamefont {Li}\ \emph {et~al.}(2025)\citenamefont {Li},
		\citenamefont {Chen},\ and\ \citenamefont {Wang}}]{z9m1-3mwb}%
	\BibitemOpen
	\bibfield  {author} {\bibinfo {author} {\bibfnamefont {B.}~\bibnamefont
			{Li}}, \bibinfo {author} {\bibfnamefont {C.}~\bibnamefont {Chen}}, \ and\
		\bibinfo {author} {\bibfnamefont {Z.}~\bibnamefont {Wang}},\ }\bibfield
	{title} {\enquote {\bibinfo {title} {Universal non-{H}ermitian transport in
				disordered systems},}\ }\href {\doibase 10.1103/z9m1-3mwb} {\bibfield
		{journal} {\bibinfo  {journal} {Phys. Rev. Lett.}\ }\textbf {\bibinfo
			{volume} {135}},\ \bibinfo {pages} {033802} (\bibinfo {year}
		{2025})}\BibitemShut {NoStop}%
	\bibitem [{\citenamefont {Wang}\ and\ \citenamefont
		{Yan}(2025)}]{PhysRevResearch.7.L022037}%
	\BibitemOpen
	\bibfield  {author} {\bibinfo {author} {\bibfnamefont {S.-X.}\ \bibnamefont
			{Wang}}\ and\ \bibinfo {author} {\bibfnamefont {Z.}~\bibnamefont {Yan}},\
	}\bibfield  {title} {\enquote {\bibinfo {title} {Enhanced sensitivity in
				non-hermitian systems at infernal points},}\ }\href {\doibase
		10.1103/PhysRevResearch.7.L022037} {\bibfield  {journal} {\bibinfo  {journal}
			{Phys. Rev. Res.}\ }\textbf {\bibinfo {volume} {7}},\ \bibinfo {pages}
		{L022037} (\bibinfo {year} {2025})}\BibitemShut {NoStop}%
	\bibitem [{\citenamefont {Gliozzi}\ \emph {et~al.}(2024)\citenamefont
		{Gliozzi}, \citenamefont {De~Tomasi},\ and\ \citenamefont
		{Hughes}}]{PhysRevLett.133.136503}%
	\BibitemOpen
	\bibfield  {author} {\bibinfo {author} {\bibfnamefont {J.}~\bibnamefont
			{Gliozzi}}, \bibinfo {author} {\bibfnamefont {G.}~\bibnamefont {De~Tomasi}},
		\ and\ \bibinfo {author} {\bibfnamefont {T.~L.}\ \bibnamefont {Hughes}},\
	}\bibfield  {title} {\enquote {\bibinfo {title} {Many-body non-{H}ermitian
				skin effect for multipoles},}\ }\href {\doibase
		10.1103/PhysRevLett.133.136503} {\bibfield  {journal} {\bibinfo  {journal}
			{Phys. Rev. Lett.}\ }\textbf {\bibinfo {volume} {133}},\ \bibinfo {pages}
		{136503} (\bibinfo {year} {2024})}\BibitemShut {NoStop}%
	\bibitem [{\citenamefont {Yoshida}\ \emph {et~al.}(2024)\citenamefont
		{Yoshida}, \citenamefont {Zhang}, \citenamefont {Neupert},\ and\
		\citenamefont {Kawakami}}]{PhysRevLett.133.076502}%
	\BibitemOpen
	\bibfield  {author} {\bibinfo {author} {\bibfnamefont {T.}~\bibnamefont
			{Yoshida}}, \bibinfo {author} {\bibfnamefont {S.-B.}\ \bibnamefont {Zhang}},
		\bibinfo {author} {\bibfnamefont {T.}~\bibnamefont {Neupert}}, \ and\
		\bibinfo {author} {\bibfnamefont {N.}~\bibnamefont {Kawakami}},\ }\bibfield
	{title} {\enquote {\bibinfo {title} {Non-{H}ermitian {M}ott skin effect},}\
	}\href {\doibase 10.1103/PhysRevLett.133.076502} {\bibfield  {journal}
		{\bibinfo  {journal} {Phys. Rev. Lett.}\ }\textbf {\bibinfo {volume} {133}},\
		\bibinfo {pages} {076502} (\bibinfo {year} {2024})}\BibitemShut {NoStop}%
	\bibitem [{\citenamefont {Yang}\ \emph {et~al.}(2025)\citenamefont {Yang},
		\citenamefont {Fang}, \citenamefont {Zhang},\ and\ \citenamefont
		{Fang}}]{Yang2025}%
	\BibitemOpen
	\bibfield  {author} {\bibinfo {author} {\bibfnamefont {A.}~\bibnamefont
			{Yang}}, \bibinfo {author} {\bibfnamefont {Z.}~\bibnamefont {Fang}}, \bibinfo
		{author} {\bibfnamefont {K.}~\bibnamefont {Zhang}}, \ and\ \bibinfo {author}
		{\bibfnamefont {C.}~\bibnamefont {Fang}},\ }\bibfield  {title} {\enquote
		{\bibinfo {title} {Tailoring bound state geometry in high-dimensional
				non-{H}ermitian systems},}\ }\href {\doibase 10.1038/s42005-025-02037-w}
	{\bibfield  {journal} {\bibinfo  {journal} {Commun. Phys.}\ }\textbf
		{\bibinfo {volume} {8}},\ \bibinfo {pages} {124} (\bibinfo {year}
		{2025})}\BibitemShut {NoStop}%
	\bibitem [{\citenamefont {Molignini}\ \emph {et~al.}(2023)\citenamefont
		{Molignini}, \citenamefont {Arandes},\ and\ \citenamefont
		{Bergholtz}}]{PhysRevResearch.5.033058}%
	\BibitemOpen
	\bibfield  {author} {\bibinfo {author} {\bibfnamefont {P.}~\bibnamefont
			{Molignini}}, \bibinfo {author} {\bibfnamefont {O.}~\bibnamefont {Arandes}},
		\ and\ \bibinfo {author} {\bibfnamefont {E.~J.}\ \bibnamefont {Bergholtz}},\
	}\bibfield  {title} {\enquote {\bibinfo {title} {Anomalous skin effects in
				disordered systems with a single non-{H}ermitian impurity},}\ }\href
	{\doibase 10.1103/PhysRevResearch.5.033058} {\bibfield  {journal} {\bibinfo
			{journal} {Phys. Rev. Res.}\ }\textbf {\bibinfo {volume} {5}},\ \bibinfo
		{pages} {033058} (\bibinfo {year} {2023})}\BibitemShut {NoStop}%
	\bibitem [{\citenamefont {Jiang}\ and\ \citenamefont
		{Lee}(2023)}]{PhysRevLett.131.076401}%
	\BibitemOpen
	\bibfield  {author} {\bibinfo {author} {\bibfnamefont {H.}~\bibnamefont
			{Jiang}}\ and\ \bibinfo {author} {\bibfnamefont {C.~H.}\ \bibnamefont
			{Lee}},\ }\bibfield  {title} {\enquote {\bibinfo {title} {Dimensional
				transmutation from non-{H}ermiticity},}\ }\href {\doibase
		10.1103/PhysRevLett.131.076401} {\bibfield  {journal} {\bibinfo  {journal}
			{Phys. Rev. Lett.}\ }\textbf {\bibinfo {volume} {131}},\ \bibinfo {pages}
		{076401} (\bibinfo {year} {2023})}\BibitemShut {NoStop}%
	\bibitem [{\citenamefont {Li}\ \emph {et~al.}(2023{\natexlab{a}})\citenamefont
		{Li}, \citenamefont {Trauzettel}, \citenamefont {Neupert},\ and\
		\citenamefont {Zhang}}]{PhysRevLett.131.116601}%
	\BibitemOpen
	\bibfield  {author} {\bibinfo {author} {\bibfnamefont {C.-A.}\ \bibnamefont
			{Li}}, \bibinfo {author} {\bibfnamefont {B.}~\bibnamefont {Trauzettel}},
		\bibinfo {author} {\bibfnamefont {T.}~\bibnamefont {Neupert}}, \ and\
		\bibinfo {author} {\bibfnamefont {S.-B.}\ \bibnamefont {Zhang}},\ }\bibfield
	{title} {\enquote {\bibinfo {title} {Enhancement of second-order
				non-{H}ermitian skin effect by magnetic fields},}\ }\href {\doibase
		10.1103/PhysRevLett.131.116601} {\bibfield  {journal} {\bibinfo  {journal}
			{Phys. Rev. Lett.}\ }\textbf {\bibinfo {volume} {131}},\ \bibinfo {pages}
		{116601} (\bibinfo {year} {2023}{\natexlab{a}})}\BibitemShut {NoStop}%
	\bibitem [{\citenamefont {Gong}\ \emph {et~al.}(2022)\citenamefont {Gong},
		\citenamefont {Bello}, \citenamefont {Malz},\ and\ \citenamefont
		{Kunst}}]{PhysRevLett.129.223601}%
	\BibitemOpen
	\bibfield  {author} {\bibinfo {author} {\bibfnamefont {Z.}~\bibnamefont
			{Gong}}, \bibinfo {author} {\bibfnamefont {M.}~\bibnamefont {Bello}},
		\bibinfo {author} {\bibfnamefont {D.}~\bibnamefont {Malz}}, \ and\ \bibinfo
		{author} {\bibfnamefont {F.~K.}\ \bibnamefont {Kunst}},\ }\bibfield  {title}
	{\enquote {\bibinfo {title} {Anomalous behaviors of quantum emitters in
				non-{H}ermitian baths},}\ }\href {\doibase 10.1103/PhysRevLett.129.223601}
	{\bibfield  {journal} {\bibinfo  {journal} {Phys. Rev. Lett.}\ }\textbf
		{\bibinfo {volume} {129}},\ \bibinfo {pages} {223601} (\bibinfo {year}
		{2022})}\BibitemShut {NoStop}%
	\bibitem [{\citenamefont {Cai}\ \emph {et~al.}(2025{\natexlab{a}})\citenamefont
		{Cai}, \citenamefont {Wang}, \citenamefont {Zhang}, \citenamefont {Liu},\
		and\ \citenamefont {Nori}}]{Cai2025}%
	\BibitemOpen
	\bibfield  {author} {\bibinfo {author} {\bibfnamefont {Z.-F.}\ \bibnamefont
			{Cai}}, \bibinfo {author} {\bibfnamefont {Y.-C.}\ \bibnamefont {Wang}},
		\bibinfo {author} {\bibfnamefont {Y.-R.}\ \bibnamefont {Zhang}}, \bibinfo
		{author} {\bibfnamefont {T.}~\bibnamefont {Liu}}, \ and\ \bibinfo {author}
		{\bibfnamefont {F.}~\bibnamefont {Nori}},\ }\bibfield  {title} {\enquote
		{\bibinfo {title} {Versatile control of nonlinear topological states in
				non-{H}ermitian systems},}\ }\href {\doibase 10.1038/s42005-025-02286-9}
	{\bibfield  {journal} {\bibinfo  {journal} {Commun. Phys.}\ }\textbf
		{\bibinfo {volume} {8}},\ \bibinfo {pages} {360} (\bibinfo {year}
		{2025}{\natexlab{a}})}\BibitemShut {NoStop}%
	\bibitem [{\citenamefont {Mo}\ \emph {et~al.}(2025)\citenamefont {Mo},
		\citenamefont {Xiao}, \citenamefont {Moessner},\ and\ \citenamefont
		{Zhao}}]{PhysRevB.111.235412}%
	\BibitemOpen
	\bibfield  {author} {\bibinfo {author} {\bibfnamefont {L.-H.}\ \bibnamefont
			{Mo}}, \bibinfo {author} {\bibfnamefont {Z.}~\bibnamefont {Xiao}}, \bibinfo
		{author} {\bibfnamefont {R.}~\bibnamefont {Moessner}}, \ and\ \bibinfo
		{author} {\bibfnamefont {H.}~\bibnamefont {Zhao}},\ }\bibfield  {title}
	{\enquote {\bibinfo {title} {Non-{H}ermitian delocalization in one dimension
				via emergent compactness},}\ }\href {\doibase 10.1103/PhysRevB.111.235412}
	{\bibfield  {journal} {\bibinfo  {journal} {Phys. Rev. B}\ }\textbf {\bibinfo
			{volume} {111}},\ \bibinfo {pages} {235412} (\bibinfo {year}
		{2025})}\BibitemShut {NoStop}%
	\bibitem [{\citenamefont {Ohnmacht}\ \emph {et~al.}(2025)\citenamefont
		{Ohnmacht}, \citenamefont {Wilhelm}, \citenamefont {Weisbrich},\ and\
		\citenamefont {Belzig}}]{PhysRevLett.134.156601}%
	\BibitemOpen
	\bibfield  {author} {\bibinfo {author} {\bibfnamefont {D.~C.}\ \bibnamefont
			{Ohnmacht}}, \bibinfo {author} {\bibfnamefont {V.}~\bibnamefont {Wilhelm}},
		\bibinfo {author} {\bibfnamefont {H.}~\bibnamefont {Weisbrich}}, \ and\
		\bibinfo {author} {\bibfnamefont {W.}~\bibnamefont {Belzig}},\ }\bibfield
	{title} {\enquote {\bibinfo {title} {Non-{H}ermitian topology in
				multiterminal superconducting junctions},}\ }\href {\doibase
		10.1103/PhysRevLett.134.156601} {\bibfield  {journal} {\bibinfo  {journal}
			{Phys. Rev. Lett.}\ }\textbf {\bibinfo {volume} {134}},\ \bibinfo {pages}
		{156601} (\bibinfo {year} {2025})}\BibitemShut {NoStop}%
	\bibitem [{\citenamefont {Shi}\ and\ \citenamefont
		{Poddubny}(2025)}]{q6wr-2rt9}%
	\BibitemOpen
	\bibfield  {author} {\bibinfo {author} {\bibfnamefont {J.}~\bibnamefont
			{Shi}}\ and\ \bibinfo {author} {\bibfnamefont {A.~N.}\ \bibnamefont
			{Poddubny}},\ }\bibfield  {title} {\enquote {\bibinfo {title} {Chiral
				dissociation of bound photon pairs for a non-{H}ermitian skin effect},}\
	}\href {\doibase 10.1103/q6wr-2rt9} {\bibfield  {journal} {\bibinfo
			{journal} {Phys. Rev. Lett.}\ }\textbf {\bibinfo {volume} {134}},\ \bibinfo
		{pages} {233602} (\bibinfo {year} {2025})}\BibitemShut {NoStop}%
	\bibitem [{\citenamefont {Hashemi}\ \emph {et~al.}(2025)\citenamefont
		{Hashemi}, \citenamefont {Pereira}, \citenamefont {Li}, \citenamefont
		{Lado},\ and\ \citenamefont {Blanco-Redondo}}]{Hashemi2025}%
	\BibitemOpen
	\bibfield  {author} {\bibinfo {author} {\bibfnamefont {A.}~\bibnamefont
			{Hashemi}}, \bibinfo {author} {\bibfnamefont {E.~L.}\ \bibnamefont
			{Pereira}}, \bibinfo {author} {\bibfnamefont {H.}~\bibnamefont {Li}},
		\bibinfo {author} {\bibfnamefont {J.~L.}\ \bibnamefont {Lado}}, \ and\
		\bibinfo {author} {\bibfnamefont {A.}~\bibnamefont {Blanco-Redondo}},\
	}\bibfield  {title} {\enquote {\bibinfo {title} {Observation of
				non-{H}ermitian topology from optical loss modulation},}\ }\href {\doibase
		10.1038/s41563-025-02278-8} {\bibfield  {journal} {\bibinfo  {journal} {Nat.
				Mater.}\ }\textbf {\bibinfo {volume} {24}},\ \bibinfo {pages} {1393}
		(\bibinfo {year} {2025})}\BibitemShut {NoStop}%
	\bibitem [{\citenamefont {Sun}\ and\ \citenamefont
		{Hu}(2025)}]{arxiv.2507.09447}%
	\BibitemOpen
	\bibfield  {author} {\bibinfo {author} {\bibfnamefont {K.}~\bibnamefont
			{Sun}}\ and\ \bibinfo {author} {\bibfnamefont {H.}~\bibnamefont {Hu}},\
	}\bibfield  {title} {\enquote {\bibinfo {title} {Lyapunov formulation of band
				theory for disordered non-{H}ermitian systems},}\ }\href
	{https://arxiv.org/abs/2507.09447} {\bibfield  {journal} {\bibinfo  {journal}
			{arXiv.2507.09447}\ } (\bibinfo {year} {2025})}\BibitemShut {NoStop}%
	\bibitem [{\citenamefont {Nakamura}\ \emph {et~al.}(2025)\citenamefont
		{Nakamura}, \citenamefont {Shiozaki}, \citenamefont {Shimomura},
		\citenamefont {Sato},\ and\ \citenamefont {Kawabata}}]{q4nh-m1jh}%
	\BibitemOpen
	\bibfield  {author} {\bibinfo {author} {\bibfnamefont {D.}~\bibnamefont
			{Nakamura}}, \bibinfo {author} {\bibfnamefont {K.}~\bibnamefont {Shiozaki}},
		\bibinfo {author} {\bibfnamefont {K.}~\bibnamefont {Shimomura}}, \bibinfo
		{author} {\bibfnamefont {M.}~\bibnamefont {Sato}}, \ and\ \bibinfo {author}
		{\bibfnamefont {K.}~\bibnamefont {Kawabata}},\ }\bibfield  {title} {\enquote
		{\bibinfo {title} {Non-{H}ermitian origin of detachable boundary states in
				topological insulators},}\ }\href {\doibase 10.1103/q4nh-m1jh} {\bibfield
		{journal} {\bibinfo  {journal} {Phys. Rev. Lett.}\ }\textbf {\bibinfo
			{volume} {135}},\ \bibinfo {pages} {096601} (\bibinfo {year}
		{2025})}\BibitemShut {NoStop}%
	\bibitem [{\citenamefont {Zhou}\ \emph {et~al.}(2025)\citenamefont {Zhou},
		\citenamefont {Yang}, \citenamefont {Zeng},\ and\ \citenamefont
		{Hu}}]{arxiv.2504.18926}%
	\BibitemOpen
	\bibfield  {author} {\bibinfo {author} {\bibfnamefont {K.}~\bibnamefont
			{Zhou}}, \bibinfo {author} {\bibfnamefont {Z.}~\bibnamefont {Yang}}, \bibinfo
		{author} {\bibfnamefont {B.}~\bibnamefont {Zeng}}, \ and\ \bibinfo {author}
		{\bibfnamefont {Y.}~\bibnamefont {Hu}},\ }\bibfield  {title} {\enquote
		{\bibinfo {title} {Critical non-{H}ermitian edge modes},}\ }\href
	{https://arxiv.org/abs/2504.18926} {\bibfield  {journal} {\bibinfo  {journal}
			{arXiv.2504.18926}\ } (\bibinfo {year} {2025})}\BibitemShut {NoStop}%
	\bibitem [{\citenamefont {Wang}\ \emph
		{et~al.}(2025{\natexlab{a}})\citenamefont {Wang}, \citenamefont {Cheng},
		\citenamefont {Zou}, \citenamefont {Ge}, \citenamefont {Zhao}, \citenamefont
		{Si}, \citenamefont {Yuan}, \citenamefont {Sun}, \citenamefont {Xue},\ and\
		\citenamefont {Zhang}}]{Wang2025}%
	\BibitemOpen
	\bibfield  {author} {\bibinfo {author} {\bibfnamefont {B.-B.}\ \bibnamefont
			{Wang}}, \bibinfo {author} {\bibfnamefont {Z.}~\bibnamefont {Cheng}},
		\bibinfo {author} {\bibfnamefont {H.-Y.}\ \bibnamefont {Zou}}, \bibinfo
		{author} {\bibfnamefont {Y.}~\bibnamefont {Ge}}, \bibinfo {author}
		{\bibfnamefont {K.-Q.}\ \bibnamefont {Zhao}}, \bibinfo {author}
		{\bibfnamefont {Q.-R.}\ \bibnamefont {Si}}, \bibinfo {author} {\bibfnamefont
			{S.-Q.}\ \bibnamefont {Yuan}}, \bibinfo {author} {\bibfnamefont {H.-X.}\
			\bibnamefont {Sun}}, \bibinfo {author} {\bibfnamefont {H.}~\bibnamefont
			{Xue}}, \ and\ \bibinfo {author} {\bibfnamefont {B.}~\bibnamefont {Zhang}},\
	}\bibfield  {title} {\enquote {\bibinfo {title} {Observation of
				disorder-induced boundary localization},}\ }\href {\doibase
		10.1073/pnas.2422154122} {\bibfield  {journal} {\bibinfo  {journal} {PNAS}\
		}\textbf {\bibinfo {volume} {122}},\ \bibinfo {pages} {e2422154122} (\bibinfo
		{year} {2025}{\natexlab{a}})}\BibitemShut {NoStop}%
	\bibitem [{\citenamefont {Esparza}\ and\ \citenamefont {Juri\ifmmode
			\check{c}\else \v{c}\fi{}i\ifmmode~\acute{c}\else
			\'{c}\fi{}}(2025)}]{dl59-vl7v}%
	\BibitemOpen
	\bibfield  {author} {\bibinfo {author} {\bibfnamefont {J.~P.}\ \bibnamefont
			{Esparza}}\ and\ \bibinfo {author} {\bibfnamefont {V.}~\bibnamefont
			{Juri\ifmmode \check{c}\else \v{c}\fi{}i\ifmmode~\acute{c}\else
				\'{c}\fi{}}},\ }\bibfield  {title} {\enquote {\bibinfo {title} {Exceptional
				magic angles in non-{H}ermitian twisted bilayer graphene},}\ }\href {\doibase
		10.1103/dl59-vl7v} {\bibfield  {journal} {\bibinfo  {journal} {Phys. Rev.
				Lett.}\ }\textbf {\bibinfo {volume} {134}},\ \bibinfo {pages} {226602}
		(\bibinfo {year} {2025})}\BibitemShut {NoStop}%
	\bibitem [{\citenamefont {Liu}\ \emph {et~al.}(2025)\citenamefont {Liu},
		\citenamefont {Liu},\ and\ \citenamefont {Xiao}}]{vs7x-clqd}%
	\BibitemOpen
	\bibfield  {author} {\bibinfo {author} {\bibfnamefont {T.-R.}\ \bibnamefont
			{Liu}}, \bibinfo {author} {\bibfnamefont {T.}~\bibnamefont {Liu}}, \ and\
		\bibinfo {author} {\bibfnamefont {M.}~\bibnamefont {Xiao}},\ }\bibfield
	{title} {\enquote {\bibinfo {title} {Anomalous non-{H}ermitian skin effect of
				chiral boundary states},}\ }\href {\doibase 10.1103/vs7x-clqd} {\bibfield
		{journal} {\bibinfo  {journal} {Phys. Rev. B}\ }\textbf {\bibinfo {volume}
			{112}},\ \bibinfo {pages} {L081112} (\bibinfo {year} {2025})}\BibitemShut
	{NoStop}%
	\bibitem [{\citenamefont {Ren}\ \emph {et~al.}(2022)\citenamefont {Ren},
		\citenamefont {Liu}, \citenamefont {Zhao}, \citenamefont {He}, \citenamefont
		{Pak}, \citenamefont {Li},\ and\ \citenamefont {Jo}}]{Ren2022}%
	\BibitemOpen
	\bibfield  {author} {\bibinfo {author} {\bibfnamefont {Z.}~\bibnamefont
			{Ren}}, \bibinfo {author} {\bibfnamefont {D.}~\bibnamefont {Liu}}, \bibinfo
		{author} {\bibfnamefont {E.}~\bibnamefont {Zhao}}, \bibinfo {author}
		{\bibfnamefont {C.}~\bibnamefont {He}}, \bibinfo {author} {\bibfnamefont
			{K.~K.}\ \bibnamefont {Pak}}, \bibinfo {author} {\bibfnamefont
			{J.}~\bibnamefont {Li}}, \ and\ \bibinfo {author} {\bibfnamefont {G.-B.}\
			\bibnamefont {Jo}},\ }\bibfield  {title} {\enquote {\bibinfo {title} {Chiral
				control of quantum states in non-{H}ermitian spin–orbit-coupled
				fermions},}\ }\href {\doibase 10.1038/s41567-021-01491-x} {\bibfield
		{journal} {\bibinfo  {journal} {Nat. Phys.}\ }\textbf {\bibinfo {volume}
			{18}},\ \bibinfo {pages} {385} (\bibinfo {year} {2022})}\BibitemShut
	{NoStop}%
	\bibitem [{\citenamefont {Naghiloo}\ \emph {et~al.}(2019)\citenamefont
		{Naghiloo}, \citenamefont {Abbasi}, \citenamefont {Joglekar},\ and\
		\citenamefont {Murch}}]{Naghiloo2019}%
	\BibitemOpen
	\bibfield  {author} {\bibinfo {author} {\bibfnamefont {M.}~\bibnamefont
			{Naghiloo}}, \bibinfo {author} {\bibfnamefont {M.}~\bibnamefont {Abbasi}},
		\bibinfo {author} {\bibfnamefont {Yogesh~N.}\ \bibnamefont {Joglekar}}, \
		and\ \bibinfo {author} {\bibfnamefont {K.~W.}\ \bibnamefont {Murch}},\
	}\bibfield  {title} {\enquote {\bibinfo {title} {Quantum state tomography
				across the exceptional point in a single dissipative qubit},}\ }\href
	{\doibase 10.1038/s41567-019-0652-z} {\bibfield  {journal} {\bibinfo
			{journal} {Nat. Phys.}\ }\textbf {\bibinfo {volume} {15}},\ \bibinfo {pages}
		{1232} (\bibinfo {year} {2019})}\BibitemShut {NoStop}%
	\bibitem [{\citenamefont {Shen}\ \emph {et~al.}(2025)\citenamefont {Shen},
		\citenamefont {Chen}, \citenamefont {Yang},\ and\ \citenamefont
		{Lee}}]{Shen2025}%
	\BibitemOpen
	\bibfield  {author} {\bibinfo {author} {\bibfnamefont {R.}~\bibnamefont
			{Shen}}, \bibinfo {author} {\bibfnamefont {T.}~\bibnamefont {Chen}}, \bibinfo
		{author} {\bibfnamefont {B.}~\bibnamefont {Yang}}, \ and\ \bibinfo {author}
		{\bibfnamefont {C.~H.}\ \bibnamefont {Lee}},\ }\bibfield  {title} {\enquote
		{\bibinfo {title} {Observation of the non-{H}ermitian skin effect and {F}ermi
				skin on a digital quantum computer},}\ }\href {\doibase
		10.1038/s41467-025-55953-4} {\bibfield  {journal} {\bibinfo  {journal} {Nat.
				Commun.}\ }\textbf {\bibinfo {volume} {16}},\ \bibinfo {pages} {1340}
		(\bibinfo {year} {2025})}\BibitemShut {NoStop}%
	\bibitem [{\citenamefont {Imhof}\ \emph {et~al.}(2018)\citenamefont {Imhof},
		\citenamefont {Berger}, \citenamefont {Bayer}, \citenamefont {Brehm},
		\citenamefont {Molenkamp}, \citenamefont {Kiessling}, \citenamefont
		{Schindler}, \citenamefont {Lee}, \citenamefont {Greiter}, \citenamefont
		{Neupert},\ and\ \citenamefont {Thomale}}]{Imhof2018}%
	\BibitemOpen
	\bibfield  {author} {\bibinfo {author} {\bibfnamefont {S.}~\bibnamefont
			{Imhof}}, \bibinfo {author} {\bibfnamefont {C.}~\bibnamefont {Berger}},
		\bibinfo {author} {\bibfnamefont {F.}~\bibnamefont {Bayer}}, \bibinfo
		{author} {\bibfnamefont {J.}~\bibnamefont {Brehm}}, \bibinfo {author}
		{\bibfnamefont {L.~W.}\ \bibnamefont {Molenkamp}}, \bibinfo {author}
		{\bibfnamefont {T.}~\bibnamefont {Kiessling}}, \bibinfo {author}
		{\bibfnamefont {F.}~\bibnamefont {Schindler}}, \bibinfo {author}
		{\bibfnamefont {C.~H.}\ \bibnamefont {Lee}}, \bibinfo {author} {\bibfnamefont
			{M.}~\bibnamefont {Greiter}}, \bibinfo {author} {\bibfnamefont
			{T.}~\bibnamefont {Neupert}}, \ and\ \bibinfo {author} {\bibfnamefont
			{R.}~\bibnamefont {Thomale}},\ }\bibfield  {title} {\enquote {\bibinfo
			{title} {Topolectrical-circuit realization of topological corner modes},}\
	}\href {https://doi.org/10.1038/s41567-018-0246-1} {\bibfield  {journal}
		{\bibinfo  {journal} {Nat. Phys.}\ }\textbf {\bibinfo {volume} {14}},\
		\bibinfo {pages} {925} (\bibinfo {year} {2018})}\BibitemShut {NoStop}%
	\bibitem [{\citenamefont {El-Ganainy}\ \emph {et~al.}(2018)\citenamefont
		{El-Ganainy}, \citenamefont {Makris}, \citenamefont {Khajavikhan},
		\citenamefont {Musslimani}, \citenamefont {Rotter},\ and\ \citenamefont
		{Christodoulides}}]{ElGanainy2018}%
	\BibitemOpen
	\bibfield  {author} {\bibinfo {author} {\bibfnamefont {R.}~\bibnamefont
			{El-Ganainy}}, \bibinfo {author} {\bibfnamefont {K.~G.}\ \bibnamefont
			{Makris}}, \bibinfo {author} {\bibfnamefont {M.}~\bibnamefont {Khajavikhan}},
		\bibinfo {author} {\bibfnamefont {Z.~H.}\ \bibnamefont {Musslimani}},
		\bibinfo {author} {\bibfnamefont {S.}~\bibnamefont {Rotter}}, \ and\ \bibinfo
		{author} {\bibfnamefont {D.~N.}\ \bibnamefont {Christodoulides}},\ }\bibfield
	{title} {\enquote {\bibinfo {title} {Non-{H}ermitian physics and {PT}
				symmetry},}\ }\href {\doibase 10.1038/nphys4323} {\bibfield  {journal}
		{\bibinfo  {journal} {Nat. Phys.}\ }\textbf {\bibinfo {volume} {14}},\
		\bibinfo {pages} {11} (\bibinfo {year} {2018})}\BibitemShut {NoStop}%
	\bibitem [{\citenamefont {Wu}\ \emph {et~al.}(2025)\citenamefont {Wu},
		\citenamefont {Hu}, \citenamefont {He}, \citenamefont {Deng}, \citenamefont
		{Huang}, \citenamefont {Ke}, \citenamefont {Deng}, \citenamefont {Lu},\ and\
		\citenamefont {Liu}}]{PhysRevLett.134.176601}%
	\BibitemOpen
	\bibfield  {author} {\bibinfo {author} {\bibfnamefont {J.}~\bibnamefont
			{Wu}}, \bibinfo {author} {\bibfnamefont {Y.}~\bibnamefont {Hu}}, \bibinfo
		{author} {\bibfnamefont {Z.}~\bibnamefont {He}}, \bibinfo {author}
		{\bibfnamefont {K.}~\bibnamefont {Deng}}, \bibinfo {author} {\bibfnamefont
			{X.}~\bibnamefont {Huang}}, \bibinfo {author} {\bibfnamefont
			{M.}~\bibnamefont {Ke}}, \bibinfo {author} {\bibfnamefont {W.}~\bibnamefont
			{Deng}}, \bibinfo {author} {\bibfnamefont {J.}~\bibnamefont {Lu}}, \ and\
		\bibinfo {author} {\bibfnamefont {Z.}~\bibnamefont {Liu}},\ }\bibfield
	{title} {\enquote {\bibinfo {title} {Hybrid-order skin effect from
				loss-induced nonreciprocity},}\ }\href {\doibase
		10.1103/PhysRevLett.134.176601} {\bibfield  {journal} {\bibinfo  {journal}
			{Phys. Rev. Lett.}\ }\textbf {\bibinfo {volume} {134}},\ \bibinfo {pages}
		{176601} (\bibinfo {year} {2025})}\BibitemShut {NoStop}%
	\bibitem [{\citenamefont {Yao}\ and\ \citenamefont
		{Wang}(2018)}]{ShunyuYao2018}%
	\BibitemOpen
	\bibfield  {author} {\bibinfo {author} {\bibfnamefont {S.}~\bibnamefont
			{Yao}}\ and\ \bibinfo {author} {\bibfnamefont {Z.}~\bibnamefont {Wang}},\
	}\bibfield  {title} {\enquote {\bibinfo {title} {Edge states and topological
				invariants of non-\uppercase{H}ermitian systems},}\ }\href
	{https://link.aps.org/doi/10.1103/PhysRevLett.121.086803} {\bibfield
		{journal} {\bibinfo  {journal} {Phys. Rev. Lett.}\ }\textbf {\bibinfo
			{volume} {121}},\ \bibinfo {pages} {086803} (\bibinfo {year}
		{2018})}\BibitemShut {NoStop}%
	\bibitem [{\citenamefont {Yokomizo}\ and\ \citenamefont
		{Murakami}(2019)}]{PhysRevLett.123.066404}%
	\BibitemOpen
	\bibfield  {author} {\bibinfo {author} {\bibfnamefont {K.}~\bibnamefont
			{Yokomizo}}\ and\ \bibinfo {author} {\bibfnamefont {S.}~\bibnamefont
			{Murakami}},\ }\bibfield  {title} {\enquote {\bibinfo {title} {Non-{B}loch
				band theory of non-{H}ermitian systems},}\ }\href {\doibase
		10.1103/PhysRevLett.123.066404} {\bibfield  {journal} {\bibinfo  {journal}
			{Phys. Rev. Lett.}\ }\textbf {\bibinfo {volume} {123}},\ \bibinfo {pages}
		{066404} (\bibinfo {year} {2019})}\BibitemShut {NoStop}%
	\bibitem [{\citenamefont {Zhang}\ \emph {et~al.}(2020)\citenamefont {Zhang},
		\citenamefont {Yang},\ and\ \citenamefont {Fang}}]{PhysRevLett.125.126402}%
	\BibitemOpen
	\bibfield  {author} {\bibinfo {author} {\bibfnamefont {K.}~\bibnamefont
			{Zhang}}, \bibinfo {author} {\bibfnamefont {Z.}~\bibnamefont {Yang}}, \ and\
		\bibinfo {author} {\bibfnamefont {C.}~\bibnamefont {Fang}},\ }\bibfield
	{title} {\enquote {\bibinfo {title} {Correspondence between winding numbers
				and skin modes in non-{H}ermitian systems},}\ }\href {\doibase
		10.1103/PhysRevLett.125.126402} {\bibfield  {journal} {\bibinfo  {journal}
			{Phys. Rev. Lett.}\ }\textbf {\bibinfo {volume} {125}},\ \bibinfo {pages}
		{126402} (\bibinfo {year} {2020})}\BibitemShut {NoStop}%
	\bibitem [{\citenamefont {Xiong}\ \emph {et~al.}(2024)\citenamefont {Xiong},
		\citenamefont {Xing},\ and\ \citenamefont {Hu}}]{arxiv.2407.01296}%
	\BibitemOpen
	\bibfield  {author} {\bibinfo {author} {\bibfnamefont {Y.}~\bibnamefont
			{Xiong}}, \bibinfo {author} {\bibfnamefont {Z.-Y.}\ \bibnamefont {Xing}}, \
		and\ \bibinfo {author} {\bibfnamefont {H.}~\bibnamefont {Hu}},\ }\bibfield
	{title} {\enquote {\bibinfo {title} {Non-{H}ermitian skin effect in arbitrary
				dimensions: non-{B}loch band theory and classification},}\ }\href
	{https://arxiv.org/abs/2407.01296} {\bibfield  {journal} {\bibinfo  {journal}
			{arXiv.2407.01296}\ } (\bibinfo {year} {2024})}\BibitemShut {NoStop}%
	\bibitem [{\citenamefont {Liu}\ \emph {et~al.}(2019)\citenamefont {Liu},
		\citenamefont {Zhang}, \citenamefont {Ai}, \citenamefont {Gong},
		\citenamefont {Kawabata}, \citenamefont {Ueda},\ and\ \citenamefont
		{Nori}}]{PhysRevLett.122.076801}%
	\BibitemOpen
	\bibfield  {author} {\bibinfo {author} {\bibfnamefont {T.}~\bibnamefont
			{Liu}}, \bibinfo {author} {\bibfnamefont {Y.-R.}\ \bibnamefont {Zhang}},
		\bibinfo {author} {\bibfnamefont {Q.}~\bibnamefont {Ai}}, \bibinfo {author}
		{\bibfnamefont {Z.}~\bibnamefont {Gong}}, \bibinfo {author} {\bibfnamefont
			{K.}~\bibnamefont {Kawabata}}, \bibinfo {author} {\bibfnamefont
			{M.}~\bibnamefont {Ueda}}, \ and\ \bibinfo {author} {\bibfnamefont
			{F.}~\bibnamefont {Nori}},\ }\bibfield  {title} {\enquote {\bibinfo {title}
			{Second-order topological phases in non-{H}ermitian systems},}\ }\href
	{\doibase 10.1103/PhysRevLett.122.076801} {\bibfield  {journal} {\bibinfo
			{journal} {Phys. Rev. Lett.}\ }\textbf {\bibinfo {volume} {122}},\ \bibinfo
		{pages} {076801} (\bibinfo {year} {2019})}\BibitemShut {NoStop}%
	\bibitem [{\citenamefont {Kunst}\ \emph {et~al.}(2018)\citenamefont {Kunst},
		\citenamefont {Edvardsson}, \citenamefont {Budich},\ and\ \citenamefont
		{Bergholtz}}]{PhysRevLett.121.026808}%
	\BibitemOpen
	\bibfield  {author} {\bibinfo {author} {\bibfnamefont {F.~K.}\ \bibnamefont
			{Kunst}}, \bibinfo {author} {\bibfnamefont {E.}~\bibnamefont {Edvardsson}},
		\bibinfo {author} {\bibfnamefont {J.~C.}\ \bibnamefont {Budich}}, \ and\
		\bibinfo {author} {\bibfnamefont {E.~J.}\ \bibnamefont {Bergholtz}},\
	}\bibfield  {title} {\enquote {\bibinfo {title} {Biorthogonal bulk-boundary
				correspondence in non-{H}ermitian systems},}\ }\href {\doibase
		10.1103/PhysRevLett.121.026808} {\bibfield  {journal} {\bibinfo  {journal}
			{Phys. Rev. Lett.}\ }\textbf {\bibinfo {volume} {121}},\ \bibinfo {pages}
		{026808} (\bibinfo {year} {2018})}\BibitemShut {NoStop}%
	\bibitem [{\citenamefont {Ling}\ \emph {et~al.}(2025)\citenamefont {Ling},
		\citenamefont {Cai},\ and\ \citenamefont {Liu}}]{PhysRevB.111.205418}%
	\BibitemOpen
	\bibfield  {author} {\bibinfo {author} {\bibfnamefont {W.-Z.}\ \bibnamefont
			{Ling}}, \bibinfo {author} {\bibfnamefont {Z.-F.}\ \bibnamefont {Cai}}, \
		and\ \bibinfo {author} {\bibfnamefont {T.}~\bibnamefont {Liu}},\ }\bibfield
	{title} {\enquote {\bibinfo {title} {Interaction-induced second-order skin
				effect},}\ }\href {\doibase 10.1103/PhysRevB.111.205418} {\bibfield
		{journal} {\bibinfo  {journal} {Phys. Rev. B}\ }\textbf {\bibinfo {volume}
			{111}},\ \bibinfo {pages} {205418} (\bibinfo {year} {2025})}\BibitemShut
	{NoStop}%
	\bibitem [{\citenamefont {Li}\ \emph {et~al.}(2024{\natexlab{a}})\citenamefont
		{Li}, \citenamefont {Cai}, \citenamefont {Liu},\ and\ \citenamefont
		{Nori}}]{arXiv:2408.12451}%
	\BibitemOpen
	\bibfield  {author} {\bibinfo {author} {\bibfnamefont {Y.}~\bibnamefont
			{Li}}, \bibinfo {author} {\bibfnamefont {Z.-F.}\ \bibnamefont {Cai}},
		\bibinfo {author} {\bibfnamefont {T.}~\bibnamefont {Liu}}, \ and\ \bibinfo
		{author} {\bibfnamefont {F.}~\bibnamefont {Nori}},\ }\bibfield  {title}
	{\enquote {\bibinfo {title} {Dissipation and interaction-controlled
				non-{H}ermitian skin effects},}\ }\href
	{https://doi.org/10.48550/arXiv.2408.12451} {\bibfield  {journal} {\bibinfo
			{journal} {arXiv:2408.12451}\ } (\bibinfo {year}
		{2024}{\natexlab{a}})}\BibitemShut {NoStop}%
	\bibitem [{\citenamefont {Gong}\ \emph {et~al.}(2018)\citenamefont {Gong},
		\citenamefont {Ashida}, \citenamefont {Kawabata}, \citenamefont {Takasan},
		\citenamefont {Higashikawa},\ and\ \citenamefont {Ueda}}]{arXiv:1802.07964}%
	\BibitemOpen
	\bibfield  {author} {\bibinfo {author} {\bibfnamefont {Z.}~\bibnamefont
			{Gong}}, \bibinfo {author} {\bibfnamefont {Y.}~\bibnamefont {Ashida}},
		\bibinfo {author} {\bibfnamefont {K.}~\bibnamefont {Kawabata}}, \bibinfo
		{author} {\bibfnamefont {K.}~\bibnamefont {Takasan}}, \bibinfo {author}
		{\bibfnamefont {S.}~\bibnamefont {Higashikawa}}, \ and\ \bibinfo {author}
		{\bibfnamefont {M.}~\bibnamefont {Ueda}},\ }\bibfield  {title} {\enquote
		{\bibinfo {title} {Topological phases of non-\uppercase{H}ermitian
				systems},}\ }\href {https://link.aps.org/doi/10.1103/PhysRevX.8.031079}
	{\bibfield  {journal} {\bibinfo  {journal} {Phys. Rev. X}\ }\textbf {\bibinfo
			{volume} {8}},\ \bibinfo {pages} {031079} (\bibinfo {year}
		{2018})}\BibitemShut {NoStop}%
	\bibitem [{\citenamefont {Yao}\ \emph {et~al.}(2018)\citenamefont {Yao},
		\citenamefont {Song},\ and\ \citenamefont {Wang}}]{YaoarXiv:1804.04672}%
	\BibitemOpen
	\bibfield  {author} {\bibinfo {author} {\bibfnamefont {S.}~\bibnamefont
			{Yao}}, \bibinfo {author} {\bibfnamefont {F.}~\bibnamefont {Song}}, \ and\
		\bibinfo {author} {\bibfnamefont {Z.}~\bibnamefont {Wang}},\ }\bibfield
	{title} {\enquote {\bibinfo {title} {Non-\uppercase{H}ermitian
				\uppercase{C}hern bands},}\ }\href
	{https://link.aps.org/doi/10.1103/PhysRevLett.121.136802} {\bibfield
		{journal} {\bibinfo  {journal} {Phys. Rev. Lett.}\ }\textbf {\bibinfo
			{volume} {121}},\ \bibinfo {pages} {136802} (\bibinfo {year}
		{2018})}\BibitemShut {NoStop}%
	\bibitem [{\citenamefont {Song}\ \emph {et~al.}(2019)\citenamefont {Song},
		\citenamefont {Yao},\ and\ \citenamefont {Wang}}]{PhysRevLett.123.170401}%
	\BibitemOpen
	\bibfield  {author} {\bibinfo {author} {\bibfnamefont {F.}~\bibnamefont
			{Song}}, \bibinfo {author} {\bibfnamefont {S.}~\bibnamefont {Yao}}, \ and\
		\bibinfo {author} {\bibfnamefont {Z.}~\bibnamefont {Wang}},\ }\bibfield
	{title} {\enquote {\bibinfo {title} {Non-{H}ermitian skin effect and chiral
				damping in open quantum systems},}\ }\href {\doibase
		10.1103/PhysRevLett.123.170401} {\bibfield  {journal} {\bibinfo  {journal}
			{Phys. Rev. Lett.}\ }\textbf {\bibinfo {volume} {123}},\ \bibinfo {pages}
		{170401} (\bibinfo {year} {2019})}\BibitemShut {NoStop}%
	\bibitem [{\citenamefont {Lee}\ \emph {et~al.}(2019)\citenamefont {Lee},
		\citenamefont {Ahn}, \citenamefont {Zhou},\ and\ \citenamefont
		{Vishwanath}}]{PhysRevLett.123.206404}%
	\BibitemOpen
	\bibfield  {author} {\bibinfo {author} {\bibfnamefont {J.~Y.}\ \bibnamefont
			{Lee}}, \bibinfo {author} {\bibfnamefont {J.}~\bibnamefont {Ahn}}, \bibinfo
		{author} {\bibfnamefont {H.}~\bibnamefont {Zhou}}, \ and\ \bibinfo {author}
		{\bibfnamefont {A.}~\bibnamefont {Vishwanath}},\ }\bibfield  {title}
	{\enquote {\bibinfo {title} {Topological correspondence between {H}ermitian
				and non-{H}ermitian systems: {A}nomalous dynamics},}\ }\href {\doibase
		10.1103/PhysRevLett.123.206404} {\bibfield  {journal} {\bibinfo  {journal}
			{Phys. Rev. Lett.}\ }\textbf {\bibinfo {volume} {123}},\ \bibinfo {pages}
		{206404} (\bibinfo {year} {2019})}\BibitemShut {NoStop}%
	\bibitem [{\citenamefont {Kawabata}\ \emph
		{et~al.}(2019{\natexlab{a}})\citenamefont {Kawabata}, \citenamefont
		{Bessho},\ and\ \citenamefont {Sato}}]{PhysRevLett.123.066405}%
	\BibitemOpen
	\bibfield  {author} {\bibinfo {author} {\bibfnamefont {K.}~\bibnamefont
			{Kawabata}}, \bibinfo {author} {\bibfnamefont {T.}~\bibnamefont {Bessho}}, \
		and\ \bibinfo {author} {\bibfnamefont {M.}~\bibnamefont {Sato}},\ }\bibfield
	{title} {\enquote {\bibinfo {title} {Classification of exceptional points and
				non-{H}ermitian topological semimetals},}\ }\href {\doibase
		10.1103/PhysRevLett.123.066405} {\bibfield  {journal} {\bibinfo  {journal}
			{Phys. Rev. Lett.}\ }\textbf {\bibinfo {volume} {123}},\ \bibinfo {pages}
		{066405} (\bibinfo {year} {2019}{\natexlab{a}})}\BibitemShut {NoStop}%
	\bibitem [{\citenamefont {Zhang}\ \emph {et~al.}(2018)\citenamefont {Zhang},
		\citenamefont {Peng}, \citenamefont {\"{O}zdemir}, \citenamefont {Pichler},
		\citenamefont {Krimer}, \citenamefont {Zhao}, \citenamefont {Nori},
		\citenamefont {Liu}, \citenamefont {Rotter},\ and\ \citenamefont
		{Yang}}]{ZhangJ2018}%
	\BibitemOpen
	\bibfield  {author} {\bibinfo {author} {\bibfnamefont {J.}~\bibnamefont
			{Zhang}}, \bibinfo {author} {\bibfnamefont {B.}~\bibnamefont {Peng}},
		\bibinfo {author} {\bibfnamefont {Ş.~K.}\ \bibnamefont {\"{O}zdemir}},
		\bibinfo {author} {\bibfnamefont {K.}~\bibnamefont {Pichler}}, \bibinfo
		{author} {\bibfnamefont {D.~O.}\ \bibnamefont {Krimer}}, \bibinfo {author}
		{\bibfnamefont {G.}~\bibnamefont {Zhao}}, \bibinfo {author} {\bibfnamefont
			{F.}~\bibnamefont {Nori}}, \bibinfo {author} {\bibfnamefont {Y.-x.}\
			\bibnamefont {Liu}}, \bibinfo {author} {\bibfnamefont {S.}~\bibnamefont
			{Rotter}}, \ and\ \bibinfo {author} {\bibfnamefont {L.}~\bibnamefont
			{Yang}},\ }\bibfield  {title} {\enquote {\bibinfo {title} {A phonon laser
				operating at an exceptional point},}\ }\href
	{http://dx.doi.org/10.1038/s41566-018-0213-5} {\bibfield  {journal} {\bibinfo
			{journal} {Nat. Photon.}\ }\textbf {\bibinfo {volume} {12}},\ \bibinfo
		{pages} {479} (\bibinfo {year} {2018})}\BibitemShut {NoStop}%
	\bibitem [{\citenamefont {Ge}\ \emph {et~al.}(2019)\citenamefont {Ge},
		\citenamefont {Zhang}, \citenamefont {Liu}, \citenamefont {Li}, \citenamefont
		{Fan},\ and\ \citenamefont {Nori}}]{PhysRevB.100.054105}%
	\BibitemOpen
	\bibfield  {author} {\bibinfo {author} {\bibfnamefont {Z.~Y.}\ \bibnamefont
			{Ge}}, \bibinfo {author} {\bibfnamefont {Y.~R.}\ \bibnamefont {Zhang}},
		\bibinfo {author} {\bibfnamefont {T.}~\bibnamefont {Liu}}, \bibinfo {author}
		{\bibfnamefont {S.~W.}\ \bibnamefont {Li}}, \bibinfo {author} {\bibfnamefont
			{H.}~\bibnamefont {Fan}}, \ and\ \bibinfo {author} {\bibfnamefont
			{F.}~\bibnamefont {Nori}},\ }\bibfield  {title} {\enquote {\bibinfo {title}
			{Topological band theory for non-{H}ermitian systems from the {D}irac
				equation},}\ }\href {\doibase 10.1103/PhysRevB.100.054105} {\bibfield
		{journal} {\bibinfo  {journal} {Phys. Rev. B}\ }\textbf {\bibinfo {volume}
			{100}},\ \bibinfo {pages} {054105} (\bibinfo {year} {2019})}\BibitemShut
	{NoStop}%
	\bibitem [{\citenamefont {Minganti}\ \emph {et~al.}(2019)\citenamefont
		{Minganti}, \citenamefont {Miranowicz}, \citenamefont {Chhajlany},\ and\
		\citenamefont {Nori}}]{PhysRevA.100.062131}%
	\BibitemOpen
	\bibfield  {author} {\bibinfo {author} {\bibfnamefont {F.}~\bibnamefont
			{Minganti}}, \bibinfo {author} {\bibfnamefont {A.}~\bibnamefont
			{Miranowicz}}, \bibinfo {author} {\bibfnamefont {R.~W.}\ \bibnamefont
			{Chhajlany}}, \ and\ \bibinfo {author} {\bibfnamefont {F.}~\bibnamefont
			{Nori}},\ }\bibfield  {title} {\enquote {\bibinfo {title} {Quantum
				exceptional points of non-{H}ermitian {H}amiltonians and {L}iouvillians:
				{T}he effects of quantum jumps},}\ }\href {\doibase
		10.1103/PhysRevA.100.062131} {\bibfield  {journal} {\bibinfo  {journal}
			{Phys. Rev. A}\ }\textbf {\bibinfo {volume} {100}},\ \bibinfo {pages}
		{062131} (\bibinfo {year} {2019})}\BibitemShut {NoStop}%
	\bibitem [{\citenamefont {Zhao}\ \emph {et~al.}(2019)\citenamefont {Zhao},
		\citenamefont {Qiao}, \citenamefont {Wu}, \citenamefont {Midya},
		\citenamefont {Longhi},\ and\ \citenamefont {Feng}}]{Zhao2019}%
	\BibitemOpen
	\bibfield  {author} {\bibinfo {author} {\bibfnamefont {H.}~\bibnamefont
			{Zhao}}, \bibinfo {author} {\bibfnamefont {X.}~\bibnamefont {Qiao}}, \bibinfo
		{author} {\bibfnamefont {T.}~\bibnamefont {Wu}}, \bibinfo {author}
		{\bibfnamefont {B.}~\bibnamefont {Midya}}, \bibinfo {author} {\bibfnamefont
			{S.}~\bibnamefont {Longhi}}, \ and\ \bibinfo {author} {\bibfnamefont
			{L.}~\bibnamefont {Feng}},\ }\bibfield  {title} {\enquote {\bibinfo {title}
			{Non-{H}ermitian topological light steering},}\ }\href {\doibase
		10.1126/science.aay1064} {\bibfield  {journal} {\bibinfo  {journal}
			{Science}\ }\textbf {\bibinfo {volume} {365}},\ \bibinfo {pages} {1163}
		(\bibinfo {year} {2019})}\BibitemShut {NoStop}%
	\bibitem [{\citenamefont {Kawabata}\ \emph
		{et~al.}(2019{\natexlab{b}})\citenamefont {Kawabata}, \citenamefont
		{Shiozaki}, \citenamefont {Ueda},\ and\ \citenamefont
		{Sato}}]{PhysRevX.9.041015}%
	\BibitemOpen
	\bibfield  {author} {\bibinfo {author} {\bibfnamefont {K.}~\bibnamefont
			{Kawabata}}, \bibinfo {author} {\bibfnamefont {K.}~\bibnamefont {Shiozaki}},
		\bibinfo {author} {\bibfnamefont {M.}~\bibnamefont {Ueda}}, \ and\ \bibinfo
		{author} {\bibfnamefont {M.}~\bibnamefont {Sato}},\ }\bibfield  {title}
	{\enquote {\bibinfo {title} {Symmetry and topology in non-{H}ermitian
				physics},}\ }\href {\doibase 10.1103/PhysRevX.9.041015} {\bibfield  {journal}
		{\bibinfo  {journal} {Phys. Rev. X}\ }\textbf {\bibinfo {volume} {9}},\
		\bibinfo {pages} {041015} (\bibinfo {year} {2019}{\natexlab{b}})}\BibitemShut
	{NoStop}%
	\bibitem [{\citenamefont {Borgnia}\ \emph {et~al.}(2020)\citenamefont
		{Borgnia}, \citenamefont {Kruchkov},\ and\ \citenamefont
		{Slager}}]{PhysRevLett.124.056802}%
	\BibitemOpen
	\bibfield  {author} {\bibinfo {author} {\bibfnamefont {D.~S.}\ \bibnamefont
			{Borgnia}}, \bibinfo {author} {\bibfnamefont {A.~J.}\ \bibnamefont
			{Kruchkov}}, \ and\ \bibinfo {author} {\bibfnamefont {R.-J.}\ \bibnamefont
			{Slager}},\ }\bibfield  {title} {\enquote {\bibinfo {title} {Non-{H}ermitian
				boundary modes and topology},}\ }\href {\doibase
		10.1103/PhysRevLett.124.056802} {\bibfield  {journal} {\bibinfo  {journal}
			{Phys. Rev. Lett.}\ }\textbf {\bibinfo {volume} {124}},\ \bibinfo {pages}
		{056802} (\bibinfo {year} {2020})}\BibitemShut {NoStop}%
	\bibitem [{\citenamefont {Martinez~Alvarez}\ \emph {et~al.}(2018)\citenamefont
		{Martinez~Alvarez}, \citenamefont {Barrios~Vargas},\ and\ \citenamefont
		{Foa~Torres}}]{PhysRevB.97.121401}%
	\BibitemOpen
	\bibfield  {author} {\bibinfo {author} {\bibfnamefont {V.~M.}\ \bibnamefont
			{Martinez~Alvarez}}, \bibinfo {author} {\bibfnamefont {J.~E.}\ \bibnamefont
			{Barrios~Vargas}}, \ and\ \bibinfo {author} {\bibfnamefont {L.~E.~F.}\
			\bibnamefont {Foa~Torres}},\ }\bibfield  {title} {\enquote {\bibinfo {title}
			{Non-\uppercase{H}ermitian robust edge states in one dimension:
				\uppercase{A}nomalous localization and eigenspace condensation at exceptional
				points},}\ }\href {https://link.aps.org/doi/10.1103/PhysRevB.97.121401}
	{\bibfield  {journal} {\bibinfo  {journal} {Phys. Rev. B}\ }\textbf {\bibinfo
			{volume} {97}},\ \bibinfo {pages} {121401} (\bibinfo {year}
		{2018})}\BibitemShut {NoStop}%
	\bibitem [{\citenamefont {Arkhipov}\ \emph {et~al.}(2020)\citenamefont
		{Arkhipov}, \citenamefont {Miranowicz}, \citenamefont {Minganti},\ and\
		\citenamefont {Nori}}]{PhysRevA.102.033715}%
	\BibitemOpen
	\bibfield  {author} {\bibinfo {author} {\bibfnamefont {I.~I.}\ \bibnamefont
			{Arkhipov}}, \bibinfo {author} {\bibfnamefont {A.}~\bibnamefont
			{Miranowicz}}, \bibinfo {author} {\bibfnamefont {F.}~\bibnamefont
			{Minganti}}, \ and\ \bibinfo {author} {\bibfnamefont {F.}~\bibnamefont
			{Nori}},\ }\bibfield  {title} {\enquote {\bibinfo {title} {Liouvillian
				exceptional points of any order in dissipative linear bosonic systems:
				{C}oherence functions and switching between $\mathcal{PT}$ and
				anti-$\mathcal{PT}$ symmetries},}\ }\href {\doibase
		10.1103/PhysRevA.102.033715} {\bibfield  {journal} {\bibinfo  {journal}
			{Phys. Rev. A}\ }\textbf {\bibinfo {volume} {102}},\ \bibinfo {pages}
		{033715} (\bibinfo {year} {2020})}\BibitemShut {NoStop}%
	\bibitem [{\citenamefont {Okuma}\ \emph {et~al.}(2020)\citenamefont {Okuma},
		\citenamefont {Kawabata}, \citenamefont {Shiozaki},\ and\ \citenamefont
		{Sato}}]{PhysRevLett.124.086801}%
	\BibitemOpen
	\bibfield  {author} {\bibinfo {author} {\bibfnamefont {N.}~\bibnamefont
			{Okuma}}, \bibinfo {author} {\bibfnamefont {K.}~\bibnamefont {Kawabata}},
		\bibinfo {author} {\bibfnamefont {K.}~\bibnamefont {Shiozaki}}, \ and\
		\bibinfo {author} {\bibfnamefont {M.}~\bibnamefont {Sato}},\ }\bibfield
	{title} {\enquote {\bibinfo {title} {Topological origin of non-{H}ermitian
				skin effects},}\ }\href {\doibase 10.1103/PhysRevLett.124.086801} {\bibfield
		{journal} {\bibinfo  {journal} {Phys. Rev. Lett.}\ }\textbf {\bibinfo
			{volume} {124}},\ \bibinfo {pages} {086801} (\bibinfo {year}
		{2020})}\BibitemShut {NoStop}%
	\bibitem [{\citenamefont {Liu}\ \emph {et~al.}(2021)\citenamefont {Liu},
		\citenamefont {He}, \citenamefont {Yang},\ and\ \citenamefont
		{Nori}}]{PhysRevLett.127.196801}%
	\BibitemOpen
	\bibfield  {author} {\bibinfo {author} {\bibfnamefont {T.}~\bibnamefont
			{Liu}}, \bibinfo {author} {\bibfnamefont {J.~J.}\ \bibnamefont {He}},
		\bibinfo {author} {\bibfnamefont {Z.}~\bibnamefont {Yang}}, \ and\ \bibinfo
		{author} {\bibfnamefont {F.}~\bibnamefont {Nori}},\ }\bibfield  {title}
	{\enquote {\bibinfo {title} {Higher-order {W}eyl-exceptional-ring
				semimetals},}\ }\href {\doibase 10.1103/PhysRevLett.127.196801} {\bibfield
		{journal} {\bibinfo  {journal} {Phys. Rev. Lett.}\ }\textbf {\bibinfo
			{volume} {127}},\ \bibinfo {pages} {196801} (\bibinfo {year}
		{2021})}\BibitemShut {NoStop}%
	\bibitem [{\citenamefont {Li}\ \emph {et~al.}(2022)\citenamefont {Li},
		\citenamefont {Liang}, \citenamefont {Wang}, \citenamefont {Lu},\ and\
		\citenamefont {Liu}}]{PhysRevLett.128.223903}%
	\BibitemOpen
	\bibfield  {author} {\bibinfo {author} {\bibfnamefont {Y.}~\bibnamefont
			{Li}}, \bibinfo {author} {\bibfnamefont {C.}~\bibnamefont {Liang}}, \bibinfo
		{author} {\bibfnamefont {C.}~\bibnamefont {Wang}}, \bibinfo {author}
		{\bibfnamefont {C.}~\bibnamefont {Lu}}, \ and\ \bibinfo {author}
		{\bibfnamefont {Y.-C.}\ \bibnamefont {Liu}},\ }\bibfield  {title} {\enquote
		{\bibinfo {title} {Gain-loss-induced hybrid skin-topological effect},}\
	}\href {\doibase 10.1103/PhysRevLett.128.223903} {\bibfield  {journal}
		{\bibinfo  {journal} {Phys. Rev. Lett.}\ }\textbf {\bibinfo {volume} {128}},\
		\bibinfo {pages} {223903} (\bibinfo {year} {2022})}\BibitemShut {NoStop}%
	\bibitem [{\citenamefont {Li}\ and\ \citenamefont
		{Xu}(2022)}]{PhysRevLett.129.093001}%
	\BibitemOpen
	\bibfield  {author} {\bibinfo {author} {\bibfnamefont {K.}~\bibnamefont
			{Li}}\ and\ \bibinfo {author} {\bibfnamefont {Y.}~\bibnamefont {Xu}},\
	}\bibfield  {title} {\enquote {\bibinfo {title} {Non-{H}ermitian absorption
				spectroscopy},}\ }\href {\doibase 10.1103/PhysRevLett.129.093001} {\bibfield
		{journal} {\bibinfo  {journal} {Phys. Rev. Lett.}\ }\textbf {\bibinfo
			{volume} {129}},\ \bibinfo {pages} {093001} (\bibinfo {year}
		{2022})}\BibitemShut {NoStop}%
	\bibitem [{\citenamefont {Zeng}\ \emph {et~al.}(2025)\citenamefont {Zeng},
		\citenamefont {Liu}, \citenamefont {Xia}, \citenamefont {Zhang},\ and\
		\citenamefont {Nori}}]{arXiv:2505.05058}%
	\BibitemOpen
	\bibfield  {author} {\bibinfo {author} {\bibfnamefont {N.}~\bibnamefont
			{Zeng}}, \bibinfo {author} {\bibfnamefont {T.}~\bibnamefont {Liu}}, \bibinfo
		{author} {\bibfnamefont {K.}~\bibnamefont {Xia}}, \bibinfo {author}
		{\bibfnamefont {Y.-R.}\ \bibnamefont {Zhang}}, \ and\ \bibinfo {author}
		{\bibfnamefont {F.}~\bibnamefont {Nori}},\ }\bibfield  {title} {\enquote
		{\bibinfo {title} {Non-{H}ermitian sensing from the perspective of
				post-selected measurements},}\ }\href {https://arxiv.org/abs/2505.05058}
	{\bibfield  {journal} {\bibinfo  {journal} {arXiv:2505.05058}\ } (\bibinfo
		{year} {2025})}\BibitemShut {NoStop}%
	\bibitem [{\citenamefont {Cai}\ \emph {et~al.}(2025{\natexlab{b}})\citenamefont
		{Cai}, \citenamefont {Wang}, \citenamefont {Liang}, \citenamefont {Liu},\
		and\ \citenamefont {Nori}}]{PhysRevAL061701}%
	\BibitemOpen
	\bibfield  {author} {\bibinfo {author} {\bibfnamefont {Z.-F.}\ \bibnamefont
			{Cai}}, \bibinfo {author} {\bibfnamefont {X.}~\bibnamefont {Wang}}, \bibinfo
		{author} {\bibfnamefont {Z.-X.}\ \bibnamefont {Liang}}, \bibinfo {author}
		{\bibfnamefont {T.}~\bibnamefont {Liu}}, \ and\ \bibinfo {author}
		{\bibfnamefont {F.}~\bibnamefont {Nori}},\ }\bibfield  {title} {\enquote
		{\bibinfo {title} {Chiral-extended photon-emitter dressed states in
				non-{H}ermitian topological baths},}\ }\href {\doibase 10.1103/8qpx-68x6}
	{\bibfield  {journal} {\bibinfo  {journal} {Phys. Rev. A}\ }\textbf {\bibinfo
			{volume} {111}},\ \bibinfo {pages} {L061701} (\bibinfo {year}
		{2025}{\natexlab{b}})}\BibitemShut {NoStop}%
	\bibitem [{\citenamefont {Zhang}\ \emph {et~al.}(2023)\citenamefont {Zhang},
		\citenamefont {Fang},\ and\ \citenamefont {Yang}}]{PhysRevLett.131.036402}%
	\BibitemOpen
	\bibfield  {author} {\bibinfo {author} {\bibfnamefont {K.}~\bibnamefont
			{Zhang}}, \bibinfo {author} {\bibfnamefont {C.}~\bibnamefont {Fang}}, \ and\
		\bibinfo {author} {\bibfnamefont {Z.}~\bibnamefont {Yang}},\ }\bibfield
	{title} {\enquote {\bibinfo {title} {Dynamical degeneracy splitting and
				directional invisibility in non-{H}ermitian systems},}\ }\href {\doibase
		10.1103/PhysRevLett.131.036402} {\bibfield  {journal} {\bibinfo  {journal}
			{Phys. Rev. Lett.}\ }\textbf {\bibinfo {volume} {131}},\ \bibinfo {pages}
		{036402} (\bibinfo {year} {2023})}\BibitemShut {NoStop}%
	\bibitem [{\citenamefont {Wang}\ \emph
		{et~al.}(2025{\natexlab{b}})\citenamefont {Wang}, \citenamefont {Liu},
		\citenamefont {Liu},\ and\ \citenamefont {Ju}}]{haowang2025}%
	\BibitemOpen
	\bibfield  {author} {\bibinfo {author} {\bibfnamefont {H.}~\bibnamefont
			{Wang}}, \bibinfo {author} {\bibfnamefont {J.}~\bibnamefont {Liu}}, \bibinfo
		{author} {\bibfnamefont {T.}~\bibnamefont {Liu}}, \ and\ \bibinfo {author}
		{\bibfnamefont {W.-B.}\ \bibnamefont {Ju}},\ }\bibfield  {title} {\enquote
		{\bibinfo {title} {Observation of impurity-induced scale-free localization in
				a disordered non-{H}ermitian electrical circuit},}\ }\href {\doibase
		10.15302/frontphys.2025.014203} {\bibfield  {journal} {\bibinfo  {journal}
			{Front. Phys.}\ }\textbf {\bibinfo {volume} {20}},\ \bibinfo {pages} {014203}
		(\bibinfo {year} {2025}{\natexlab{b}})}\BibitemShut {NoStop}%
	\bibitem [{\citenamefont {Li}\ \emph {et~al.}(2024{\natexlab{b}})\citenamefont
		{Li}, \citenamefont {Liu},\ and\ \citenamefont {Liu}}]{arXiv:2403.07459}%
	\BibitemOpen
	\bibfield  {author} {\bibinfo {author} {\bibfnamefont {X.}~\bibnamefont
			{Li}}, \bibinfo {author} {\bibfnamefont {J.}~\bibnamefont {Liu}}, \ and\
		\bibinfo {author} {\bibfnamefont {T.}~\bibnamefont {Liu}},\ }\bibfield
	{title} {\enquote {\bibinfo {title} {Localization-delocalization transitions
				in non-{H}ermitian {Aharonov-Bohm} cages},}\ }\href
	{https://doi.org/10.48550/arXiv.2403.07459} {\bibfield  {journal} {\bibinfo
			{journal} {Front. Phys.}\ }\textbf {\bibinfo {volume} {19}},\ \bibinfo
		{pages} {33211} (\bibinfo {year} {2024}{\natexlab{b}})}\BibitemShut {NoStop}%
	\bibitem [{\citenamefont {Jin}\ \emph {et~al.}(2025)\citenamefont {Jin},
		\citenamefont {Liu}, \citenamefont {Wang}, \citenamefont {Zhang},
		\citenamefont {Huang}, \citenamefont {Wei}, \citenamefont {Ju}, \citenamefont
		{Yang}, \citenamefont {Liu},\ and\ \citenamefont {Nori}}]{arXiv:2311.03777}%
	\BibitemOpen
	\bibfield  {author} {\bibinfo {author} {\bibfnamefont {W.-W.}\ \bibnamefont
			{Jin}}, \bibinfo {author} {\bibfnamefont {J.}~\bibnamefont {Liu}}, \bibinfo
		{author} {\bibfnamefont {X.}~\bibnamefont {Wang}}, \bibinfo {author}
		{\bibfnamefont {Y.-R.}\ \bibnamefont {Zhang}}, \bibinfo {author}
		{\bibfnamefont {X.}~\bibnamefont {Huang}}, \bibinfo {author} {\bibfnamefont
			{X.}~\bibnamefont {Wei}}, \bibinfo {author} {\bibfnamefont {W.}~\bibnamefont
			{Ju}}, \bibinfo {author} {\bibfnamefont {Z.}~\bibnamefont {Yang}}, \bibinfo
		{author} {\bibfnamefont {T.}~\bibnamefont {Liu}}, \ and\ \bibinfo {author}
		{\bibfnamefont {F.}~\bibnamefont {Nori}},\ }\bibfield  {title} {\enquote
		{\bibinfo {title} {Anderson delocalization in strongly coupled disordered
				non-{H}ermitian chains},}\ }\href {\doibase 10.1103/lpm2-vcb4} {\bibfield
		{journal} {\bibinfo  {journal} {Phys. Rev. Lett.}\ }\textbf {\bibinfo
			{volume} {135}},\ \bibinfo {pages} {076602} (\bibinfo {year}
		{2025})}\BibitemShut {NoStop}%
	\bibitem [{\citenamefont {Parto}\ \emph {et~al.}(2023)\citenamefont {Parto},
		\citenamefont {Leefmans}, \citenamefont {Williams}, \citenamefont {Nori},\
		and\ \citenamefont {Marandi}}]{Parto2023}%
	\BibitemOpen
	\bibfield  {author} {\bibinfo {author} {\bibfnamefont {M.}~\bibnamefont
			{Parto}}, \bibinfo {author} {\bibfnamefont {C.}~\bibnamefont {Leefmans}},
		\bibinfo {author} {\bibfnamefont {J.}~\bibnamefont {Williams}}, \bibinfo
		{author} {\bibfnamefont {F.}~\bibnamefont {Nori}}, \ and\ \bibinfo {author}
		{\bibfnamefont {A.}~\bibnamefont {Marandi}},\ }\bibfield  {title} {\enquote
		{\bibinfo {title} {Non-{A}belian effects in dissipative photonic topological
				lattices},}\ }\href {http://dx.doi.org/10.1038/s41467-023-37065-z} {\bibfield
		{journal} {\bibinfo  {journal} {Nat. Commun.}\ }\textbf {\bibinfo {volume}
			{14}},\ \bibinfo {pages} {1440} (\bibinfo {year} {2023})}\BibitemShut
	{NoStop}%
	\bibitem [{\citenamefont {Wang}\ \emph {et~al.}(2024)\citenamefont {Wang},
		\citenamefont {Song},\ and\ \citenamefont {Wang}}]{PhysRevX.14.021011}%
	\BibitemOpen
	\bibfield  {author} {\bibinfo {author} {\bibfnamefont {H.-Y.}\ \bibnamefont
			{Wang}}, \bibinfo {author} {\bibfnamefont {F.}~\bibnamefont {Song}}, \ and\
		\bibinfo {author} {\bibfnamefont {Z.}~\bibnamefont {Wang}},\ }\bibfield
	{title} {\enquote {\bibinfo {title} {Amoeba formulation of non-{B}loch band
				theory in arbitrary dimensions},}\ }\href {\doibase
		10.1103/PhysRevX.14.021011} {\bibfield  {journal} {\bibinfo  {journal} {Phys.
				Rev. X}\ }\textbf {\bibinfo {volume} {14}},\ \bibinfo {pages} {021011}
		(\bibinfo {year} {2024})}\BibitemShut {NoStop}%
	\bibitem [{\citenamefont {Guo}\ \emph {et~al.}(2024)\citenamefont {Guo},
		\citenamefont {Su}, \citenamefont {Wang}, \citenamefont {Li}, \citenamefont
		{Wang}, \citenamefont {Ruan}, \citenamefont {Du}, \citenamefont {Zheng},
		\citenamefont {Chen},\ and\ \citenamefont {Hu}}]{Guo2024}%
	\BibitemOpen
	\bibfield  {author} {\bibinfo {author} {\bibfnamefont {C.-X.}\ \bibnamefont
			{Guo}}, \bibinfo {author} {\bibfnamefont {L.}~\bibnamefont {Su}}, \bibinfo
		{author} {\bibfnamefont {Y.}~\bibnamefont {Wang}}, \bibinfo {author}
		{\bibfnamefont {L.}~\bibnamefont {Li}}, \bibinfo {author} {\bibfnamefont
			{J.}~\bibnamefont {Wang}}, \bibinfo {author} {\bibfnamefont {X.}~\bibnamefont
			{Ruan}}, \bibinfo {author} {\bibfnamefont {Y.}~\bibnamefont {Du}}, \bibinfo
		{author} {\bibfnamefont {D.}~\bibnamefont {Zheng}}, \bibinfo {author}
		{\bibfnamefont {S.}~\bibnamefont {Chen}}, \ and\ \bibinfo {author}
		{\bibfnamefont {H.}~\bibnamefont {Hu}},\ }\bibfield  {title} {\enquote
		{\bibinfo {title} {Scale-tailored localization and its observation in
				non-{H}ermitian electrical circuits},}\ }\href {\doibase
		10.1038/s41467-024-53434-8} {\bibfield  {journal} {\bibinfo  {journal} {Nat.
				Commun.}\ }\textbf {\bibinfo {volume} {15}},\ \bibinfo {pages} {9120}
		(\bibinfo {year} {2024})}\BibitemShut {NoStop}%
	\bibitem [{\citenamefont {Liang}\ \emph {et~al.}(2025)\citenamefont {Liang},
		\citenamefont {Ou}, \citenamefont {Li},\ and\ \citenamefont
		{Xu}}]{1bvp-p2cz}%
	\BibitemOpen
	\bibfield  {author} {\bibinfo {author} {\bibfnamefont {H.-Q.}\ \bibnamefont
			{Liang}}, \bibinfo {author} {\bibfnamefont {Z.}~\bibnamefont {Ou}}, \bibinfo
		{author} {\bibfnamefont {L.}~\bibnamefont {Li}}, \ and\ \bibinfo {author}
		{\bibfnamefont {G.-F.}\ \bibnamefont {Xu}},\ }\bibfield  {title} {\enquote
		{\bibinfo {title} {Intrinsic perturbation induced anomalous higher-order
				boundary states in non-{H}ermitian systems},}\ }\href {\doibase
		10.1103/1bvp-p2cz} {\bibfield  {journal} {\bibinfo  {journal} {Phys. Rev. B}\
		}\textbf {\bibinfo {volume} {111}},\ \bibinfo {pages} {L241112} (\bibinfo
		{year} {2025})}\BibitemShut {NoStop}%
	\bibitem [{\citenamefont {Li}\ \emph {et~al.}(2020)\citenamefont {Li},
		\citenamefont {Lee}, \citenamefont {Mu},\ and\ \citenamefont
		{Gong}}]{Li2020}%
	\BibitemOpen
	\bibfield  {author} {\bibinfo {author} {\bibfnamefont {L.}~\bibnamefont
			{Li}}, \bibinfo {author} {\bibfnamefont {C.~H.}\ \bibnamefont {Lee}},
		\bibinfo {author} {\bibfnamefont {S.}~\bibnamefont {Mu}}, \ and\ \bibinfo
		{author} {\bibfnamefont {J.}~\bibnamefont {Gong}},\ }\bibfield  {title}
	{\enquote {\bibinfo {title} {Critical non-{H}ermitian skin effect},}\ }\href
	{\doibase 10.1038/s41467-020-18917-4} {\bibfield  {journal} {\bibinfo
			{journal} {Nat. Commun.}\ }\textbf {\bibinfo {volume} {11}},\ \bibinfo
		{pages} {5491} (\bibinfo {year} {2020})}\BibitemShut {NoStop}%
	\bibitem [{\citenamefont {Li}\ \emph {et~al.}(2023{\natexlab{b}})\citenamefont
		{Li}, \citenamefont {Wang}, \citenamefont {Song},\ and\ \citenamefont
		{Wang}}]{PhysRevB.108.L161409}%
	\BibitemOpen
	\bibfield  {author} {\bibinfo {author} {\bibfnamefont {B.}~\bibnamefont
			{Li}}, \bibinfo {author} {\bibfnamefont {H.-R.}\ \bibnamefont {Wang}},
		\bibinfo {author} {\bibfnamefont {F.}~\bibnamefont {Song}}, \ and\ \bibinfo
		{author} {\bibfnamefont {Z.}~\bibnamefont {Wang}},\ }\bibfield  {title}
	{\enquote {\bibinfo {title} {Scale-free localization and $\mathcal{PT}$
				symmetry breaking from local non-hermiticity},}\ }\href {\doibase
		10.1103/PhysRevB.108.L161409} {\bibfield  {journal} {\bibinfo  {journal}
			{Phys. Rev. B}\ }\textbf {\bibinfo {volume} {108}},\ \bibinfo {pages}
		{L161409} (\bibinfo {year} {2023}{\natexlab{b}})}\BibitemShut {NoStop}%
	\bibitem [{\citenamefont {Cai}\ \emph {et~al.}(2024)\citenamefont {Cai},
		\citenamefont {Liu},\ and\ \citenamefont {Yang}}]{PhysRevA.109.063329}%
	\BibitemOpen
	\bibfield  {author} {\bibinfo {author} {\bibfnamefont {Z.-F.}\ \bibnamefont
			{Cai}}, \bibinfo {author} {\bibfnamefont {T.}~\bibnamefont {Liu}}, \ and\
		\bibinfo {author} {\bibfnamefont {Z.}~\bibnamefont {Yang}},\ }\bibfield
	{title} {\enquote {\bibinfo {title} {Non-{H}ermitian skin effect in
				periodically driven dissipative ultracold atoms},}\ }\href {\doibase
		10.1103/PhysRevA.109.063329} {\bibfield  {journal} {\bibinfo  {journal}
			{Phys. Rev. A}\ }\textbf {\bibinfo {volume} {109}},\ \bibinfo {pages}
		{063329} (\bibinfo {year} {2024})}\BibitemShut {NoStop}%
	\bibitem [{\citenamefont {Kawabata}\ \emph {et~al.}(2023)\citenamefont
		{Kawabata}, \citenamefont {Numasawa},\ and\ \citenamefont
		{Ryu}}]{PhysRevX.13.021007}%
	\BibitemOpen
	\bibfield  {author} {\bibinfo {author} {\bibfnamefont {K.}~\bibnamefont
			{Kawabata}}, \bibinfo {author} {\bibfnamefont {T.}~\bibnamefont {Numasawa}},
		\ and\ \bibinfo {author} {\bibfnamefont {S.}~\bibnamefont {Ryu}},\ }\bibfield
	{title} {\enquote {\bibinfo {title} {Entanglement phase transition induced
				by the non-{H}ermitian skin effect},}\ }\href {\doibase
		10.1103/PhysRevX.13.021007} {\bibfield  {journal} {\bibinfo  {journal} {Phys.
				Rev. X}\ }\textbf {\bibinfo {volume} {13}},\ \bibinfo {pages} {021007}
		(\bibinfo {year} {2023})}\BibitemShut {NoStop}%
	\bibitem [{\citenamefont {Xue}\ \emph {et~al.}(2022)\citenamefont {Xue},
		\citenamefont {Hu}, \citenamefont {Song},\ and\ \citenamefont
		{Wang}}]{PhysRevLett.128.120401}%
	\BibitemOpen
	\bibfield  {author} {\bibinfo {author} {\bibfnamefont {W.-T.}\ \bibnamefont
			{Xue}}, \bibinfo {author} {\bibfnamefont {Y.-M.}\ \bibnamefont {Hu}},
		\bibinfo {author} {\bibfnamefont {F.}~\bibnamefont {Song}}, \ and\ \bibinfo
		{author} {\bibfnamefont {Z.}~\bibnamefont {Wang}},\ }\bibfield  {title}
	{\enquote {\bibinfo {title} {Non-{H}ermitian edge burst},}\ }\href {\doibase
		10.1103/PhysRevLett.128.120401} {\bibfield  {journal} {\bibinfo  {journal}
			{Phys. Rev. Lett.}\ }\textbf {\bibinfo {volume} {128}},\ \bibinfo {pages}
		{120401} (\bibinfo {year} {2022})}\BibitemShut {NoStop}%
	\bibitem [{\citenamefont {Xue}\ \emph {et~al.}(2025)\citenamefont {Xue},
		\citenamefont {Song}, \citenamefont {Hu},\ and\ \citenamefont
		{Wang}}]{arxiv.2503.13671}%
	\BibitemOpen
	\bibfield  {author} {\bibinfo {author} {\bibfnamefont {W.-T.}\ \bibnamefont
			{Xue}}, \bibinfo {author} {\bibfnamefont {F.}~\bibnamefont {Song}}, \bibinfo
		{author} {\bibfnamefont {Y.-M.}\ \bibnamefont {Hu}}, \ and\ \bibinfo {author}
		{\bibfnamefont {Z.}~\bibnamefont {Wang}},\ }\bibfield  {title} {\enquote
		{\bibinfo {title} {Non-{B}loch edge dynamics of non-{H}ermitian lattices},}\
	}\href {https://arxiv.org/abs/2503.13671} {\bibfield  {journal} {\bibinfo
			{journal} {arXiv.2503.13671}\ } (\bibinfo {year} {2025})}\BibitemShut
	{NoStop}%
	\bibitem [{\citenamefont {Yang}\ and\ \citenamefont
		{Lee}(2024)}]{PhysRevLett.133.136602}%
	\BibitemOpen
	\bibfield  {author} {\bibinfo {author} {\bibfnamefont {M.}~\bibnamefont
			{Yang}}\ and\ \bibinfo {author} {\bibfnamefont {C.~H.}\ \bibnamefont {Lee}},\
	}\bibfield  {title} {\enquote {\bibinfo {title} {Percolation-induced
				$\mathcal{P}\mathcal{T}$ symmetry breaking},}\ }\href {\doibase
		10.1103/PhysRevLett.133.136602} {\bibfield  {journal} {\bibinfo  {journal}
			{Phys. Rev. Lett.}\ }\textbf {\bibinfo {volume} {133}},\ \bibinfo {pages}
		{136602} (\bibinfo {year} {2024})}\BibitemShut {NoStop}%
	\bibitem [{\citenamefont {Yuce}\ and\ \citenamefont
		{Ramezani}(2023)}]{PhysRevB.107.L140302}%
	\BibitemOpen
	\bibfield  {author} {\bibinfo {author} {\bibfnamefont {C.}~\bibnamefont
			{Yuce}}\ and\ \bibinfo {author} {\bibfnamefont {H.}~\bibnamefont
			{Ramezani}},\ }\bibfield  {title} {\enquote {\bibinfo {title}
			{Non-{H}ermitian edge burst without skin localization},}\ }\href {\doibase
		10.1103/PhysRevB.107.L140302} {\bibfield  {journal} {\bibinfo  {journal}
			{Phys. Rev. B}\ }\textbf {\bibinfo {volume} {107}},\ \bibinfo {pages}
		{L140302} (\bibinfo {year} {2023})}\BibitemShut {NoStop}%
	\bibitem [{\citenamefont {Xiao}\ \emph {et~al.}(2024)\citenamefont {Xiao},
		\citenamefont {Xue}, \citenamefont {Song}, \citenamefont {Hu}, \citenamefont
		{Yi}, \citenamefont {Wang},\ and\ \citenamefont
		{Xue}}]{PhysRevLett.133.070801}%
	\BibitemOpen
	\bibfield  {author} {\bibinfo {author} {\bibfnamefont {L.}~\bibnamefont
			{Xiao}}, \bibinfo {author} {\bibfnamefont {W.-T.}\ \bibnamefont {Xue}},
		\bibinfo {author} {\bibfnamefont {F.}~\bibnamefont {Song}}, \bibinfo {author}
		{\bibfnamefont {Y.-M.}\ \bibnamefont {Hu}}, \bibinfo {author} {\bibfnamefont
			{W.}~\bibnamefont {Yi}}, \bibinfo {author} {\bibfnamefont {Z.}~\bibnamefont
			{Wang}}, \ and\ \bibinfo {author} {\bibfnamefont {P.}~\bibnamefont {Xue}},\
	}\bibfield  {title} {\enquote {\bibinfo {title} {Observation of
				non-{H}ermitian edge burst in quantum dynamics},}\ }\href {\doibase
		10.1103/PhysRevLett.133.070801} {\bibfield  {journal} {\bibinfo  {journal}
			{Phys. Rev. Lett.}\ }\textbf {\bibinfo {volume} {133}},\ \bibinfo {pages}
		{070801} (\bibinfo {year} {2024})}\BibitemShut {NoStop}%
	\bibitem [{\citenamefont {Yang}\ and\ \citenamefont
		{Fang}(2025)}]{arxiv.2503.11505}%
	\BibitemOpen
	\bibfield  {author} {\bibinfo {author} {\bibfnamefont {T.-H.}\ \bibnamefont
			{Yang}}\ and\ \bibinfo {author} {\bibfnamefont {C.}~\bibnamefont {Fang}},\
	}\bibfield  {title} {\enquote {\bibinfo {title} {Real-time edge dynamics of
				non-{H}ermitian lattices},}\ }\href {https://arxiv.org/abs/2503.11505}
	{\bibfield  {journal} {\bibinfo  {journal} {arXiv.2503.11505}\ } (\bibinfo
		{year} {2025})}\BibitemShut {NoStop}%
	\bibitem [{\citenamefont {Zhang}\ \emph {et~al.}(2022)\citenamefont {Zhang},
		\citenamefont {Yang},\ and\ \citenamefont {Fang}}]{Zhang2022}%
	\BibitemOpen
	\bibfield  {author} {\bibinfo {author} {\bibfnamefont {K.}~\bibnamefont
			{Zhang}}, \bibinfo {author} {\bibfnamefont {Z.}~\bibnamefont {Yang}}, \ and\
		\bibinfo {author} {\bibfnamefont {C.}~\bibnamefont {Fang}},\ }\bibfield
	{title} {\enquote {\bibinfo {title} {Universal non-{H}ermitian skin effect in
				two and higher dimensions},}\ }\href {\doibase 10.1038/s41467-022-30161-6}
	{\bibfield  {journal} {\bibinfo  {journal} {Nat. Commun.}\ }\textbf {\bibinfo
			{volume} {13}},\ \bibinfo {pages} {2496} (\bibinfo {year}
		{2022})}\BibitemShut {NoStop}%
	\bibitem [{\citenamefont {Wang}\ \emph {et~al.}(2023)\citenamefont {Wang},
		\citenamefont {Hu}, \citenamefont {Wang}, \citenamefont {Ma},\ and\
		\citenamefont {Ding}}]{PhysRevLett.131.207201}%
	\BibitemOpen
	\bibfield  {author} {\bibinfo {author} {\bibfnamefont {W.}~\bibnamefont
			{Wang}}, \bibinfo {author} {\bibfnamefont {M.}~\bibnamefont {Hu}}, \bibinfo
		{author} {\bibfnamefont {X.}~\bibnamefont {Wang}}, \bibinfo {author}
		{\bibfnamefont {G.}~\bibnamefont {Ma}}, \ and\ \bibinfo {author}
		{\bibfnamefont {K.}~\bibnamefont {Ding}},\ }\bibfield  {title} {\enquote
		{\bibinfo {title} {Experimental realization of geometry-dependent skin effect
				in a reciprocal two-dimensional lattice},}\ }\href {\doibase
		10.1103/PhysRevLett.131.207201} {\bibfield  {journal} {\bibinfo  {journal}
			{Phys. Rev. Lett.}\ }\textbf {\bibinfo {volume} {131}},\ \bibinfo {pages}
		{207201} (\bibinfo {year} {2023})}\BibitemShut {NoStop}%
	\bibitem [{\citenamefont {Zhou}\ \emph {et~al.}(2023)\citenamefont {Zhou},
		\citenamefont {Wu}, \citenamefont {Pu}, \citenamefont {Lu}, \citenamefont
		{Huang}, \citenamefont {Deng}, \citenamefont {Ke},\ and\ \citenamefont
		{Liu}}]{Zhou2023}%
	\BibitemOpen
	\bibfield  {author} {\bibinfo {author} {\bibfnamefont {Q.}~\bibnamefont
			{Zhou}}, \bibinfo {author} {\bibfnamefont {J.}~\bibnamefont {Wu}}, \bibinfo
		{author} {\bibfnamefont {Z.}~\bibnamefont {Pu}}, \bibinfo {author}
		{\bibfnamefont {J.}~\bibnamefont {Lu}}, \bibinfo {author} {\bibfnamefont
			{X.}~\bibnamefont {Huang}}, \bibinfo {author} {\bibfnamefont
			{W.}~\bibnamefont {Deng}}, \bibinfo {author} {\bibfnamefont {M.}~\bibnamefont
			{Ke}}, \ and\ \bibinfo {author} {\bibfnamefont {Z.}~\bibnamefont {Liu}},\
	}\bibfield  {title} {\enquote {\bibinfo {title} {Observation of
				geometry-dependent skin effect in non-{H}ermitian phononic crystals with
				exceptional points},}\ }\href {\doibase 10.1038/s41467-023-40236-7}
	{\bibfield  {journal} {\bibinfo  {journal} {Nat. Commun.}\ }\textbf {\bibinfo
			{volume} {14}},\ \bibinfo {pages} {4569} (\bibinfo {year}
		{2023})}\BibitemShut {NoStop}%
	\bibitem [{\citenamefont {Wan}\ \emph {et~al.}(2023)\citenamefont {Wan},
		\citenamefont {Zhang}, \citenamefont {Li}, \citenamefont {Yang},\ and\
		\citenamefont {Yang}}]{Wan2023}%
	\BibitemOpen
	\bibfield  {author} {\bibinfo {author} {\bibfnamefont {T.}~\bibnamefont
			{Wan}}, \bibinfo {author} {\bibfnamefont {K.}~\bibnamefont {Zhang}}, \bibinfo
		{author} {\bibfnamefont {J.}~\bibnamefont {Li}}, \bibinfo {author}
		{\bibfnamefont {Z.}~\bibnamefont {Yang}}, \ and\ \bibinfo {author}
		{\bibfnamefont {Z.}~\bibnamefont {Yang}},\ }\bibfield  {title} {\enquote
		{\bibinfo {title} {Observation of the geometry-dependent skin effect and
				dynamical degeneracy splitting},}\ }\href {\doibase
		10.1016/j.scib.2023.09.013} {\bibfield  {journal} {\bibinfo  {journal} {Sci.
				Bull.}\ }\textbf {\bibinfo {volume} {68}},\ \bibinfo {pages} {2330} (\bibinfo
		{year} {2023})}\BibitemShut {NoStop}%
	\bibitem [{\citenamefont {Zhang}\ \emph {et~al.}(2025)\citenamefont {Zhang},
		\citenamefont {Shu},\ and\ \citenamefont {Sun}}]{cwwd-bclc}%
	\BibitemOpen
	\bibfield  {author} {\bibinfo {author} {\bibfnamefont {K.}~\bibnamefont
			{Zhang}}, \bibinfo {author} {\bibfnamefont {C.}~\bibnamefont {Shu}}, \ and\
		\bibinfo {author} {\bibfnamefont {K.}~\bibnamefont {Sun}},\ }\bibfield
	{title} {\enquote {\bibinfo {title} {Algebraic non-{H}ermitian skin effect
				and generalized {F}ermi surface formula in arbitrary dimensions},}\ }\href
	{\doibase 10.1103/cwwd-bclc} {\bibfield  {journal} {\bibinfo  {journal}
			{Phys. Rev. X}\ }\textbf {\bibinfo {volume} {15}},\ \bibinfo {pages} {031039}
		(\bibinfo {year} {2025})}\BibitemShut {NoStop}%
	\bibitem [{\citenamefont {M.~Li}(2025)}]{arXiv:2501.13440}%
	\BibitemOpen
	\bibfield  {author} {\bibinfo {author} {\bibfnamefont {K.~Ding}\ \bibnamefont
			{M.~Li}, \bibfnamefont {J.~Lin}},\ }\bibfield  {title} {\enquote {\bibinfo
			{title} {Algebraic skin effect in two-dimensional non-{H}ermitian
				metamaterials},}\ }\href {https://doi.org/10.48550/arXiv.2501.13440}
	{\bibfield  {journal} {\bibinfo  {journal} {arXiv:2501.13440}\ } (\bibinfo
		{year} {2025})}\BibitemShut {NoStop}%
	\bibitem [{\citenamefont {C.~Shu}(2024)}]{arXiv:2409.13623}%
	\BibitemOpen
	\bibfield  {author} {\bibinfo {author} {\bibfnamefont {K.~Sun}\ \bibnamefont
			{C.~Shu}, \bibfnamefont {K.~Zhang}},\ }\bibfield  {title} {\enquote {\bibinfo
			{title} {Ultra spectral sensitivity and non-local bi-impurity bound states
				from quasi-long-range non-{H}ermitian skin modes},}\ }\href
	{https://doi.org/10.48550/arXiv.2409.13623} {\bibfield  {journal} {\bibinfo
			{journal} {arXiv:2409.13623}\ } (\bibinfo {year} {2024})}\BibitemShut
	{NoStop}%
	\bibitem [{\citenamefont {Wang}\ and\ \citenamefont
		{Clerk}(2019)}]{PhysRevA.99.063834}%
	\BibitemOpen
	\bibfield  {author} {\bibinfo {author} {\bibfnamefont {Y.-X.}\ \bibnamefont
			{Wang}}\ and\ \bibinfo {author} {\bibfnamefont {A.~A.}\ \bibnamefont
			{Clerk}},\ }\bibfield  {title} {\enquote {\bibinfo {title} {Non-{H}ermitian
				dynamics without dissipation in quantum systems},}\ }\href {\doibase
		10.1103/PhysRevA.99.063834} {\bibfield  {journal} {\bibinfo  {journal} {Phys.
				Rev. A}\ }\textbf {\bibinfo {volume} {99}},\ \bibinfo {pages} {063834}
		(\bibinfo {year} {2019})}\BibitemShut {NoStop}%
	\bibitem [{\citenamefont {McDonald}\ \emph {et~al.}(2018)\citenamefont
		{McDonald}, \citenamefont {Pereg-Barnea},\ and\ \citenamefont
		{Clerk}}]{PhysRevX.8.041031}%
	\BibitemOpen
	\bibfield  {author} {\bibinfo {author} {\bibfnamefont {A.}~\bibnamefont
			{McDonald}}, \bibinfo {author} {\bibfnamefont {T.}~\bibnamefont
			{Pereg-Barnea}}, \ and\ \bibinfo {author} {\bibfnamefont {A.~A.}\
			\bibnamefont {Clerk}},\ }\bibfield  {title} {\enquote {\bibinfo {title}
			{Phase-dependent chiral transport and effective non-{H}ermitian dynamics in a
				bosonic {K}itaev-{M}ajorana chain},}\ }\href {\doibase
		10.1103/PhysRevX.8.041031} {\bibfield  {journal} {\bibinfo  {journal} {Phys.
				Rev. X}\ }\textbf {\bibinfo {volume} {8}},\ \bibinfo {pages} {041031}
		(\bibinfo {year} {2018})}\BibitemShut {NoStop}%
	\bibitem [{\citenamefont {Flynn}\ \emph {et~al.}(2020)\citenamefont {Flynn},
		\citenamefont {Cobanera},\ and\ \citenamefont {Viola}}]{Flynn2020}%
	\BibitemOpen
	\bibfield  {author} {\bibinfo {author} {\bibfnamefont {V.~P.}\ \bibnamefont
			{Flynn}}, \bibinfo {author} {\bibfnamefont {E.}~\bibnamefont {Cobanera}}, \
		and\ \bibinfo {author} {\bibfnamefont {L.}~\bibnamefont {Viola}},\ }\bibfield
	{title} {\enquote {\bibinfo {title} {Deconstructing effective
				non-{H}ermitian dynamics in quadratic bosonic {H}amiltonians},}\ }\href
	{\doibase 10.1088/1367-2630/ab9e87} {\bibfield  {journal} {\bibinfo
			{journal} {New J. Phys.}\ }\textbf {\bibinfo {volume} {22}},\ \bibinfo
		{pages} {083004} (\bibinfo {year} {2020})}\BibitemShut {NoStop}%
	\bibitem [{\citenamefont {Wang}\ \emph {et~al.}(2022)\citenamefont {Wang},
		\citenamefont {Zhu}, \citenamefont {Wang}, \citenamefont {Zhang},\ and\
		\citenamefont {Chong}}]{PhysRevB.106.024301}%
	\BibitemOpen
	\bibfield  {author} {\bibinfo {author} {\bibfnamefont {Q.}~\bibnamefont
			{Wang}}, \bibinfo {author} {\bibfnamefont {C.}~\bibnamefont {Zhu}}, \bibinfo
		{author} {\bibfnamefont {Y.}~\bibnamefont {Wang}}, \bibinfo {author}
		{\bibfnamefont {B.}~\bibnamefont {Zhang}}, \ and\ \bibinfo {author}
		{\bibfnamefont {Y.~D.}\ \bibnamefont {Chong}},\ }\bibfield  {title} {\enquote
		{\bibinfo {title} {Amplification of quantum signals by the non-{H}ermitian
				skin effect},}\ }\href {\doibase 10.1103/PhysRevB.106.024301} {\bibfield
		{journal} {\bibinfo  {journal} {Phys. Rev. B}\ }\textbf {\bibinfo {volume}
			{106}},\ \bibinfo {pages} {024301} (\bibinfo {year} {2022})}\BibitemShut
	{NoStop}%
	\bibitem [{\citenamefont {Okuma}(2022)}]{PhysRevB.105.224301}%
	\BibitemOpen
	\bibfield  {author} {\bibinfo {author} {\bibfnamefont {N.}~\bibnamefont
			{Okuma}},\ }\bibfield  {title} {\enquote {\bibinfo {title}
			{Boundary-dependent dynamical instability of bosonic {G}reen's function:
				Dissipative {B}ogoliubov--de {G}ennes {H}amiltonian and its application to
				non-{H}ermitian skin effect},}\ }\href {\doibase 10.1103/PhysRevB.105.224301}
	{\bibfield  {journal} {\bibinfo  {journal} {Phys. Rev. B}\ }\textbf {\bibinfo
			{volume} {105}},\ \bibinfo {pages} {224301} (\bibinfo {year}
		{2022})}\BibitemShut {NoStop}%
	\bibitem [{\citenamefont {Luo}\ \emph {et~al.}(2022)\citenamefont {Luo},
		\citenamefont {Zhang},\ and\ \citenamefont {Du}}]{PhysRevLett.128.173602}%
	\BibitemOpen
	\bibfield  {author} {\bibinfo {author} {\bibfnamefont {X.-W.}\ \bibnamefont
			{Luo}}, \bibinfo {author} {\bibfnamefont {C.}~\bibnamefont {Zhang}}, \ and\
		\bibinfo {author} {\bibfnamefont {S.}~\bibnamefont {Du}},\ }\bibfield
	{title} {\enquote {\bibinfo {title} {Quantum squeezing and sensing with
				pseudo-anti-parity-time symmetry},}\ }\href {\doibase
		10.1103/PhysRevLett.128.173602} {\bibfield  {journal} {\bibinfo  {journal}
			{Phys. Rev. Lett.}\ }\textbf {\bibinfo {volume} {128}},\ \bibinfo {pages}
		{173602} (\bibinfo {year} {2022})}\BibitemShut {NoStop}%
	\bibitem [{\citenamefont {Lv}\ and\ \citenamefont
		{Zhou}(2024)}]{PhysRevD.110.084039}%
	\BibitemOpen
	\bibfield  {author} {\bibinfo {author} {\bibfnamefont {C.}~\bibnamefont
			{Lv}}\ and\ \bibinfo {author} {\bibfnamefont {Q.}~\bibnamefont {Zhou}},\
	}\bibfield  {title} {\enquote {\bibinfo {title} {Complexity geometry in
				{H}ermitian and non-{H}ermitian quantum dynamics},}\ }\href {\doibase
		10.1103/PhysRevD.110.084039} {\bibfield  {journal} {\bibinfo  {journal}
			{Phys. Rev. D}\ }\textbf {\bibinfo {volume} {110}},\ \bibinfo {pages}
		{084039} (\bibinfo {year} {2024})}\BibitemShut {NoStop}%
	\bibitem [{\citenamefont {Flynn}\ \emph {et~al.}(2021)\citenamefont {Flynn},
		\citenamefont {Cobanera},\ and\ \citenamefont
		{Viola}}]{PhysRevLett.127.245701}%
	\BibitemOpen
	\bibfield  {author} {\bibinfo {author} {\bibfnamefont {V.~P.}\ \bibnamefont
			{Flynn}}, \bibinfo {author} {\bibfnamefont {E.}~\bibnamefont {Cobanera}}, \
		and\ \bibinfo {author} {\bibfnamefont {L.}~\bibnamefont {Viola}},\ }\bibfield
	{title} {\enquote {\bibinfo {title} {Topology by dissipation: {M}ajorana
				bosons in metastable quadratic {M}arkovian dynamics},}\ }\href {\doibase
		10.1103/PhysRevLett.127.245701} {\bibfield  {journal} {\bibinfo  {journal}
			{Phys. Rev. Lett.}\ }\textbf {\bibinfo {volume} {127}},\ \bibinfo {pages}
		{245701} (\bibinfo {year} {2021})}\BibitemShut {NoStop}%
	\bibitem [{\citenamefont {Wan}\ and\ \citenamefont
		{L\"u}(2023)}]{PhysRevLett.130.203605}%
	\BibitemOpen
	\bibfield  {author} {\bibinfo {author} {\bibfnamefont {L.-L.}\ \bibnamefont
			{Wan}}\ and\ \bibinfo {author} {\bibfnamefont {X.-Y.}\ \bibnamefont {L\"u}},\
	}\bibfield  {title} {\enquote {\bibinfo {title} {Quantum-squeezing-induced
				point-gap topology and skin effect},}\ }\href {\doibase
		10.1103/PhysRevLett.130.203605} {\bibfield  {journal} {\bibinfo  {journal}
			{Phys. Rev. Lett.}\ }\textbf {\bibinfo {volume} {130}},\ \bibinfo {pages}
		{203605} (\bibinfo {year} {2023})}\BibitemShut {NoStop}%
	\bibitem [{\citenamefont {Yokomizo}\ and\ \citenamefont
		{Murakami}(2021)}]{PhysRevB.103.165123}%
	\BibitemOpen
	\bibfield  {author} {\bibinfo {author} {\bibfnamefont {K.}~\bibnamefont
			{Yokomizo}}\ and\ \bibinfo {author} {\bibfnamefont {S.}~\bibnamefont
			{Murakami}},\ }\bibfield  {title} {\enquote {\bibinfo {title} {Non-{B}loch
				band theory in bosonic {B}ogoliubov--de {G}ennes systems},}\ }\href {\doibase
		10.1103/PhysRevB.103.165123} {\bibfield  {journal} {\bibinfo  {journal}
			{Phys. Rev. B}\ }\textbf {\bibinfo {volume} {103}},\ \bibinfo {pages}
		{165123} (\bibinfo {year} {2021})}\BibitemShut {NoStop}%
	\bibitem [{\citenamefont {Wanjura}\ \emph {et~al.}(2020)\citenamefont
		{Wanjura}, \citenamefont {Brunelli},\ and\ \citenamefont
		{Nunnenkamp}}]{Wanjura2020}%
	\BibitemOpen
	\bibfield  {author} {\bibinfo {author} {\bibfnamefont {C.~C.}\ \bibnamefont
			{Wanjura}}, \bibinfo {author} {\bibfnamefont {M.}~\bibnamefont {Brunelli}}, \
		and\ \bibinfo {author} {\bibfnamefont {A.}~\bibnamefont {Nunnenkamp}},\
	}\bibfield  {title} {\enquote {\bibinfo {title} {Topological framework for
				directional amplification in driven-dissipative cavity arrays},}\ }\href
	{\doibase 10.1038/s41467-020-16863-9} {\bibfield  {journal} {\bibinfo
			{journal} {Nat. Commun.}\ }\textbf {\bibinfo {volume} {11}},\ \bibinfo
		{pages} {3149} (\bibinfo {year} {2020})}\BibitemShut {NoStop}%
	\bibitem [{\citenamefont {Lieu}(2018)}]{PhysRevB.98.115135}%
	\BibitemOpen
	\bibfield  {author} {\bibinfo {author} {\bibfnamefont {S.}~\bibnamefont
			{Lieu}},\ }\bibfield  {title} {\enquote {\bibinfo {title} {Topological
				symmetry classes for non-{H}ermitian models and connections to the bosonic
				{B}ogoliubov--de {G}ennes equation},}\ }\href {\doibase
		10.1103/PhysRevB.98.115135} {\bibfield  {journal} {\bibinfo  {journal} {Phys.
				Rev. B}\ }\textbf {\bibinfo {volume} {98}},\ \bibinfo {pages} {115135}
		(\bibinfo {year} {2018})}\BibitemShut {NoStop}%
	\bibitem [{\citenamefont {Yang}(2020)}]{arxiv.2012.03333}%
	\BibitemOpen
	\bibfield  {author} {\bibinfo {author} {\bibfnamefont {Z.}~\bibnamefont
			{Yang}},\ }\bibfield  {title} {\enquote {\bibinfo {title} {Non-perturbative
				breakdown of {B}loch's theorem and {H}ermitian skin effects},}\ }\href
	{https://arxiv.org/abs/2012.03333} {\bibfield  {journal} {\bibinfo  {journal}
			{arXiv.2012.03333}\ } (\bibinfo {year} {2020})}\BibitemShut {NoStop}%
	\bibitem [{\citenamefont {Arandes}\ and\ \citenamefont
		{Bergholtz}(2025)}]{PhysRevResearch.7.013309}%
	\BibitemOpen
	\bibfield  {author} {\bibinfo {author} {\bibfnamefont {O.}~\bibnamefont
			{Arandes}}\ and\ \bibinfo {author} {\bibfnamefont {E.~J.}\ \bibnamefont
			{Bergholtz}},\ }\bibfield  {title} {\enquote {\bibinfo {title} {Quantum
				sensing with driven-dissipative {Su-Schrieffer-Heeger} lattices},}\ }\href
	{\doibase 10.1103/PhysRevResearch.7.013309} {\bibfield  {journal} {\bibinfo
			{journal} {Phys. Rev. Res.}\ }\textbf {\bibinfo {volume} {7}},\ \bibinfo
		{pages} {013309} (\bibinfo {year} {2025})}\BibitemShut {NoStop}%
	\bibitem [{\citenamefont {Estake}\ \emph {et~al.}(2025)\citenamefont {Estake},
		\citenamefont {Vishnu},\ and\ \citenamefont {Roy}}]{arxiv.2508.14560}%
	\BibitemOpen
	\bibfield  {author} {\bibinfo {author} {\bibfnamefont {K.~B.}\ \bibnamefont
			{Estake}}, \bibinfo {author} {\bibfnamefont {T.~R.}\ \bibnamefont {Vishnu}},
		\ and\ \bibinfo {author} {\bibfnamefont {D.}~\bibnamefont {Roy}},\ }\bibfield
	{title} {\enquote {\bibinfo {title} {From chiral topological dynamics to
				chiral topological amplification: {R}eal vs imaginary parameters in a
				{H}ermitian bosonic chain},}\ }\href {https://arxiv.org/abs/2508.14560}
	{\bibfield  {journal} {\bibinfo  {journal} {arXiv.2508.14560}\ } (\bibinfo
		{year} {2025})}\BibitemShut {NoStop}%
	\bibitem [{\citenamefont {Bestler}\ \emph {et~al.}(2025)\citenamefont
		{Bestler}, \citenamefont {Dikopoltsev},\ and\ \citenamefont
		{Zilberberg}}]{arxiv.2505.02776}%
	\BibitemOpen
	\bibfield  {author} {\bibinfo {author} {\bibfnamefont {M.}~\bibnamefont
			{Bestler}}, \bibinfo {author} {\bibfnamefont {A.}~\bibnamefont
			{Dikopoltsev}}, \ and\ \bibinfo {author} {\bibfnamefont {O.}~\bibnamefont
			{Zilberberg}},\ }\bibfield  {title} {\enquote {\bibinfo {title}
			{Non-{H}ermitian topology and skin modes in the continuum via parametric
				processes},}\ }\href {https://arxiv.org/abs/2505.02776} {\bibfield  {journal}
		{\bibinfo  {journal} {arXiv.2505.02776}\ } (\bibinfo {year}
		{2025})}\BibitemShut {NoStop}%
	\bibitem [{\citenamefont {Slim}\ \emph {et~al.}(2024)\citenamefont {Slim},
		\citenamefont {Wanjura}, \citenamefont {Brunelli}, \citenamefont {del Pino},
		\citenamefont {Nunnenkamp},\ and\ \citenamefont {Verhagen}}]{Slim2024}%
	\BibitemOpen
	\bibfield  {author} {\bibinfo {author} {\bibfnamefont {J.~J.}\ \bibnamefont
			{Slim}}, \bibinfo {author} {\bibfnamefont {C.~C.}\ \bibnamefont {Wanjura}},
		\bibinfo {author} {\bibfnamefont {M.}~\bibnamefont {Brunelli}}, \bibinfo
		{author} {\bibfnamefont {J.}~\bibnamefont {del Pino}}, \bibinfo {author}
		{\bibfnamefont {A.}~\bibnamefont {Nunnenkamp}}, \ and\ \bibinfo {author}
		{\bibfnamefont {E.}~\bibnamefont {Verhagen}},\ }\bibfield  {title} {\enquote
		{\bibinfo {title} {Optomechanical realization of the bosonic {K}itaev
				chain},}\ }\href {\doibase 10.1038/s41586-024-07174-w} {\bibfield  {journal}
		{\bibinfo  {journal} {Nature}\ }\textbf {\bibinfo {volume} {627}},\ \bibinfo
		{pages} {767} (\bibinfo {year} {2024})}\BibitemShut {NoStop}%
	\bibitem [{\citenamefont {Pocklington}\ \emph {et~al.}(2023)\citenamefont
		{Pocklington}, \citenamefont {Wang},\ and\ \citenamefont
		{Clerk}}]{PhysRevLett.130.123602}%
	\BibitemOpen
	\bibfield  {author} {\bibinfo {author} {\bibfnamefont {A.}~\bibnamefont
			{Pocklington}}, \bibinfo {author} {\bibfnamefont {Y.-X.}\ \bibnamefont
			{Wang}}, \ and\ \bibinfo {author} {\bibfnamefont {A.~A.}\ \bibnamefont
			{Clerk}},\ }\bibfield  {title} {\enquote {\bibinfo {title} {Dissipative
				pairing interactions: Quantum instabilities, topological light, and
				volume-law entanglement},}\ }\href {\doibase 10.1103/PhysRevLett.130.123602}
	{\bibfield  {journal} {\bibinfo  {journal} {Phys. Rev. Lett.}\ }\textbf
		{\bibinfo {volume} {130}},\ \bibinfo {pages} {123602} (\bibinfo {year}
		{2023})}\BibitemShut {NoStop}%
	\bibitem [{\citenamefont {Abdo}\ \emph {et~al.}(2013)\citenamefont {Abdo},
		\citenamefont {Kamal},\ and\ \citenamefont {Devoret}}]{PhysRevB.87.014508}%
	\BibitemOpen
	\bibfield  {author} {\bibinfo {author} {\bibfnamefont {B.}~\bibnamefont
			{Abdo}}, \bibinfo {author} {\bibfnamefont {A.}~\bibnamefont {Kamal}}, \ and\
		\bibinfo {author} {\bibfnamefont {M.}~\bibnamefont {Devoret}},\ }\bibfield
	{title} {\enquote {\bibinfo {title} {Nondegenerate three-wave mixing with the
				{J}osephson ring modulator},}\ }\href {\doibase 10.1103/PhysRevB.87.014508}
	{\bibfield  {journal} {\bibinfo  {journal} {Phys. Rev. B}\ }\textbf {\bibinfo
			{volume} {87}},\ \bibinfo {pages} {014508} (\bibinfo {year}
		{2013})}\BibitemShut {NoStop}%
	\bibitem [{\citenamefont {Busnaina}\ \emph {et~al.}(2024)\citenamefont
		{Busnaina}, \citenamefont {Shi}, \citenamefont {McDonald}, \citenamefont
		{Dubyna}, \citenamefont {Nsanzineza}, \citenamefont {Hung}, \citenamefont
		{Chang}, \citenamefont {Clerk},\ and\ \citenamefont {Wilson}}]{Busnaina2024}%
	\BibitemOpen
	\bibfield  {author} {\bibinfo {author} {\bibfnamefont {J.~H.}\ \bibnamefont
			{Busnaina}}, \bibinfo {author} {\bibfnamefont {Z.}~\bibnamefont {Shi}},
		\bibinfo {author} {\bibfnamefont {A.}~\bibnamefont {McDonald}}, \bibinfo
		{author} {\bibfnamefont {D.}~\bibnamefont {Dubyna}}, \bibinfo {author}
		{\bibfnamefont {I.}~\bibnamefont {Nsanzineza}}, \bibinfo {author}
		{\bibfnamefont {J.~S.~C.}\ \bibnamefont {Hung}}, \bibinfo {author}
		{\bibfnamefont {C.~W.~S.}\ \bibnamefont {Chang}}, \bibinfo {author}
		{\bibfnamefont {A.~A.}\ \bibnamefont {Clerk}}, \ and\ \bibinfo {author}
		{\bibfnamefont {C.~M.}\ \bibnamefont {Wilson}},\ }\bibfield  {title}
	{\enquote {\bibinfo {title} {Quantum simulation of the bosonic {K}itaev
				chain},}\ }\href {\doibase 10.1038/s41467-024-47186-8} {\bibfield  {journal}
		{\bibinfo  {journal} {Nat. Commun.}\ }\textbf {\bibinfo {volume} {15}},\
		\bibinfo {pages} {3065} (\bibinfo {year} {2024})}\BibitemShut {NoStop}%
	\bibitem [{\citenamefont {Villiers}\ \emph {et~al.}(2024)\citenamefont
		{Villiers}, \citenamefont {Smith}, \citenamefont {Petrescu}, \citenamefont
		{Borgognoni}, \citenamefont {Delbecq}, \citenamefont {Sarlette},
		\citenamefont {Mirrahimi}, \citenamefont {Campagne-Ibarcq}, \citenamefont
		{Kontos},\ and\ \citenamefont {Leghtas}}]{PRXQuantum.5.020306}%
	\BibitemOpen
	\bibfield  {author} {\bibinfo {author} {\bibfnamefont {M.}~\bibnamefont
			{Villiers}}, \bibinfo {author} {\bibfnamefont {W.C.}\ \bibnamefont {Smith}},
		\bibinfo {author} {\bibfnamefont {A.}~\bibnamefont {Petrescu}}, \bibinfo
		{author} {\bibfnamefont {A.}~\bibnamefont {Borgognoni}}, \bibinfo {author}
		{\bibfnamefont {M.}~\bibnamefont {Delbecq}}, \bibinfo {author} {\bibfnamefont
			{A.}~\bibnamefont {Sarlette}}, \bibinfo {author} {\bibfnamefont
			{M.}~\bibnamefont {Mirrahimi}}, \bibinfo {author} {\bibfnamefont
			{P.}~\bibnamefont {Campagne-Ibarcq}}, \bibinfo {author} {\bibfnamefont
			{T.}~\bibnamefont {Kontos}}, \ and\ \bibinfo {author} {\bibfnamefont
			{Z.}~\bibnamefont {Leghtas}},\ }\bibfield  {title} {\enquote {\bibinfo
			{title} {Dynamically enhancing qubit-photon interactions with
				antisqueezing},}\ }\href {\doibase 10.1103/PRXQuantum.5.020306} {\bibfield
		{journal} {\bibinfo  {journal} {PRX Quantum}\ }\textbf {\bibinfo {volume}
			{5}},\ \bibinfo {pages} {020306} (\bibinfo {year} {2024})}\BibitemShut
	{NoStop}%
	\bibitem [{\citenamefont {Safavi-Naeini}\ \emph {et~al.}(2013)\citenamefont
		{Safavi-Naeini}, \citenamefont {Gr\"{o}blacher}, \citenamefont {Hill},
		\citenamefont {Chan}, \citenamefont {Aspelmeyer},\ and\ \citenamefont
		{Painter}}]{SafaviNaeini2013}%
	\BibitemOpen
	\bibfield  {author} {\bibinfo {author} {\bibfnamefont {A.~H.}\ \bibnamefont
			{Safavi-Naeini}}, \bibinfo {author} {\bibfnamefont {S.}~\bibnamefont
			{Gr\"{o}blacher}}, \bibinfo {author} {\bibfnamefont {J.~T.}\ \bibnamefont
			{Hill}}, \bibinfo {author} {\bibfnamefont {J.}~\bibnamefont {Chan}}, \bibinfo
		{author} {\bibfnamefont {M.}~\bibnamefont {Aspelmeyer}}, \ and\ \bibinfo
		{author} {\bibfnamefont {O.}~\bibnamefont {Painter}},\ }\bibfield  {title}
	{\enquote {\bibinfo {title} {Squeezed light from a silicon micromechanical
				resonator},}\ }\href {\doibase 10.1038/nature12307} {\bibfield  {journal}
		{\bibinfo  {journal} {Nature}\ }\textbf {\bibinfo {volume} {500}},\ \bibinfo
		{pages} {185} (\bibinfo {year} {2013})}\BibitemShut {NoStop}%
	\bibitem [{\citenamefont {Marti}\ \emph {et~al.}(2024)\citenamefont {Marti},
		\citenamefont {von L\"{u}pke}, \citenamefont {Joshi}, \citenamefont {Yang},
		\citenamefont {Bild}, \citenamefont {Omahen}, \citenamefont {Chu},\ and\
		\citenamefont {Fadel}}]{Marti2024}%
	\BibitemOpen
	\bibfield  {author} {\bibinfo {author} {\bibfnamefont {S.}~\bibnamefont
			{Marti}}, \bibinfo {author} {\bibfnamefont {U.}~\bibnamefont {von
				L\"{u}pke}}, \bibinfo {author} {\bibfnamefont {O.}~\bibnamefont {Joshi}},
		\bibinfo {author} {\bibfnamefont {Y.}~\bibnamefont {Yang}}, \bibinfo {author}
		{\bibfnamefont {M.}~\bibnamefont {Bild}}, \bibinfo {author} {\bibfnamefont
			{A.}~\bibnamefont {Omahen}}, \bibinfo {author} {\bibfnamefont
			{Y.}~\bibnamefont {Chu}}, \ and\ \bibinfo {author} {\bibfnamefont
			{M.}~\bibnamefont {Fadel}},\ }\bibfield  {title} {\enquote {\bibinfo {title}
			{Quantum squeezing in a nonlinear mechanical oscillator},}\ }\href {\doibase
		10.1038/s41567-024-02545-6} {\bibfield  {journal} {\bibinfo  {journal} {Nat.
				Phys.}\ }\textbf {\bibinfo {volume} {20}},\ \bibinfo {pages} {1448} (\bibinfo
		{year} {2024})}\BibitemShut {NoStop}%
	\bibitem [{\citenamefont {Vaidya}\ \emph {et~al.}(2020)\citenamefont {Vaidya},
		\citenamefont {Morrison}, \citenamefont {Helt}, \citenamefont {Shahrokshahi},
		\citenamefont {Mahler}, \citenamefont {Collins}, \citenamefont {Tan},
		\citenamefont {Lavoie}, \citenamefont {Repingon}, \citenamefont {Menotti},
		\citenamefont {Quesada}, \citenamefont {Pooser}, \citenamefont {Lita},
		\citenamefont {Gerrits}, \citenamefont {Nam},\ and\ \citenamefont
		{Vernon}}]{Vaidya2020}%
	\BibitemOpen
	\bibfield  {author} {\bibinfo {author} {\bibfnamefont {V.~D.}\ \bibnamefont
			{Vaidya}}, \bibinfo {author} {\bibfnamefont {B.}~\bibnamefont {Morrison}},
		\bibinfo {author} {\bibfnamefont {L.~G.}\ \bibnamefont {Helt}}, \bibinfo
		{author} {\bibfnamefont {R.}~\bibnamefont {Shahrokshahi}}, \bibinfo {author}
		{\bibfnamefont {D.~H.}\ \bibnamefont {Mahler}}, \bibinfo {author}
		{\bibfnamefont {M.~J.}\ \bibnamefont {Collins}}, \bibinfo {author}
		{\bibfnamefont {K.}~\bibnamefont {Tan}}, \bibinfo {author} {\bibfnamefont
			{J.}~\bibnamefont {Lavoie}}, \bibinfo {author} {\bibfnamefont
			{A.}~\bibnamefont {Repingon}}, \bibinfo {author} {\bibfnamefont
			{M.}~\bibnamefont {Menotti}}, \bibinfo {author} {\bibfnamefont
			{N.}~\bibnamefont {Quesada}}, \bibinfo {author} {\bibfnamefont {R.~C.}\
			\bibnamefont {Pooser}}, \bibinfo {author} {\bibfnamefont {A.~E.}\
			\bibnamefont {Lita}}, \bibinfo {author} {\bibfnamefont {T.}~\bibnamefont
			{Gerrits}}, \bibinfo {author} {\bibfnamefont {S.~W.}\ \bibnamefont {Nam}}, \
		and\ \bibinfo {author} {\bibfnamefont {Z.}~\bibnamefont {Vernon}},\
	}\bibfield  {title} {\enquote {\bibinfo {title} {Broadband
				quadrature-squeezed vacuum and nonclassical photon number correlations from a
				nanophotonic device},}\ }\href {\doibase 10.1126/sciadv.aba9186} {\bibfield
		{journal} {\bibinfo  {journal} {Sci. Adv.}\ }\textbf {\bibinfo {volume}
			{6}},\ \bibinfo {pages} {eaba9186} (\bibinfo {year} {2020})}\BibitemShut
	{NoStop}%
	\bibitem [{\citenamefont {Nehra}\ \emph {et~al.}(2022)\citenamefont {Nehra},
		\citenamefont {Sekine}, \citenamefont {Ledezma}, \citenamefont {Guo},
		\citenamefont {Gray}, \citenamefont {Roy},\ and\ \citenamefont
		{Marandi}}]{Nehra2022}%
	\BibitemOpen
	\bibfield  {author} {\bibinfo {author} {\bibfnamefont {R.}~\bibnamefont
			{Nehra}}, \bibinfo {author} {\bibfnamefont {R.}~\bibnamefont {Sekine}},
		\bibinfo {author} {\bibfnamefont {L.}~\bibnamefont {Ledezma}}, \bibinfo
		{author} {\bibfnamefont {Q.}~\bibnamefont {Guo}}, \bibinfo {author}
		{\bibfnamefont {R.~M.}\ \bibnamefont {Gray}}, \bibinfo {author}
		{\bibfnamefont {A.}~\bibnamefont {Roy}}, \ and\ \bibinfo {author}
		{\bibfnamefont {A.}~\bibnamefont {Marandi}},\ }\bibfield  {title} {\enquote
		{\bibinfo {title} {Few-cycle vacuum squeezing in nanophotonics},}\ }\href
	{\doibase 10.1126/science.abo6213} {\bibfield  {journal} {\bibinfo  {journal}
			{Science}\ }\textbf {\bibinfo {volume} {377}},\ \bibinfo {pages} {1333}
		(\bibinfo {year} {2022})}\BibitemShut {NoStop}%
	\bibitem [{\citenamefont {Shindou}\ \emph {et~al.}(2013)\citenamefont
		{Shindou}, \citenamefont {Matsumoto}, \citenamefont {Murakami},\ and\
		\citenamefont {Ohe}}]{PhysRevB.87.174427}%
	\BibitemOpen
	\bibfield  {author} {\bibinfo {author} {\bibfnamefont {R.}~\bibnamefont
			{Shindou}}, \bibinfo {author} {\bibfnamefont {R.}~\bibnamefont {Matsumoto}},
		\bibinfo {author} {\bibfnamefont {S.}~\bibnamefont {Murakami}}, \ and\
		\bibinfo {author} {\bibfnamefont {J.}~\bibnamefont {Ohe}},\ }\bibfield
	{title} {\enquote {\bibinfo {title} {Topological chiral magnonic edge mode in
				a magnonic crystal},}\ }\href {\doibase 10.1103/PhysRevB.87.174427}
	{\bibfield  {journal} {\bibinfo  {journal} {Phys. Rev. B}\ }\textbf {\bibinfo
			{volume} {87}},\ \bibinfo {pages} {174427} (\bibinfo {year}
		{2013})}\BibitemShut {NoStop}%
	\bibitem [{\citenamefont {Yokomizo}\ and\ \citenamefont
		{Murakami}(2023)}]{PhysRevB.107.195112}%
	\BibitemOpen
	\bibfield  {author} {\bibinfo {author} {\bibfnamefont {K.}~\bibnamefont
			{Yokomizo}}\ and\ \bibinfo {author} {\bibfnamefont {S.}~\bibnamefont
			{Murakami}},\ }\bibfield  {title} {\enquote {\bibinfo {title} {Non-{B}loch
				bands in two-dimensional non-{H}ermitian systems},}\ }\href {\doibase
		10.1103/PhysRevB.107.195112} {\bibfield  {journal} {\bibinfo  {journal}
			{Phys. Rev. B}\ }\textbf {\bibinfo {volume} {107}},\ \bibinfo {pages}
		{195112} (\bibinfo {year} {2023})}\BibitemShut {NoStop}%
	\bibitem [{\citenamefont {Mac\'e}\ \emph {et~al.}(2019)\citenamefont {Mac\'e},
		\citenamefont {Alet},\ and\ \citenamefont
		{Laflorencie}}]{PhysRevLett.123.180601}%
	\BibitemOpen
	\bibfield  {author} {\bibinfo {author} {\bibfnamefont {N.}~\bibnamefont
			{Mac\'e}}, \bibinfo {author} {\bibfnamefont {F.}~\bibnamefont {Alet}}, \ and\
		\bibinfo {author} {\bibfnamefont {N.}~\bibnamefont {Laflorencie}},\
	}\bibfield  {title} {\enquote {\bibinfo {title} {Multifractal scalings across
				the many-body localization transition},}\ }\href {\doibase
		10.1103/PhysRevLett.123.180601} {\bibfield  {journal} {\bibinfo  {journal}
			{Phys. Rev. Lett.}\ }\textbf {\bibinfo {volume} {123}},\ \bibinfo {pages}
		{180601} (\bibinfo {year} {2019})}\BibitemShut {NoStop}%
\end{thebibliography}
%

\end{document}